\documentclass[aps, superscriptaddress,12pt]{revtex4-1}
\usepackage[colorlinks=true,allcolors=blue!92!black]{hyperref}
\usepackage{amssymb}
\usepackage{amsmath}
\usepackage{subcaption}
\usepackage{amsfonts}
\usepackage{xcolor}
\usepackage{epsfig}
\usepackage{color}
\usepackage{bbm}
\usepackage{bm}
\usepackage{tabularx}
\usepackage{multirow}
\usepackage{mathtools}
\usepackage{xfrac}
\usepackage{amsfonts}
\usepackage{latexsym}
\usepackage{amsmath,amssymb}
\usepackage{mathrsfs}
\usepackage{setspace}
\usepackage{color,xcolor}
\usepackage{bm}
\usepackage{slashed}
\usepackage{url}
\usepackage{empheq}
\usepackage{textcomp}
\usepackage{dcolumn}
\usepackage{ulem}
\usepackage{cancel}
\usepackage{braket}
\usepackage[title]{appendix}
\usepackage{lipsum}
\usepackage{ulem}
\usepackage{cancel}
\usepackage{color}
\numberwithin{equation}{section}

\def\beq{\begin{equation}}
	\def\eeq{\end{equation}}
\def\bea{\begin{eqnarray}}
	\def\eea{\end{eqnarray}}

\def\tmes{t_{\textrm{mes}}}
\def\tmic{t_{\textrm{mic}}}
\newcommand{\q}{\mathbf{q }}
\newcommand{\p}{\mathbf{ p}}
\newcommand{\x}{\mathbf{ x}}
\def\k{\textbf{k}}

\newcommand{\tr}[1]{\mathrm{Tr}#1}



\begin{document}

	\title{ \Large Open Quantum System Approach to the Gravitational Decoherence of Spin-1/2 Particles}
	
	\author{M. Sharifian}
%	\email[]{mohammadsharifian@ph.iut.ac.ir}
	\affiliation{Department of Physics, Isfahan University of Technology, Isfahan 84156-83111, Iran}
	\affiliation{ICRANet-Isfahan, Isfahan University of Technology, 84156-83111, Iran}
	
	\author{M. Zarei}
%	\email[]{m.zarei@iut.ac.ir}
	\affiliation{Department of Physics, Isfahan University of Technology, Isfahan 84156-83111, Iran}
	\affiliation{ICRANet-Isfahan, Isfahan University of Technology, 84156-83111, Iran}
	
	\author{M. Abdi}
%	\email[]{mehabdi@gmail.com}
	\affiliation{Wilczek Quantum Center, School of Physics and Astronomy, Shanghai Jiao Tong University, Shanghai 200240, China}
	\affiliation{Department of Physics, Isfahan University of Technology, Isfahan 84156-83111, Iran}

	\author{N. Bartolo}
%	\email[]{nicola.bartolo@pd.infn.it}
	\affiliation{Dipartimento di Fisica e Astronomia “Galileo Galilei” Universita` di Padova, 35131 Padova, Italy}
	\affiliation{INFN, Sezione di Padova, 35131 Padova, Italy}
	\affiliation{INAF - Osservatorio Astronomico di Padova, I-35122 Padova, Italy}

	\author{S. Matarrese}
%	\email[]{sabino.matarrese@pd.infn.it}
	\affiliation{Dipartimento di Fisica e Astronomia “Galileo Galilei” Universita` di Padova, 35131 Padova, Italy}
	\affiliation{INFN, Sezione di Padova, 35131 Padova, Italy}
	\affiliation{INAF - Osservatorio Astronomico di Padova, I-35122 Padova, Italy}
	\affiliation{Gran Sasso Science Institute, I-67100 L'Aquila, Italy}

%	\date{\today}
	
\begin{abstract}
		\baselineskip=6mm
		
This paper investigates the decoherence effect resulting from the interaction of squeezed gravitational waves with a system of massive particles in spatial superposition. This paper investigates the decoherence effect resulting from the interaction of squeezed gravitational waves with a system of massive particles in spatial superposition. We first employ the open quantum system approach to obtain the established decoherence in a spatial superposition of massive objects induced by squeezed gravitational waves. Subsequently, we focus on the spin-1/2 particle system, and our analysis reveals that the decoherence rate depends on both the squeezing strength and the squeezing angle of the gravitational waves. Our results demonstrate that squeezed gravitational waves with squeezing strengths of $r_p\geq1.2$ and a squeezing angle of $\varphi_p=\pi/2$ can induce a 1\% decoherence within 1~s free falling of a cloud of spin-1/2 particles.
This investigation sheds light on the relationship between squeezed gravitational waves and the coherence of spatial superposition states in systems of massive particles and their spin. The dependence of decoherence on squeezing strength and, in the case of spin-$1/2$ particles, on the squeezing angle paves the way for further exploration and understanding of the quantum-gravity connection. We suggest that such an experimental setup could also be employed to eventually investigate the level of squeezing effect (and hence quantum-related properties) of gravitational waves produced in the early universe from inflation. 
	\end{abstract}
	
	\maketitle
	\vspace{1.0cm}
	
	%\end{center}
	%\end{titlepage}
	
	\pagestyle{plain}
	\setcounter{page}{1}
	\newcounter{bean}
	\baselineskip18pt
	
	\setcounter{tocdepth}{2}
	
	\newpage
	\tableofcontents
	\newpage
	
	\section{Introduction}
	
Several theoretical proposals have been made to detect the quantum aspects of the stochastic gravitational-wave background (SGWB), which is believed to arise from the superposition of numerous independent and unresolved gravitational wave (GW) sources. These sources can have cosmological origins, such as various inflationary models, first-order phase transitions, cosmic string models, or astrophysical origins, resulting from the superposition of waves generated by astrophysical sources like supernovae (see \cite{Christensen:2018iqi,Guzzetii2016,Caprini2018} for a review).

The energy associated with gravitons (which are the quantum counterparts of the SGWB) is expected to be minuscule, making them difficult to observe directly using current technology \cite{Allen1999}. That said, indirect detection methods rather study their effects on other quantum systems. For example, the noise induced by gravitons in gravitational wave detectors or the effects of gravitons on quantum states can be used to indirectly probe the existence and influence of gravitons. Gravitons can induce decoherence in certain quantum systems with which they interact. When a quantum system comes into contact with gravitons or experiences a gravitational wave, its quantum coherence can be disrupted.  
Recently, there has been significant interest in detecting the induced decoherence of macroscopic entangled states as an indirect probe of gravitons \cite{ Parikh:2020nrd, Parikh:2020kfh, Parikh:2020fhy, Kanno:2020usf,bassi2017gravitational,anastopoulos2013master,Riedel:2013yca, Suzuki:2015nva, Guerreiro:2019vbq,blencowe2013effective,lamine2006ultimate}. One proposal by Kanno et al. \cite{Kanno:2021gpt} involves detecting the decay of entanglement between massive mirrors in an interferometric GW detector. They suggest generating entanglement by sending a single photon through the interferometer, creating a spatially non-local excitation. The decay of this entangled state, resulting from its coupling to a noisy environment of gravitons, provides indirect evidence for the quantum nature of GW.
		
In a similar study, Parikh et al. proposed a method to investigate gravitons through the stochastic modification of the geodesic deviation equation for a pair of freely falling masses \cite{Parikh:2020nrd, Parikh:2020kfh, Parikh:2020fhy}. This proposal is based on the concept that freely falling bodies experience noise due to their interaction with gravitons. In their method, the graviton environment induces decoherence in the quantum dynamics of the system through a non-unitary evolution master equation that describes the open quantum system (OQS). The noise predicted by \cite{Parikh:2020nrd, Parikh:2020kfh, Parikh:2020fhy, Kanno:2020usf} depends on the gravitational quantum state, and the authors estimated the noise correlators and their effect on the motion of the detector's mirrors for quantized gravitational waves in various states, such as vacuum, coherent, thermal, and squeezed states, using the Feynman-Vernon influence functionals technique \cite{Feynman:1963fq}. The authors demonstrated that coherent states recover the classical phenomenology of gravitational waves.

Squeezed states are a well-known quantum mechanical phenomenon that is even prepared in various quantum optical experiments \cite{Scully}. These states have also been proposed for primordial gravitational waves \cite{Grishchuk:1990bj}, and their observation would provide evidence of the quantum nature of primordial GWs. The power spectral density of strain resulting from quantum fluctuations originating from the squeezed quantum state of GWs is computed to have an exponential enhancement factor proportional to the squeezing strength. This suggests that significant quantum fluctuation effects could be detected by future GW detectors \cite{Parikh:2020nrd, Parikh:2020kfh, Parikh:2020fhy}.
	
The concept of quantum squeezed states is closely intertwined with the the so-called problem of quantum-to-classical transition of primordial perturbations during cosmic inflation \cite{Guzzetii2016}. 
The prevailing view, based on the theory of cosmic inflation and our understanding of quantum mechanics, is that quantum fluctuations in the early universe, including those of the tensor field (which we associate with gravitons), were stretched to macroscopic scales due to the accelerated expansion of space during inflation. 
These tensor fluctuations can be described as being in a squeezed state. This is a purely quantum mechanical effect. 
However, as these fluctuations continued to stretch and became classical perturbations, they formed the seeds for the structure in the universe we observe today. How such a transition from quantum to classical perturbations took place and whether the squeezed state of gravitons has entirely classicalized is very subtle issue and is linked to deeper questions about the interpretation of quantum mechanics.

 Some interpretations conclude that the quantum-to-classical transition was completed by the end of inflation and that the squeezed state has fully classicalized when the modes become sub-Hubble again (see, e.g. \cite{Allen1999,Lesgourgues1997,sudarsky2011,Kiefer2008,Kiefer2007,Kiefer2007P}). However such an interpretation has been debated in recent years and clarified. Indeed other recent interpretations suggest that quantum correlations might still be present \cite{Campo2006,Martin2016,Martin2017,Martin2018,Martin2018JCAP,Martin2023,Berera2021,Berera2022,Colas2023,Burgess2022,Daddi2023} (for a recent review see \cite{Micheli2022}). They show the conditions for a complete decoherence of our universe, and hence how it would be eventually possible to detect the signature of quantumness in cosmological observations.
 
 Detecting the quantum properties, such as the squeezed nature of gravitons, would require new observational techniques or new physical effects.
Various theoretical proposals have been developed to detect quantum aspects of gravity, as outlined in references \cite{Anastopolos2020, Anastopolos2021, Giacomini2022, Belenchia2018, Streiter2021, Matsumura2020, Matsumura2021}. These proposals explore scenarios in which two quantum systems are coupled through gravity to detect the presence of quantum correlations. The central concept is that quantum entanglement between objects that do not directly interact can only increase through a quantum mediator, believed to be the gravitational field in these scenarios. While these proposals suggest that quantum mechanics may play a role in gravity, they do not provide a comprehensive understanding of the quantum structure of gravity. Recent proposals, such as those described in \cite{Bose:2017nin, Marletto:2017kzi, Krisnanda:2019glc, Rijavec:2020qxd, Guerreiro:2021qgk}, have focused on using two quantum systems coupled through gravity to mediate quantum correlations as a means of detecting quantum aspects of gravity.
	
References \cite{Bose:2017nin, Marletto:2017kzi} propose a test to detect quantum effects in gravity by employing two systems that interact with each other only via the gravitational field. If two quantum systems do not directly interact with each other but become entangled after local interaction through the gravitational field, it suggests that the gravitational field is quantum. While the proposal by \cite{Bose:2017nin, Marletto:2017kzi} demonstrates some non-classicality in gravity, it does not provide a detailed understanding of the quantum structure of gravity.

Nonetheless, there are practical and fundamental debates regarding the feasibility of achieving sufficient spatial superposition for massive objects~\cite{Penrose1996on,Bassi2014collapse} or using them as a probe for quantum gravity~\cite{Bronstein2012republication,Gorelik2005from,Gunnink2023gravitational}. For instance, Bronstein, challenged the validity of quantum gravity tests, focusing his argument on the backreaction of the ``quantum" system on the ``classical" detector in the weak field limit, posing a conflict with the equivalence principle~\cite{Bronstein2012republication,Torrieri2022the}. In cases where the mass of the detector is comparable to that of the test particles, dephasing occurs due to detector recoil. Conversely, a heavier detector introduces a more substantial interaction than that between the two test particles. It may be thought that using electromagnetic beam splitters and reflectors with devices that are far from the test particles will lessen the impact of the detector. However the electromagnetic light also carries momentum and the further masses are the longer pulses of
light have to be and the more momentum they will transfer~\cite{Torrieri2022the}.
	
In this work, we focus on the decoherence arising from the quantum effects of gravitational waves. Decoherence is a phenomenon in quantum theory where coherence and, thus, interference effects are lost \cite{Breuer2001Destruction,Breuer:2002pc}. By analyzing the reduced density operator of a mixed state that includes both system and environmental states, we can observe that the state of the system loses phase, causing the off-diagonal terms to decay. Therefore, observing the decay of coherence due to interaction with a noisy bath of gravitons could indirectly provide evidence of the quantum nature of the gravitational field.  
	
The applicability of the quantum Boltzmann equation (QBE) \cite{Zarei:2021dpb} is limited to situations where the coupling between the system and environment is weak, and the memory effects of the environment can be neglected. In this study, we use QBE to examine interference experiments with massive object and massive spin-1/2 particles and explore the loss of coherence resulting from their interaction with graviton noise in laboratory-scale setups. 
	
The paper is organized as follows: Section \ref{sec::marcov} discusses Markovian and Non-Markovian QBE and derives the dissipative term, which will be used in the subsequent sections. The mesoscopic and microscopic time scales, which will be used to describe the time evolution of the density matrix, are also defined. Section \ref{sec::massiveobjectcoup} is devoted to the derivation of the interaction Hamiltonian between the massive object and squeezed gravitational wave. The massive object in this section is prepared in spatial superposition. In Section \ref{sec::massivedecoherence} the decoherence of the aforementioned massive object system due to interaction with squeezed gravitational waves is derived and a descriptive setup for understanding the amount of decoherence effect in different squeezing strengths is stated. In Section \ref{sec::fermion} the decoherence effect due to the interaction of a massive spin-1/2 object with squeezed gravitational waves is derived and a minimum squeezing strength for imposing a 1\% decoherence in a descriptive setup is obtained. 
	
%%%%%%%%%%%%%%%%%%%%%%%%%%%%%%%%%%%%%%%%%%%%
	
\section{Markovian and Non-Markovian quantum Boltzmann equation }\label{sec::marcov}

In this work, we employ the QBE to describe the dynamics of decoherence induced by the interaction between the quantum system and the gravitational field \cite{Zarei:2021dpb}. The QBE is a powerful tool for analyzing OQS \cite{Breuer:2002pc}, where the time evolution of the reduced density matrix of the system is of interest as it is coupled to an environment.
	
The QBE assumes that the initial state of the system and environment is a product state, and in our case we take the environment in a squeezed state. It provides a convenient way to calculate the time evolution of the reduced density matrix of the system under the influence of the environment. However, in many cases, determining the exact time evolution of the density matrix analytically is not possible, and approximation schemes are used to derive master equations for the approximate evolution of the reduced density matrix.
	
The Markovian QBE is based on two core approximations known as the Born and the Markov approximations. The Born approximation assumes that the coupling between the system and environment is weak and can be treated perturbatively. The Markov approximation assumes that the memory effects of the environment are negligible, and any self-correlations within the environment created by the coupling to the system decay rapidly compared to the system's dynamics timescale. In the next section, we will discuss Markovian and non-Markovian QBEs and derive the dissipative term to investigate the decoherence dynamics of the quantum system under the influence of gravitational waves.
	
Although the Born approximation is applicable in many physical situations, memory effects cannot always be neglected. When significant memory effects are present in the environment, the evolution of the reduced density operator becomes strongly dependent on the history of the entire system-environment combination. In such cases, non-Markovian quantum master equations (non-MaQBE) that rely on retarded-time kernels and integrations over the system's history must be solved. In \cite{Zarei:2021dpb}, we have extended the non-MaQBE formalism to describe irreversible processes. This type of equation is encountered in situations such as decoherence or damping effects in a quantum system of gravitons due to interaction with a viscous medium \cite{Zarei:2021dpb}.
	
For a system described by a density matrix $\rho^\mathcal{S}_{ij}(t_{\textrm{mes}})$, the evolution of the reduced density matrix of the system is given by a non-Markovian quantum Boltzmann equation (non-MaQBE), as expressed in the following equation \cite{Zarei:2021dpb}
\begin{eqnarray}
	\label{boltzmanneq3}
	\frac{d}{dt_{\textrm{mes}}}\left<\rho^\mathcal{S}_{ij}(t_{\textrm{mes}})\right>= D_{ij}[\rho^\mathcal{S}(t_{\textrm{mes}})]~,
\end{eqnarray}
where $\left<\cdots\right>=\text{Tr}[\rho^\mathcal{S}\cdots]$, the dissipator $D_{ij}[\rho^\mathcal{S}(t_{\textrm{mes}})]$ is given by the following expression
\begin{eqnarray}
	D_{ij}[\rho^\mathcal{S}(t_{\textrm{mes}})]= -\int_{0}^{\tau}dt_{\textrm{mic}}\left<\left[H_{\textrm{int}}(t_{\textrm{mes}}),\left[ H^{0}_{\textrm{int}}(-t_{\textrm{mic}}) ,\hat{\mathcal{N}}^S_{ij}(t_{\textrm{mes}}-t_{\textrm{mic}})\right]\right]\right>_\textrm{c}~,
	\label{damping1}
\end{eqnarray}
in which $\tau$ is the experiment time, $\hat{\mathcal{N}}^S_{ij}$ is the number operator associated with the system's degrees of freedom, and $H_{\textrm{int}}$ is the interaction Hamiltonian with superscript 0 denoting that it is a functional of the free field.
In this context, we introduce two time scales: $\tmic$ and $\tmes$, which respectively represent the interaction time scale of individual particles and the time scale on which the macroscopic system evolves. The subscript ``c" denotes the connected part of the correlation functions.  
	
In dissipative processes, the microscopic interaction Hamiltonian is not invariant under time reversal operation, i.e., $H_{\textrm{int}}(t)\neq H_{\textrm{int}}(-t)$. However, in practice, the time-reversed Hamiltonian can be represented as $H_{\textrm{int}}(-t)=H^{0\dag}_{\textrm{int}}(t)$. Consequently, the dissipator is expressed as follows
	\begin{eqnarray} \label{D1}
		D_{ij}[\rho^\mathcal{S}(t_{\textrm{mes}})]= -\int_{0}^{\tau}dt_{\textrm{mic}}\left<\left[H_{\textrm{int}}(t_{\textrm{mes}}),\left[ H^{0\dag}_{\textrm{int}}(t_{\textrm{mic}}) ,\hat{\mathcal{N}}^S_{ij}(t_{\textrm{mes}}-t_{\textrm{mic}})\right]\right]\right>_\textrm{c}~.
		\label{damping2}
	\end{eqnarray}
	The dissipator typically consists of two types of terms
	\beq
	D_{ij}\propto \Gamma^{\textrm{in}}_{ij}[\rho^\mathcal{S}(\k)]-\Gamma^{\textrm{out}}_{ij}[\rho^\mathcal{S}(\k')]~,
	\eeq
	Where $\mathbf{k}$ and $\mathbf{k}'$ are the transport momenta.
	The first term, $\Gamma^{\textrm{in}}$, corresponds to scattering-out processes ($a + b \rightarrow c + d$), while the second term, $\Gamma^{\textrm{out}}$, corresponds conversely to scattering-in processes ($c + d \rightarrow a + b$). For scattering-out processes, the collision term is proportional not only to the distribution function of the initial states (species $a$ and $b$) but also to the Pauli blocking function of the final states (species $c$ and $d$). Conversely, for scattering-in processes, this relationship is reversed. In the case of reactions that are invariant under time reversal, $\Gamma^{\textrm{in}}$ and $\Gamma^{\textrm{out}}$ satisfy the detailed balance relation
	\beq \label{detailed}
	\Gamma_{ij}^{\textrm{in}}=e^{-\beta \Delta}\Gamma_{ij}^{\textrm{out}}~,
	\eeq
	Where $\beta$ is the Boltzmann factor and $\Delta=|\mathbf{k}|-|\mathbf{k}'|$. The detailed balance condition holds true for elastic scattering under equilibrium conditions. However, it should be noted that the condition \eqref{detailed} is only valid in the presence of time-reversal symmetry, indicating that the scattering is reciprocal.
	
	In the subsequent analysis, we will determine the dissipative term $D_{ij}[\rho^\mathcal{S}(t_{\textrm{mes}})]$ for two quantum systems composed of massive objects and spin-1/2 particles. By applying this method, we will calculate the decoherence rate for these quantum systems.

	%%%%%%%%%%%%%%%%%%%%%%%%%%%%%%%%%%%%%%%%%%%%%
	
	\section{Coupling with Massive Object}
	\label{sec::massiveobjectcoup}
	
	In this section, we provide a concise overview of the key components of our OQS. We consider a massive body in a spatial superposition and utilize the QBE to illustrate how the interaction between this system and its environment of gravitons leads to decoherence. We prepare a spatial superposition of two distinct states representing the location of a massive free-falling particle. By employing this superposition, we determine the density matrix of the system. The interaction Hamiltonian between the system and its environment is expressed in terms of creation and annihilation operators associated with their respective degrees of freedom. Any occurrence of decoherence is manifested by the gradual decay of the off-diagonal elements of the system's density matrix over time. In the following, we employ the QBE to calculate the time evolution of these off-diagonal terms and derive the precise form of the decoherence rate.
	
	%%%%%%%%%%%%%%%%%%%%%%%%%%%%%%%%%%%%%%%%%%%%
	
	\subsection{Environment of gravitons}
	\label{sec::massiveobjectcoup.3}
	
	In the linearized approach, the graviton field $h_{\mu\nu}$ is defined as a small perturbation of the space-time metric $g_{\mu\nu}$ around the flat Minkowski background. The metric, which describes gravitational waves in the transverse traceless gauge, can be expressed as follows
	\begin{eqnarray}
		ds^2=-dt^2+(\delta_{ij}+\kappa\, h_{ij}) dx^idx^j\,\,,
		\label{metric}
	\end{eqnarray}
where $\delta_{ij}$ is the Kronecker delta and $\kappa^2 = 16\pi G$, with $G$ representing Newton's constant related to the reduced Planck's mass $M_\text{P}$ through $G = 8/M_\text{P}^2$. The indices $i$ and $j$ take values from $1$ to $3$ and $h_{ij}$ is in
the transverse traceless gauge, $h_{\;i}^i=0$ and $\nabla^ih_{ij}=0$.  Substituting the metric given in Eq.~\eqref{metric} into the Einstein-Hilbert action yields the quadratic Hamiltonian density
	\begin{eqnarray}
		\mathcal{H}_g =\frac{1}{4}
		\left[\,\dot{h}^{ij} \dot{h}_{ij}  - \nabla^kh^{ij}\,     \nabla_kh_{ij}
		\,\right]  \label{G_action}\, ,
	\end{eqnarray}
where a dot denotes the derivative with respect to  time.
	We can expand the graviton field $h_{ij}( \x, t)$ in terms of the Fourier modes as
	\bea
	\!\!\!\!\!\!\!\!\!\! h_{ij}(\x,t ) &=& h^+_{ij}(\x,t ) +h^-_{ij}(\x,t ) \nonumber \\ &=&
	\sum_{ s}\int \frac{d^3p}{(2\pi)^32p^0} \left[ u^s_{\p}(t)\,e^{i \p \cdot {\x}} e_{ij}^{(s)}(\p) b^{(s)}(\mathbf{ p})+
	u^{s\ast}_{\k}(t)\,e^{-i \p \cdot {\x}} e_{ij}^{(s)}(\p) b^{\dag (s)}(\mathbf{ p})
	\right ]~,
	\label{fourierh}
	\eea
	where the operators $h^+$ and $h^-$ are associated with the absorption and creation of gravitons, respectively, and are linearly related to the corresponding annihilation and creation operators. Additionally, we have introduced the polarization tensor $e^s_{ij}({\bf k})$, which is normalized such that $e^{*s}_{ij}({\bf k}) e^{s'}_{ij} ({\bf k})= \delta^{ss'}$, where the index $s$ denotes the linear polarization modes $s=+,\times$. The creation and annihilation operators satisfy the standard commutation relations as follows
	\begin{align}
		\left[\, b_s({\bf p})\,,\,b^\dag_{s'}({\bf p'})\,\right]
		&=2p^0\delta^3(\p-\p')\delta_{ss'}~,
		\label{eq:quantization1}
	\end{align} 
	and $u_\p(t)$ denotes a mode function properly normalized as \cite{Kanno:2020usf}
	\begin{align}
		\dot u_\p(t)u_\p^*(t)-u_\p(t)\dot u_\p^*(t)=-i\,.
	\end{align} 
	The Minkowski vacuum is defined by $\ket{0}$ with $b_s({\bf p}) \ket{0} =0$ and choosing the mode function as 
	\begin{align}
		u_\p(t)=e^{-ip^0t}~.
		\label{minkowski}
	\end{align} 
\quad Physically, squeezed vacuum states are conjectured to arise from inflationary scenarios \cite{grishchuk1990squeezed,albrecht1994inflation,kanno2021squeezed}. The definition of the squeezed vacuum state $\ket{\zeta}$ is given by the action of a squeezing operator $\hat{S}(\zeta)$ on the vacuum state $\ket{0}$, i.e., $\ket{\zeta} = \hat{S}(\zeta)\ket{0}$. The squeezing operator is defined as
	\begin{eqnarray}
		\hat S(\zeta)\equiv \exp\left[\frac{1}{V}\sum_{{\bf p},s}
		\left(\zeta^*_p\, b_s({\bf p}) b_s(-{\bf p})
		+ \zeta_p \,b_s^\dag  ({\bf p}) { b}_s^\dag (-{\bf p})  \right) \right]  \,,   
	\end{eqnarray}
	where $b_s$ and $b_s^\dagger$ are respectively the annihilation and creation operators of the graviton, and $\zeta_p\equiv r_p\exp[i\varphi_p]$ is a complex squeezing parameter with $r_p$ and $\varphi_p$ as its strength and angle, respectively.
	The squeezing operator is a unitary operator and satisfies the following transformation properties
\begin{eqnarray}
	&&\hspace{-1.5cm}
	\hat S^\dag(\zeta)\, b_s({\bf p})\,\hat S(\zeta)
	= b_s({\bf p})\cosh r_p
	- b_s^\dag(-{\bf p})e^{i\varphi_p}\sinh r_p~,\nonumber\\
	&&\hspace{-1.5cm}
	\hat S^\dag(\zeta)\, b_s^\dag({-\bf p})\,\hat S(\zeta)
	= b_s^\dag({-\bf p})\cosh r_p
	- b_s({\bf p})e^{-i\varphi_p}\sinh r_p~.
	\label{eq:transf1}
\end{eqnarray}
	These relations demonstrate that applying the squeezing operator to the creation and annihilation operators results in a linear combination of annihilation and creation operators, where the coefficients are determined by the hyperbolic functions $\cosh r_p$ and $\sinh r_p$ respectively.	Therefore, the vacuum expectation value of the transformed operators is given by
	\begin{eqnarray}
		&&\hspace{-1.5cm}
		\langle 0|\hat S^\dag(\zeta)\, b_s({\bf p})\,\hat S(\zeta)|0\rangle
		=\cosh r_p- e^{i\varphi_p}\sinh r_p~,\nonumber\\
		&&\hspace{-1.5cm}
		\langle 0| \hat{S}^\dag(\zeta)\,b_s^\dag({-\bf p})\,\hat S(\zeta)|0\rangle
		=\cosh r_p
		-e^{-i\varphi_p}\sinh r_p~.
	\end{eqnarray}
These relations indicate that the squeezing operator transforms the vacuum state of the field into a squeezed state, which is distinguished by a non-zero expectation value of the transformed annihilation and creation operators.
Moreover, as it is shown in \cite{albrecht1994inflation} the transformation of the graviton quantum field under the action of the squeezing operator is equal to the replacement of the mode functions by their squeezed state counterpart as in  \eqref{eq:transf1} and the mode function in the squeezed state is given in terms of that in the Minkowski space in Eq.~(\ref{minkowski}), such as
\begin{eqnarray}\label{usq}
	u^{\rm sq}_\p(t)\equiv u_\p(t)\cosh r_p-e^{-i\varphi_p}{u^*_\p}(t)\sinh r_p\,.
\end{eqnarray}
In the following, we use the squeezed mode function  of Eq.~\eqref{usq} in the interaction Hamiltonian between the squeezed gravitational wave and the spatial superposition system.

	%%%%%%%%%%%%%%%%%%%%%%%%%%%%%%%%%%%%%%%%%%%%%%%%%%
	%%%%%%%%%%%%%%%%%%%%%%%%%%%%%%%%%%%%%%%%%%%%%%%%%%

	\subsection{Spatial Superposition System of Massive Bodies }
	\label{sec::massivecoup.1}
	
	Here, we consider two masses in free fall, where one of them initially exists in a coherent spatial superposition. According to Einstein's equivalence principle, a single particle remains unaffected by a gravitational wave. In this system, the mass $M$ is supermassive, and thus we choose it as our spatial coordinate origin. While the center of mass frame could have been selected, we simplify the problem by adopting $M$ as the origin. To establish a convenient coordinate system, we introduce a Fermi normal coordinate system along the geodesic of mass $m$ \cite{Parikh:2020fhy, Kanno:2020usf}. The deviation of the geodesic of the second particle, denoted as $m$, is described by the vector $\bm{\xi}$. Finally, we derive the action for the geodesic deviation up to second order in $\bm{\xi}$ as follows
	\beq
	S_m=\int dt\left [ \frac{m}{2}\dot{\bm{\xi}}\cdot\dot{\bm{\xi}}+\frac{m\kappa}{4}\ddot{h}_{ij}\xi_i\xi_j\right]~,
	\eeq
	where the first term can be interpreted as the kinetic term of the mass $m$, while the second term describes the interaction between the mass $m$ and the graviton environment, which will be discussed in more detail. To account for quantum effects, we now consider the deviation vector $\bm{\xi}$ as an operator, denoted as $\hat{\bm{\xi}}$. As a result, the free component of the Hamiltonian is given by
	\cite{Parikh:2020fhy,Kanno:2020usf}
	\begin{eqnarray}
		H_0  =  \frac{m}{2}\dot{\hat{\bm{\xi} }}\cdot \dot{\hat{\bm{\xi} }}~.
	\end{eqnarray}
	The deviation operators can be expanded as 
	\begin{eqnarray}
		\hat{\xi}^i(t) =\hat{\xi}^{i\,+}(t)+\hat{\xi}^{i\,-}(t)= \sum_r\left[\xi_r^i(t) \hat{a}^{(r)}+\xi_r^{i\ast}(t)\hat{a}^{\dagger (r)}\right]~,
		\label{fourierxi}
	\end{eqnarray}
	where $r=1,2$ and $i=x,y,z$, and $\hat{a}^{(r)}$ and $\hat{a}^{\dagger (r)}$ are dimensionless ladder operators that satisfy the canonical commutation relation
	\beq
	\left[\hat{a}^{ (r)},\hat{a}^{\dagger (r')}\right]=\delta_{rr'}~.
	\eeq
	Here, $\hat{\xi}^{i\,+}$ and $\hat{\xi}^{i\,-}$ are linear functions of the annihilation and creation operators, respectively. The mass is assumed to be in a superposition state of two paths. The initial superposition state can be represented as follows
	\beq
	\ket{\psi(t_i)}=\ket{\vec{\xi}^{\,1}(t_i)}+\ket{\vec{\xi}^{\,2}(t_i)}~,
	\eeq
	where $\ket{\vec{\xi}^{\,r}}$ represents the quantum states of the massive test masses, and the polarization-like state space of the system is spanned by a pair of basis vectors: $\ket{\vec{\xi}^{\,1}}$ and $\ket{\vec{\xi}^{\,2}}$. Alternatively, the superposition state can be expressed in terms of the density matrix $\hat{\rho}^\mathcal{S}(t_i) = \ket{\psi(t_i)}\bra{\psi(t_i)}$, which can be written as
	\beq
	\hat{\rho}^\mathcal{S}(t_i)= \ket{\vec{\xi}^{\,1}(t_i)}\bra{\vec{\xi}^{\,1}(t_i)}+ \ket{\vec{\xi}^{\,1}(t_i)}\bra{\vec{\xi}^{\,2}(t_i)}+ \ket{\vec{\xi}^{\,2}(t_i)}\bra{\vec{\xi}^{\,1}(t_i)}+ \ket{\vec{\xi}^{\,2}(t_i)}\bra{\vec{\xi}^{\,2}(t_i)}~.
	\eeq
Additionally, one can define quantum-mechanical operators in the linear basis that correspond to each Bloch vector component as
	\bea\label{eq:stokes1}
	\hat{S}^0 &=&\ket{\vec{\xi}^{\,1}}\bra{\vec{\xi}^{\,1}}+\ket{\vec{\xi}^{\,2}}\bra{\vec{\xi}^{\,2} }~, \\
	\hat{S}^1 &=&\ket{\vec{\xi}^{\,1}}\bra{\vec{\xi}^{\,1}}-\ket{\vec{\xi}^{\,2}}\bra{\vec{\xi}^{\,2} }~,\\
	\hat{S}^2 &=&\ket{\vec{\xi}^{\,1}}\bra{\vec{\xi}^{\,2}}+\ket{\vec{\xi}^{\,2}}\bra{\vec{\xi}^{\,1} }~,\\
	\hat{S}^3 &=& i\ket{\vec{\xi}^{\,2}}\bra{\vec{\xi}^{\,1}}-i\ket{\vec{\xi}^{\,1}}\bra{\vec{\xi}^{\,2} }~.\label{eq:stokes4}
	\eea
The Eqs.~\eqref{eq:stokes1}-\eqref{eq:stokes4} allow to build the analogues of standard Stokes parameters. The expectation value of the Stokes parameter $I$ can be expressed in terms of the elements of the density matrix, $\rho^\mathcal{S}$, as follows
\beq
	\label{IStokes}
	\left<\hat{S}^0 \right>=\tr(\hat{\rho}^\mathcal{S}\hat{I})=\rho^\mathcal{S}_{11}+\rho^\mathcal{S}_{22}~,
\eeq
and similarly, the other three parameters $Q^S$, $U^S$, and $V^S$ can be expressed in the same way. Thus, the polarization matrix of the system can be described in terms of the following three parameters
\begin{eqnarray} \label{rhoS}
		\rho^\mathcal{S} =\frac{1}{2} \left(\begin{array}{cc}1+S^1 & S^2-iS^3 \\ S^2+iS^3 & 1-S^1\end{array}\right)~, \label{matrixS}
\end{eqnarray}
where we have normalized $S^0=1$ and
\bea
\label{Stokes1}
S^1= \rho^\mathcal{S}_{11}-\rho^\mathcal{S}_{22}~,~~~~~~~~~S^2 =\rho^\mathcal{S}_{12}+\rho^\mathcal{S}_{21}~,~~~~~~~~~\textrm{and}~~~~~~~ -iS^3 = \rho^\mathcal{S}_{12}-\rho^\mathcal{S}_{21}~.
\eea
The diagonal elements of the density matrix represent the probabilities of the particle taking each path, while the off-diagonal elements describe the quantum coherence of the state. The density operator can be expressed in terms of the Stokes parameters and ladder operators as follows
\beq
\hat{\rho}^\mathcal{S} (t_i)=\sum_{ij} \rho^\mathcal{S} _{rr'}(t)\hat{a}^{\dag (r)}\hat{a}^{ (r')}~.
\eeq
The quantum-mechanical operators in the linear basis corresponding to each element can be expressed as
\begin{eqnarray}
	\hat{\rho}^\mathcal{S} = \rho^\mathcal{S}_{11}\ket{\vec{\xi}^{\,1}}\bra{\vec{\xi}^{\,1}}+\rho^\mathcal{S}_{12}\ket{\vec{\xi}^{\,1}}\bra{\vec{\xi}^{\,2}}
	+\rho^\mathcal{S}_{21}\ket{\vec{\xi}^{\,2}}\bra{\vec{\xi}^{\,1}}+\rho^\mathcal{S}_{22}\ket{\vec{\xi}^{\,2}}\bra{\vec{\xi}^{\,2}}~.
\end{eqnarray}
In our context, we will focus on the decoherence effect that induces the decay of the off-diagonal elements. This phenomenon leads to the loss of coherence, transforming a coherent superposition of the $\ket{\vec{\xi}^{\,1}}$ and $\ket{\vec{\xi}^{\,2}}$ states into a statistical mixture. On the other hand, an incoherent state is a statistical mixture that cannot exhibit interference effects, which are characteristic of coherent superpositions.

%%%%%%%%%%%%%%%%%%%%%%%%

%%%%%%%%%%%%%%%%%%%%%%%%%%%%%%%%%%%%%%%%%%%%%
\subsection{Interaction Hamiltonian}
\label{sec::massivecoup.2}

\begin{figure*}[t]
	\centering
	\captionsetup[subfigure]{oneside,margin={1.5cm,2cm}}
	\begin{subfigure}{0.45\textwidth}
		\includegraphics[width=\linewidth]{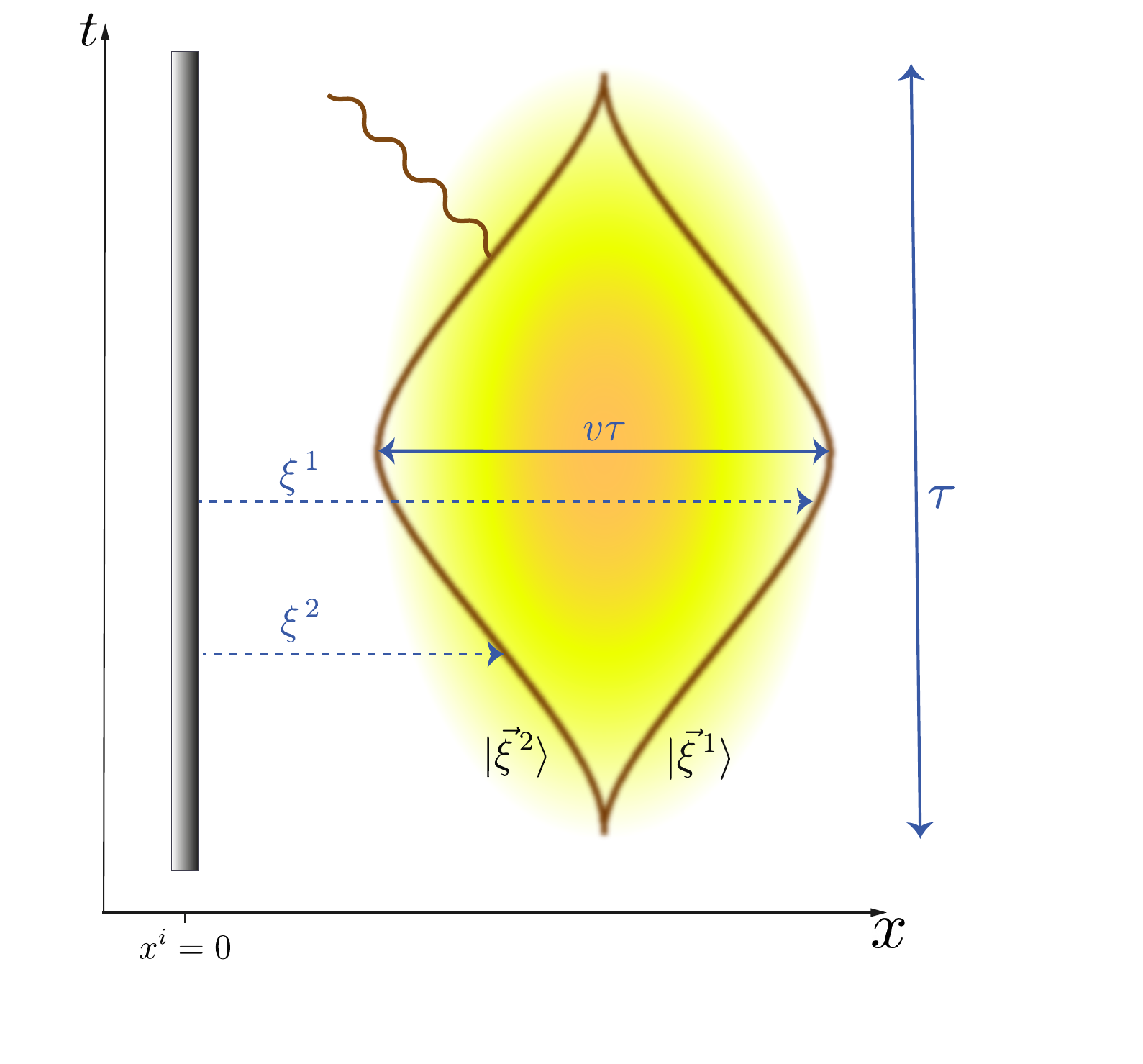}
		\vspace{-1cm} 
		\caption{}\label{Fig:absorption}
	\end{subfigure}
	\begin{subfigure}{0.45\textwidth}
		\includegraphics[width=\linewidth]{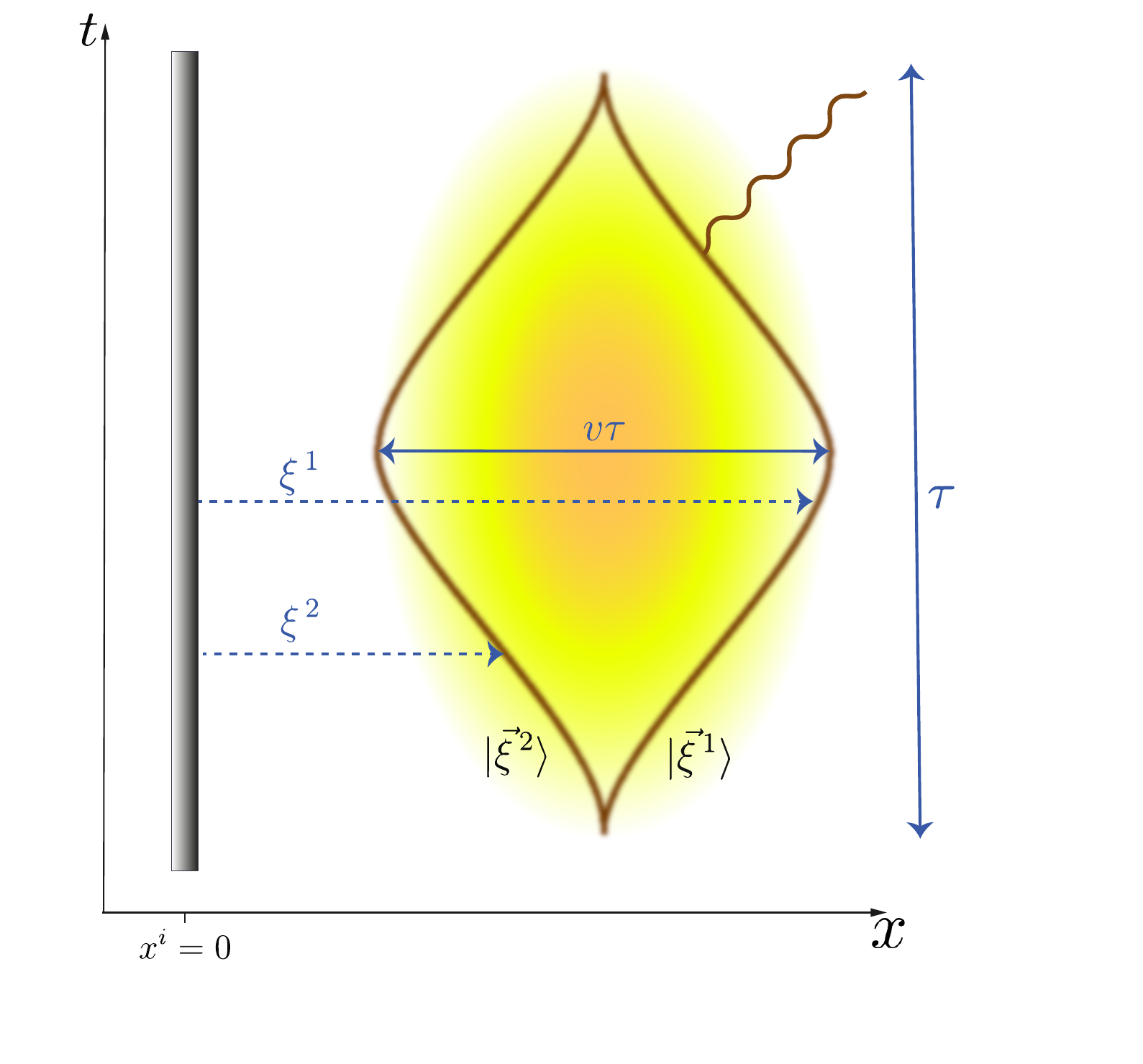}
		\vspace{-1cm}
		\caption{}\label{Fig:emission}
	\end{subfigure}
	\caption{(a) The absorption of a graviton noise by the system. (b)  The emission process. }\label{fig:spin0setup}
\end{figure*}

Decoherence arises when a quantum system undergoes unwanted interactions with its environment. In our analysis, we will explore two possible interactions between the mass $m$ and the gravitons. These interactions are represented by the diagrams depicted in Figures \eqref{Fig:absorption} and \eqref{Fig:emission}. Figure \eqref{Fig:absorption} illustrates the absorption of a graviton, which induces decoherence in the superposition and records with-path information. Figure \eqref{Fig:emission}, on the other hand, depicts a contributing diagram for the emission of bremsstrahlung gravitons. The interaction Hamiltonians associated with absorption and emission can be expressed as follows
\begin{eqnarray}
	H_{\textrm{abs}}(t) = \frac{m}{4}\kappa
	\,\ddot{h}^+_{ij}(0,t )\hat{\xi}^{i-} \hat{\xi}^{j+}~,
	\label{Hintbr}
\end{eqnarray}
and
\begin{eqnarray}
	H_{\textrm{emi}}(t) = \frac{m}{4}\kappa
	\,\ddot{h}^-_{ij}(0,t )\hat{\xi}^{i-} \hat{\xi}^{j+}~,
	\label{Hintab}
\end{eqnarray}
where the superscripts ``abs" and ``emi" denote the absorption and emission bremsstrahlung processes. The Fourier-space interaction Hamiltonians can be obtained by substituting Eqs.~\eqref{fourierh} and \eqref{fourierxi} into equations \eqref{Hintbr} and \eqref{Hintab}. The resulting interaction Hamiltonians are
\begin{eqnarray}
	H_{\textrm{abs}}(t) = 
	\frac{\kappa m}{4} \sum_{s,r,r'}
	\int \frac{d^3 p}{(2\pi)^32p^0}\, \ddot{u}^{\textrm{ab}(s)}_{\p}(t)e_{ij}^{\textrm{ab}(s)}(\p) \xi^{i\ast}_r(t)\xi^j_{r'}(t) b^{(s)}(\mathbf{ p})
	\hat{a}^{(r)\dag} \hat{a}^{(r')}~,
\end{eqnarray}
and
\begin{eqnarray}
	H_{\textrm{emi}}(t) =  \frac{\kappa m}{4} \sum_{s,r,r'}
	\int \frac{d^3 p}{(2\pi)^32p^0}\, \ddot{u}^{\textrm{em}(s)\ast}_{\p}(t)e_{ij}^{\textrm{em}(s)}(\p) \xi^{i\ast}_r(t)\xi^j_{r'}(t) b^{\dag (s)}(\mathbf{ p})
	\hat{a}^{(r)\dag} \hat{a}^{(r')}~.
\end{eqnarray}
The interaction of a quantum system with a noisy environment leads to irreversible decoherence. To accurately describe this phenomenon, we employ the QBE. Notably, we will show that these decoherence phenomena demonstrate non-Markovian behavior.

\section{Decoherence Rate of the Massive Object Superposition}
\label{sec::massivedecoherence}
Entanglement with the environment is widely recognized as one of the primary causes of decoherence in a quantum system. In this section, we derive coupled differential equations for the time evolution of the system's density matrix elements, enabling us to calculate the gravitationally-induced decoherence rate.

When the system frequently interacts with the environment, and the state of the environment is not actively observed, in general the off-diagonal terms in the system's density operator rapidly decay in a preferred basis. This basis is typically spatially localized and depends on the nature of the system-environment coupling. 

We begin by calculating the damping term \eqref{damping2}, followed by the interaction term
\begin{eqnarray}\label{eq:Hintmassive}
	H_{\textrm{int}}(t)= H_{\textrm{emi}}(t)+H_{\textrm{abs}}(t)~.
\end{eqnarray}
We substitute the interaction Hamiltonian into Eq.~\eqref{damping2}, resulting in the following form for the dissipator
\begin{eqnarray}
	&&D_{ij}[\rho^\mathcal{S}(t_{\textrm{mes}})]=-\int_{0}^{\tau}dt_{\textrm{mic}}\left<\left[H_{\textrm{int}}(t_{\textrm{mes}}),\left[H^{0\dag}_{\textrm{int}}(t_{\textrm{mic}}) ,\hat{\mathcal{N}}^S_{ij}(t_{\textrm{mes}}-t_{\textrm{mic}})\right]\right]\right>_\textrm{c} \nonumber \\&&
	=-\frac{\kappa^2m^2}{8} \sum_{s_1,r_1,r'_1}\sum_{s_2,r_2,r'_2}
	\int \frac{d^3 p_1}{(2\pi)^32p_1^0}\,\int \frac{d^3 p_2}{(2\pi)^32p_2^0}\,
	\textrm{Re} \left [\ddot{u}^{\rm sq}_{\p_1}(\tmic)  \ddot{u}^{\rm sq\,\ast}_{\p_2}(\tmes) \right]
	\nonumber \\&& \times e_{m_1n_1}^{(s_1)}(\p_1)e_{m_2n_2}^{(s_2)\ast}(\p_2)
	\xi^{m_1\ast}_{r_1}(\tmic)\xi^{n_1}_{r'_1}(\tmic)  \xi^{m_2\ast}_{r_2}(\tmes)\xi^{n_2}_{r'_2}(\tmes) 
	\left< b^{(s_2\dag)}(\mathbf{ p_2})b^{(s_1)}(\mathbf{ p_1})\right>_\textrm{c}
	\nonumber \\&& \times \left[
	\left<\hat{a}^{(r_1)\dag} \hat{a}^{(r'_1)}\hat{a}^{(r'_2)\dag}\hat{a}^{(r_2)} \hat{a}^{(i)\dag} \hat{a}^{(j)} \right> _\textrm{c}- \left<\hat{a}^{(r_1)\dag} \hat{a}^{(r'_1)}\hat{a}^{(i)\dag} \hat{a}^{(j)}\hat{a}^{(r'_2)\dag}\hat{a}^{(r_2)}  \right>_\textrm{c}
	\right. \nonumber  \\  &&  \left.  -
	\left<\hat{a}^{(r'_2)\dag}\hat{a}^{(r_2)} \hat{a}^{(i)\dag} \hat{a}^{(j)}\hat{a}^{(r_1)\dag} \hat{a}^{(r'_1)} \right>_\textrm{c} +\left< \hat{a}^{(i)\dag} \hat{a}^{(j)}\hat{a}^{(r'_2)\dag}\hat{a}^{(r_2)} \hat{a}^{(r_1)\dag} \hat{a}^{(r'_1)} \right>_\textrm{c}~
	\right ]~.
\end{eqnarray}
Using the following expectation values \cite{Zarei:2021dpb}
\begin{eqnarray} \label{EVGW}
	\left<\hat{a}^{(r_1)\dag} \hat{a}^{(r'_1)}\hat{a}^{(r'_2)\dag}\hat{a}^{(r_2)} \hat{a}^{(i)\dag} \hat{a}^{(j)} \right>_\textrm{c} &\simeq& \rho^\mathcal{S}_{jr_1}\delta^{ir_2}\delta^{r'_2r'_1}~, \nonumber\\
	\left<\hat{a}^{(r_1)\dag} \hat{a}^{(r'_1)}\hat{a}^{(i)\dag} \hat{a}^{(j)}\hat{a}^{(r'_2)\dag}\hat{a}^{(r_2)}  \right>_\textrm{c} &\simeq& \rho^\mathcal{S}_{r_2r_1}\delta^{ir'_1}\delta^{r'_2j}~, \nonumber\\
	\left<\hat{a}^{(r'_2)\dag}\hat{a}^{(r_2)} \hat{a}^{(i)\dag} \hat{a}^{(j)}\hat{a}^{(r_1)\dag} \hat{a}^{(r'_1)} \right>_\textrm{c} &\simeq&  \rho^\mathcal{S}_{r'_1r'_2}\delta^{r_1j}\delta^{ir_2}~, \nonumber\\
	\left< \hat{a}^{(i)\dag} \hat{a}^{(j)}\hat{a}^{(r'_2)\dag}\hat{a}^{(r_2)} \hat{a}^{(r_1)\dag} \hat{a}^{(r'_1)} \right>_\textrm{c} &\simeq&  \rho^\mathcal{S}_{r'_1i}\delta^{r_1r_2}\delta^{r'_2j}~,\nonumber\\
\end{eqnarray}
and
\begin{eqnarray} \label{EVS}
	\left< b^{(s_2\dag)}(\mathbf{ p_2})b^{(s_1)}(\mathbf{ p_1})\right>_\textrm{c} = (2\pi)^32p^0_1\delta^3(\p_1-\p_2)\rho^g_{s_1s_2}(\p_1)~.
\end{eqnarray}
we find the damping dissipator term as
\begin{eqnarray}\label{dissipator1}
	&&D_{ij}[\rho^\mathcal{S}(t_{\textrm{mes}})]=-\int_{0}^{\tau}dt_{\textrm{mic}}\left<\left[H_{\textrm{int}}(t_{\textrm{mes}}),\left[ H^{0\dag}_{\textrm{int}}(t_{\textrm{mic}}) ,\hat{\mathcal{N}}^S_{ij}(t_{\textrm{mes}}-t_{\textrm{mic}})\right]\right]\right>_\textrm{c} \nonumber \\&&
	=-\frac{\kappa^2m^2}{8} \sum_{s_1,r_1,r'_1}\sum_{s_2,r_2,r'_2}
	\int \frac{d^3 p_1}{(2\pi)^32p_1^0}\,\int \frac{d^3 p_2}{(2\pi)^32p_2^0}\int_{0}^{\tau}dt_{\textrm{mic}}\,\textrm{Re} \left [\ddot{u}^{\rm sq}_{\p_1}(\tmic)  \ddot{u}^{\rm sq\,\ast}_{\p_2}(\tmes) \right]
	\nonumber \\&& \times e_{m_1n_1}^{(s_1)}(\p_1)e_{m_2n_2}^{(s_2)\ast}(\p_2)
	\xi^{m_1\ast}_{r_1}(\tmic)\xi^{n_1}_{r'_1}(\tmic)  \xi^{m_2\ast}_{r_2}(\tmes)\xi^{n_2}_{r'_2}(\tmes) 
	(2\pi)^32p^0_1\delta^3(\p_1-\p_2)\rho^g_{s_1s_2}(\p_1)
	\nonumber \\&& \times \left[ \rho^\mathcal{S}_{jr_1}\delta^{ir_2}\delta^{r'_2r'_1}
	-\rho^\mathcal{S}_{r_2r_1}\delta^{ir'_1}\delta^{r'_2j}
	-\rho^\mathcal{S}_{r'_1r'_2}\delta^{r_1j}\delta^{ir_2}
	+ \rho^\mathcal{S}_{r'_1i}\delta^{r_1r_2}\delta^{r'_2j}
	\right ]~,\nonumber\\
\end{eqnarray}
where in the second line we have neglected the non-local time dependence $\tmes-\tmic$ in $\hat{\mathcal{N}}^\mathcal{S}_{ij}$. We now substitute \eqref{dissipator1} into \eqref{boltzmanneq3} and integrate over $p_1$ to find 
\begin{eqnarray}
	\frac{d}{dt_{\textrm{mes}}}\rho^\mathcal{S}_{ij}(t_{\textrm{mes}})&=&
		-\frac{\kappa^2m^2}{8} \sum_{s_1,r_1,r'_1}\sum_{s_2,r_2,r'_2}
	\int \frac{d^3 p_2}{(2\pi)^32p_2^0}\int_{0}^{\tau}dt_{\textrm{mic}}\, \textrm{Re} \left [\ddot{u}^{\rm sq}_{\p_2}(\tmic)  \ddot{u}^{\rm sq\,\ast}_{\p_2}(\tmes) \right]
	\nonumber \\&& \times e_{m_1n_1}^{(s_1)}(\p_2)e_{m_2n_2}^{(s_2)\ast}(\p_2)
	\xi^{m_1\ast}_{r_1}(\tmic)\xi^{n_1}_{r'_1}(\tmic)  \xi^{m_2\ast}_{r_2}(\tmes)\xi^{n_2}_{r'_2}(\tmes) 
	\rho^g_{s_1s_2}(\p_2)
	\nonumber \\&& \times \left[\rho^\mathcal{S}_{jr_1}\delta^{ir_2}\delta^{r'_2r'_1}
	-\rho^\mathcal{S}_{r_2r_1}\delta^{ir'_1}\delta^{r'_2j}
	-\rho^\mathcal{S}_{r'_1r'_2}\delta^{r_1j}\delta^{ir_2}
	+ \rho^\mathcal{S}_{r'_1i}\delta^{r_1r_2}\delta^{r'_2j}
	\right ]~.\nonumber\\
\end{eqnarray}
As mentioned previously, a two-level system can be effectively described using Bloch vector components to represent its state. In this section, our focus is on the dephasing channel, also known as the phase-damping channel. This particular scenario illustrates the decoherence that arises in the system as a result of its interaction with graviton noise.
We examine a scenario where $S_3$ is initially zero so remains zero for all forward times. Therefore, the time evolution of the diagonal and off-diagonal elements of the system's density matrix can be described in terms of the components $S_1$ and $S_2$. Assuming $\bm{\xi}_r=(\xi_r,0,0)$ and utilizing \eqref{Stokes1}, we can derive the following expressions
\begin{eqnarray} \label{dotQ}
	\frac{d}{dt_{\textrm{mes}}}S_1(t_{\textrm{mes}})&=&\Gamma_{11} S_1(\tmes)+\Gamma_{12}\,S_2(\tmes)~, \\
	\frac{d}{dt_{\textrm{mes}}}S_2(t_{\textrm{mes}})&=&\Gamma_{22} S_2(\tmes)+\Gamma_{21}\,S_1(\tmes)~,\label{dotU}
\end{eqnarray}
where the coupling coefficients of two coupled differential equations are given as the following 
\begin{eqnarray} \label{gammaQQ}
	\Gamma_{11}&=&
	-\frac{\kappa^2m^2}{16} 
	\int \frac{d^3 p_2}{(2\pi)^3p_2^0}\int_{0}^{\tau}dt_{\textrm{mic}}\, \textrm{Re} \left [\ddot{u}^{\rm sq}_{\p_2}(\tmic)  \ddot{u}^{\rm sq\,\ast}_{\p_2}(\tmes) \right]I^g(\p_2)
	\nonumber \\&& \times 
	4\xi_{1}(\tmes)\xi_{2}(\tmes)\xi_{1}(\tmic)\xi_{2}(\tmic)~,
\end{eqnarray}
\begin{eqnarray} \label{gammaQU}
	\Gamma_{12}&=&
	-\frac{\kappa^2m^2}{16} 
	\int \frac{d^3 p_2}{(2\pi)^3p_2^0}\int_{0}^{\tau}dt_{\textrm{mic}}\,\textrm{Re} \left [\ddot{u}^{\rm sq}_{\p_2}(\tmic)  \ddot{u}^{\rm sq\,\ast}_{\p_2}(\tmes) \right]I^g(\p_2)
	\nonumber \\&& \times 
	2\left[(\xi_{2}(\tmes))^2- (\xi_{1}(\tmes))^2\right]\xi_{1}(\tmic)\xi_{2}(\tmic)~,
\end{eqnarray}
\begin{eqnarray} \label{gammaUQ}
	\Gamma_{21}&=&
	-\frac{\kappa^2m^2}{16} 
	\int \frac{d^3 p_2}{(2\pi)^3p_2^0}\int_{0}^{\tau}dt_{\textrm{mic}}\,\textrm{Re} \left [\ddot{u}^{\rm sq}_{\p_2}(\tmic)  \ddot{u}^{\rm sq\,\ast}_{\p_2}(\tmes) \right]I^g(\p_2)
	\nonumber \\&& \times 2
	\left[(\xi_{2}(\tmic)^2- (\xi_{1}(\tmic))^2\right]\xi_{1}(\tmes)\xi_{2}(\tmes)~,
\end{eqnarray}
and 
\begin{eqnarray} \label{gammaUU}
	\Gamma_{22}&=&
	-\frac{\kappa^2m^2}{16} 
	\int \frac{d^3 p_2}{(2\pi)^3p_2^0}\int_{0}^{\tau}dt_{\textrm{mic}}\, \textrm{Re} \left [\ddot{u}^{\rm sq}_{\p_2}(\tmic)  \ddot{u}^{\rm sq\,\ast}_{\p_2}(\tmes) \right]I^g(\p_2)
	\nonumber \\&& \times 
	\left[(\xi_{1}(\tmes))^2- (\xi_{2}(\tmes))^2\right]\left[(\xi_{1}(\tmic))^2- (\xi_{2}(\tmic))^2\right]~,
\end{eqnarray}
in which $\Gamma_{22}$ characterizes the rate of decoherence, and $I^g(\mathbf{p})$ denotes the intensity Stokes parameter of the graviton, as defined in Eq.~\eqref{IGW0}.
The decoherence damping parameter $\Gamma_{22}$, obtained through our non-Markovian approach, aligns with the findings of Kanno et al. using the influence functional method \cite{Kanno:2020usf}.
Utilizing Eq.~\eqref{usq}, we can express the following relation
\begin{eqnarray}
	\!\!\textrm{Re} \left [\ddot{u}^{\rm sq}_{\p_2}(\tmic)  \ddot{u}^{\rm sq\,\ast}_{\p_2}(\tmes) \right]&=&
	|\p_2|^4\cos[|\p_2|(\tmic-\tmes)]\cosh(2r_p)
	\nonumber \\&-&
	|\p_2|^4\cos[|\p_2|(\tmic+\tmes)-\varphi_p]\sinh(2r_p))~,
\end{eqnarray}
where in the Minkowski vacuum ($r_p\rightarrow 0$) we get
\begin{eqnarray}
	\textrm{Re} \left [\ddot{u}^{\rm sq}_{\p_2}(\tmic)  \ddot{u}^{\rm sq\,\ast}_{\p_2}(\tmes) \right]=
	|\p_2|^4\cos[|\p_2|(\tmic-\tmes)]~.
\end{eqnarray}
Regarding the terms $\Gamma_{ij}$, it is conceivable to consider a nonlocal-in-time kernel $M(\tmic,\tmes)$. The presence of this kernel implies that Eqs.~\eqref{dotQ} and \eqref{dotU} describe non-Markovian dynamics.

%%%%%%%%%%%%%%%%%%%%%%%%%%%%%%%%%%%%%%%%%%%%%%

\subsection{A closed loop example  }
\label{section4.2}

As mentioned earlier, the experimental setup involves preparing a massive particle in a quantum superposition at different spatial locations and subsequently recombining the particle to assess its quantum coherence.

In this scenario, we will focus on a specific interference device depicted schematically in Fig.~\eqref{fig:spin0setup}. A mass in free fall can traverse two distinct world lines, $\xi_1$ and $\xi_2$, until it reaches a detector at the end, where its coherence is measured. These paths give rise to a quantum superposition consisting of four straight world line segments with the following configuration \cite{Kanno:2020usf, Breuer:2002pc}
\bea \label{xi1}
\xi_1(t) =
\begin{cases}
	vt+\xi & {\rm for} \quad 0<t\leq \tau/2 \,,\\
	v(\tau-t)+\xi & {\rm for} \quad \tau/2<t<\tau \,,\label{eq:model}
\end{cases} 
\eea
and
\bea \label{xi2}
\xi_2(t) =
\begin{cases}
	- vt+\xi & {\rm for} \quad 0<t\leq \tau/2 \,,\\
	-v(\tau-t)+\xi & {\rm for} \quad \tau/2<t<\tau \,,
\end{cases} 
\eea
where in this setup, we define $\xi=(\xi_{1}(t)+\xi_{2}(t))/2$, in which $\xi_1(t)$ and $\xi_2(t)$ represent the positions of the particle along the respective world lines. The particle's velocity is constant and denoted by $v$, with $0<v<1$. The velocity only changes at $t=\tau/2$ in this particular configuration and at $t=\tau$ the loop can be closed, and correlations can be measured. In Appendix~\ref{appendix:MassiveGammas}, we calculate the values of $\Gamma_{ij}$ for the trajectory described above.

\subsection{Behavior of Decaying Solution }
\label{massivebehavior}
\begin{figure}[t]\centering	
	\vspace{-1cm}
	\includegraphics[width=.9\textwidth]{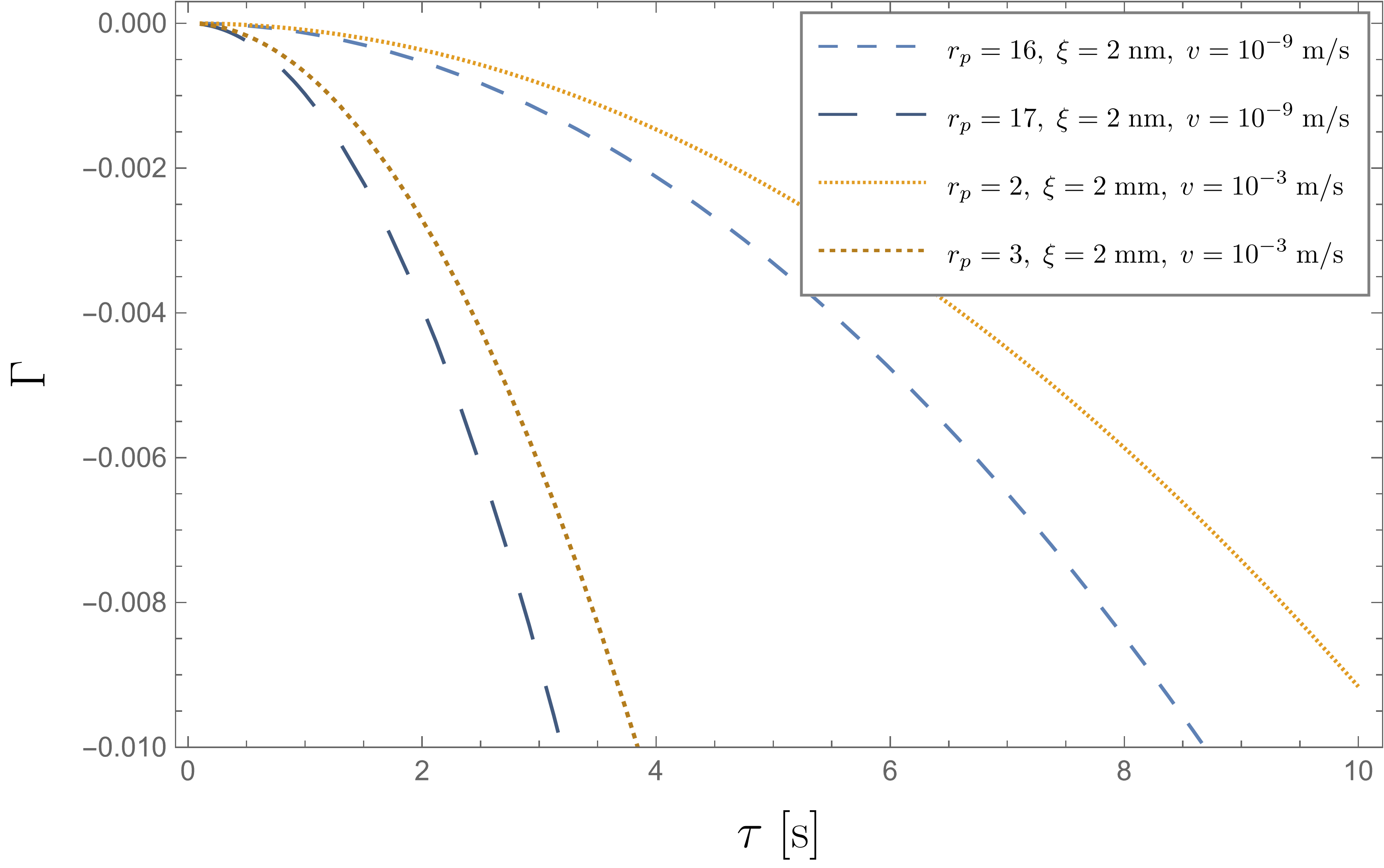}
	\caption{Decoherence rate of the closed loop system of a massive object, specifically a BEC consisting of $10^8$ rubidium ($^{87}\text{Rb}$) atoms with a mass of $1.4\times10^{-17}$~g, within a maximum experiment time of 10 s. The dashed blue curves correspond to an initial spatial distance of $\xi=2~\text{nm}$ and indicate probing with squeezing strengths greater than or equal to $r_p\geq16$. The dotted orange lines represent probing with squeezing strengths greater than or equal to $r_p\geq2$ and an initial spatial distance of $\xi=2$~mm. The squeezing angle, denoted as $\varphi_p$, has a negligible effect in this scenario but is assumed to be $\varphi_p=\pi/2$. }
	\label{Fig:MassiveObject}
\end{figure}

To find a decaying solution for $S_2(t_{\textrm{mes}})$, the system is prepared such that  $\rho^\mathcal{S}_{11}=\rho^\mathcal{S}_{22}$, resulting in $S_1=0$, and $\rho^\mathcal{S}_{12}=\rho^\mathcal{S}_{21}$, which enforces $S_3=0$ on the system. Under certain conditions, $\Gamma_{22}$ dominates over $\Gamma_{21}$, leading to the primary influence of $\Gamma_{22}$ on the evolution of $S_2$ (see Appendix~\ref{appendix:dom} for further details). Consequently, the contribution from $S_1(t_{\textrm{mes}})$ becomes negligible, allowing us to determine the dynamics of the density matrix by solving the differential equation
\begin{eqnarray} \label{dotS2}
	\frac{d}{dt_{\textrm{mes}}}S_2(t_{\textrm{mes}})&=&\Gamma_{22}(\tmes) S_2(\tmes)~.
\end{eqnarray}
Using Eq. \eqref{gammaUU2} one finds the system's dimensionless decoherence rate as 
\begin{eqnarray} \label{gammaDimLess}
	\Gamma&\equiv&\int_{0}^{\tau}\Gamma_{22}(\tmes)\,d\tmes=-\pi\kappa^2m^2 f^2_{\textrm{ref}}A^2_{-1} \xi^2 v^2\tau^2\big(0.09\,\sinh[2r_p]\,\cos[\varphi_p]+0.69\cosh(2 r_p)\big)~.\nonumber\\
\end{eqnarray}
The off-diagonal element $S_2(\tmes)$ undergoes decay over time due to its interaction with the environment, a phenomenon known as decoherence. While the squeezing strength $r_p$ can influence the decoherence rate given by Eq. \eqref{gammaDimLess}, the squeezing angle $\varphi_p$ has a negligible effect on it.

However, to elucidate the decoherence rate $\Gamma$, we begin by considering a specific scenario where the squeezing angle is set to $\varphi_p=\pi/2$. In addition, we consider a test mass with a mass of $1.4\times10^{-17}$ grams, equivalent to the mass of $10^8$ rubidium atoms ($^{87}\text{Rb}$) within a Bose-Einstein condensate (BEC) \cite{D2022}.

 The experiment is carried out over a 10-second duration with a velocity of $v=10^{-9}$~m/s, as well as with a velocity of $v=10^{-3}$~m/s, as outlined in the reference \cite{D2022} for atom interferometers. In accordance with the work of Thrane et al. \cite{thrane2013sensitivity}, we adopt a reference frequency of $f_{\textrm{ref}}=100~\text{Hz}$ for ground-based gravitational wave detectors.

We investigate two distinct values for the spatial distance, which are linked to the velocity and experiment duration via the condition described by Eq.~\eqref{eq:condition}: $\xi$ assumes values of $\xi=2.05$~nm and $\xi=2.05$~mm. The value $A_{-1}/M_\text{P}=10^{-10}$ corresponds to the characteristic strain of stochastic gravitational waves.
In the realm of conventional interferometers, the preference is to employ objects characterized by low mass and low velocity. This choice is aimed at extending the duration of the superposition state, thereby enhancing the efficacy of the interferometer \cite{D2022}.

In Fig.~\eqref{Fig:MassiveObject}, we depict the 1\% decoherence rate over an experimental duration $\tau\leq10~\text{s}$. The dashed lines within the diagram correspond to two different squeezing strengths, namely $r_p=16$ and $r_p=17$, at a velocity of $v=10^{-9}~\text{m/s}$. Similarly, another set of dotted lines represents the squeezing strengths $r_p=2$ and $r_p=3$ at a velocity of $v=10^{-3}$~m/s. In order to achieve a 1\% decoherence within a maximum experimental period of $\tau=10~\text{s}$, the minimum squeezing strengths required are $r_p\geq15.86$ for $v=10^{-9}~\text{m/s}$ and $r_p\geq2.1$ for $v=10^{-3}~\text{m/s}$.

By establishing a connection between the squeezing strength and the frequency of primordial gravitational waves, we have the relation
\begin{eqnarray} \label{squeezevsfreq}
e^{r_p}\sim \left(\frac{f_\text{c}}{f}\right)^2,
\end{eqnarray}
where in the context of grand unified theory inflation, it is noted that $f_\text{c}\sim10^8~\text{Hz}$ \cite{grishchuk1990squeezed,albrecht1994inflation,Kanno:2020usf}. To detect squeezing strengths $r_p\geq15.86$, it is anticipated that the system would be influenced by gravitational waves with the frequency range of $f\leq36~\text{Hz}$. 
Moreover, the Appendix~\ref{appendix:opfreq} identifies the frequency range that exerts the most significant impact on this system. We have shown that gravitational waves (GWs) within the interval of $0 \leq f \leq 6/\tau$ have the predominant influence on this collection of massive objects. For instance, GWs falling within the range of $0 \leq f \leq 0.6$ Hz exhibit the most pronounced effect on experiments lasting 10 seconds. As the experimental duration is extended, this frequency range becomes even more refined, rendering lower frequency GWs more discernible.

In the following section, we demonstrate that a spin-1/2 system can experience the decoherence effect induced by squeezed gravitational waves, and importantly, the decoherence rate remains independent of the choice of origin.

\section{Coupling With a Spin-1/2 System}\label{sec::fermion}
\begin{figure}[t]
	\includegraphics[height=7cm]{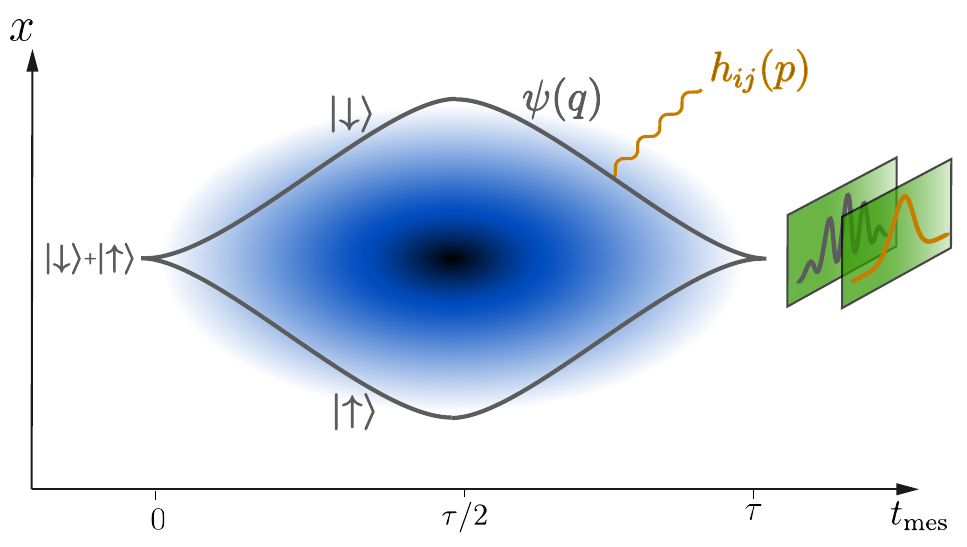}\centering
	\caption{Schematic of the interaction between a fermionic atom and a graviton inside an interferometer.}
	\label{fig:fermionsys}
\end{figure}
Now, we turn to investigate the effect of gravitationally induced decoherence on a fermion system due to bremsstrahlung effect. In a study, the decoherence impact of electromagnetic bremsstrahlung on a fermion system has been discussed \cite{Breuer2001Destruction,Breuer:2002pc}. The framework used in that study has an electron in spatial superposition, and using the decoherence functional it has been demonstrated that the electromagnetic bremsstrahlung causes a fundamental decoherence which briefly predominates. The effect of gravitational bremsstrahlung on a spin-1/2 system is something we also want to look into. We can conceptualize the fermion as an atom with spin-$1/2$ that interacts with a graviton environment. As illustrated in Fig.~\eqref{fig:fermionsys}, our system has two levels and can exist in any quantum superposition of two independent spin states. Despite being macroscopic, we effectively describe it using a microscopic Dirac spinor.
The interaction Hamiltonian between gravitons and the spin-$1/2$ particle is expressed as \cite{Zarei:2021dpb}
\begin{eqnarray}
	H_\text{fg}=\frac{\kappa}{2}\int d^3x\, h_{ij}(x)\partial^i\bar{\psi}(x)\gamma^j\psi(x)~,
\end{eqnarray}
in which the spinor field $\psi$ represents the spin-$1/2$ particles, and $\gamma^i$ are the Dirac matrices. In order to analyze the effects on the spin-1/2 system, we utilize the path of the superposition given by Eqs.~\eqref{xi1} and \eqref{xi2}. To facilitate this analysis, we decompose the spinor field into its annihilation and creation components, denoted by $\psi^+(x)$ and $\psi^-(x)$, respectively. We can then expand these components as follows
\begin{eqnarray} \label{psi1}
	\psi^+(x)&=&\int \frac{d^3q}{(2\pi)^3}\sum_{r}\chi_r(t)\,\hat{a}_r(\q)\,e^{i(q^0t-\bf{q}\cdot\bf{x})}~,\\
	 \label{psi2}
	\psi^-(x)&=&\int \frac{d^3q'}{(2\pi)^3}\sum_{r}\bar{\chi}_r(t)\,\hat{a}^\dag_r(\q')\,e^{-i(q'^0t-\bf{q}'\cdot\bf{x})}~,
\end{eqnarray}
where the spin is labeled by $r=1,2$, and $\chi_r$ represents the non-relativistic free particle spinor. The creation and annihilation operators, denoted by $a_r^\dagger(\mathbf{q})$ and $a_r(\mathbf{q})$ respectively, satisfy the following anti-commutation relation
\begin{align}
	\left\{\, \hat{a}_r({\bf q})\,,\,\hat{a}^\dag_{r'}({\bf q'})\,\right\}
	&=(2\pi)^3\delta^3(\bf q-\bf q')\delta_{rr'}~.
\end{align} 
The spinor $\chi_r$ is presented as
\begin{eqnarray}
	\chi_1(t)=\left(\begin{array}{c}1 \\0 \\\frac{q_z}{2m} \\\frac{q_x}{2m}\end{array}\right)~,~~~~~~~~~~~~~~~~~~
	\chi_2(t)=\left(\begin{array}{c}0 \\1 \\\frac{q_x}{2m} \\-\frac{q_z}{2m}\end{array}\right)~,
\end{eqnarray}
where we assume that a particle is following the trajectory given by Eqs.~\eqref{xi1} and \eqref{xi2}, where $q_y=0$ and $q_x$ is defined as follows
\begin{eqnarray}
	q_x=(-1)^r\big(-1+2\,\Theta(t-\tau/2)\,\big)mv~,
\end{eqnarray}
where $\Theta(t)$ denotes the step function. The effective interaction Hamiltonian that describes the absorption of bremsstrahlung gravitons by fermions can be expressed in terms of the S-matrix element as follows
\begin{eqnarray} \label{Hint1}
	H_{\mathrm{abs}}(t)=\frac{\kappa}{2}\int d^3x\, h^+_{ij}(\x,t)\,\bar{\psi}^-(x)\gamma^j\partial^i \psi^+(x)~,
\end{eqnarray}
and the emission one as
\begin{eqnarray} \label{Hemifermion}
	H_{\mathrm{emi}}(t)=\frac{\kappa}{2}\int d^3x\, h^-_{ij}(\x,t)\,\bar{\psi}^-(x)\gamma^j\partial^i \psi^+(x)~,
\end{eqnarray}
which in total make the interaction Hamiltonian using Eq.~\eqref{eq:Hintmassive}. 
In Appendix \ref{appendix:ferden} there is a detailed calculation of the density matrix elements.
Using Eq.~\eqref{eq:rhofer} as the density matrix components and the relations of \eqref{Stokes1}, the resulting coupling coefficients of the coupled differential equations~\eqref{dotQ} and \eqref{dotU} are as follows
\begin{eqnarray}
	\Gamma_{11}=\Gamma_{12}=\Gamma_{21}=0,
\end{eqnarray}
\begin{eqnarray}\label{eq:fermionGamma}
	\Gamma_{22}&=&-\kappa^2\,m^2\,v^4
	\,\int \frac{d^3 p_2}{(2\pi)^32p_2^0}\,\int_{0}^{\tau}dt_{\textrm{mic}}\,
	\textrm{Re} \left[u^{\rm sq}_{\p_2}(\tmes)  u^{\rm sq\,\ast}_{\p_2}(\tmic) \right]\,I^g(\p_2).
\end{eqnarray}
Employing the mode functions of squeezed states in Eq.~\eqref{usq} leads to
\begin{eqnarray}
	\textrm{Re} \left[u^{\rm sq}_{\p_2}(\tmes)  u^{\rm sq\,\ast}_{\p_2}(\tmic) \right]&=&
	\cos[|\p_2|(\tmic-\tmes)]\cosh(2r_p)
	\nonumber \\&-&
	\cos[|\p_2|(\tmic+\tmes)-\varphi_p]\sinh(2r_p)~.
\end{eqnarray}
Taking $|\p_2|=2\pi f$ and integrating over $t_{\text{mic}}$ the decoherence rate is
\begin{eqnarray}
	\Gamma_{22}&=&\frac{\kappa^2f_{\text{ref}}^2A^2_{-1}v^4\,m^2}{8\,\pi^2}\int_{\Omega_\text{IR}}^\infty \frac{df}{f^4}\Big\{ \cosh[2r_p](\sin[2\pi f(\tmes-\tau)]-\sin[2\pi f \tmes])\nonumber\\
	&+&\sinh[2r_p](\sin[2\pi f(\tmes+\tau)-\varphi_p]-\sin[2\pi f \tmes-\varphi_p])\Big\},
\end{eqnarray}
where $\Omega_\text{IR}$ is the IR cutoff frequency. In this context, we focus on processes that occur over a finite time scale, rather than transitions between asymptotic states. This implies that the decoherence rate does not contain infrared divergences. The reason for this is that the rate pertains to a process that transpires within a finite time interval between the splitting of the wave packet at $t=0$ and its recombination at $t=\tau$. As a result, there is a natural frequency resolution inherent in the process, given by~\cite{Breuer2001Destruction,Breuer:2002pc}
\begin{eqnarray}
	\Omega_{\text{IR}}=\frac{1}{\tau}.
\end{eqnarray}
Then the dimensionless decoherence rate based on Eq.~\eqref{dotU} is
\begin{eqnarray}
	\Gamma&\equiv&\int_{0}^{\tau}\Gamma_{22}(\tmes)\,d\tmes=\frac{\kappa^2f_{\text{ref}}^2A^2_{-1}v^4\,m^2\tau^2}{48\,\pi\, \Omega_{\text{IR}}^2}\nonumber\\
	&&\times\;\bigg\{\sinh[2r_p]\bigg(\cos[\varphi_p]\Big(3-4\pi^2\tau^2\Omega_{\text{IR}}^2\big(-7-\text{Ci}[2\,\pi\,\tau\,\Omega_{\text{IR}}]+8\text{Ci}[4\,\pi\,\tau\,\Omega_{\text{IR}}]\big)\Big)\nonumber\\
	&&+\;\sin[\varphi_p]\big(\pi\tau\Omega_{\text{IR}}(18-14\pi^2\tau\Omega_{\text{IR}})\big)\bigg)-\cosh[2r_p]\big(3-4\pi^2\tau^2\Omega_{\text{IR}}^2(-1+\text{Ci}[2\,\pi\,\tau\,\Omega_{\text{IR}}])\big)\bigg\}~,\nonumber\\
\end{eqnarray}
where $\text{Ci}$ denotes the cosine integral function. The above expression reveals that the spin-1/2 system is influenced by squeezed gravitational waves. Unlike the previous section, the decoherence rate of the fermion depends on the squeezing strength and also in a non-negligible way from the squeezing angle. In the following, we choose specific parameter values to provide an illustrative understanding of this phenomenon and examine the behavior of the decoherence rate.

\begin{figure}[t]
	\vspace{-1cm}
	\includegraphics[height=9cm]{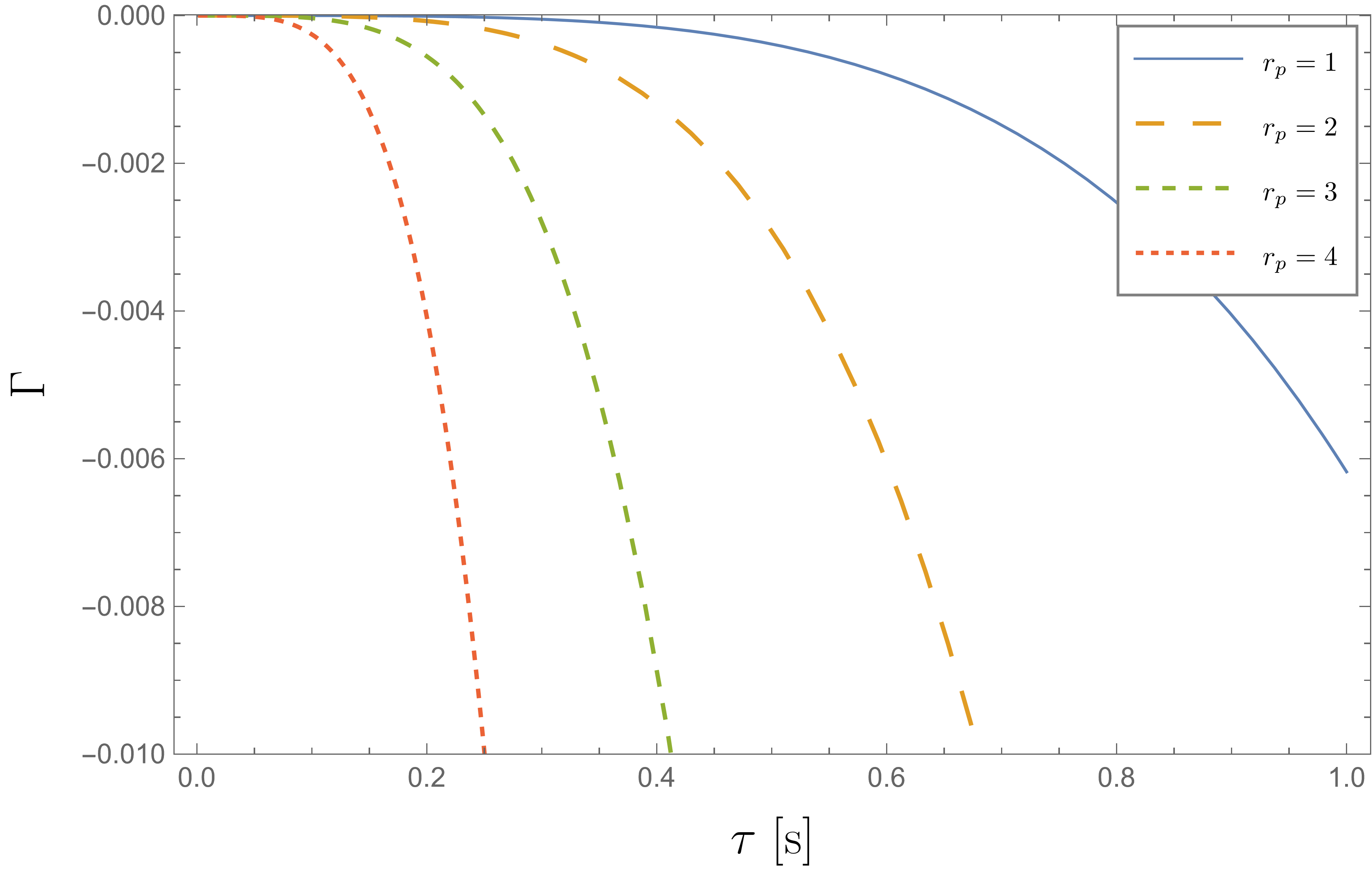}\centering
	\vspace{-.1cm}
	\caption{Decoherence rate of the massive spin-1/2 particle system ($10^4$ atoms of $^{40}\text{K}$ in an atom interferometer) as a function of experiment time. The squeezing strengths considered are $r_p=1,~2,~3,$ and $4$, with a fixed squeezing angle of $\varphi_p=\pi/2$.}
	\label{Fig:S2Fermion}
\end{figure}

\subsection{Behavior of decaying solution }

A practical example to observe this phenomenon is a Stern-Gerlach interferometer, which establishes a correlation between the spin and trajectory of particles \cite{D2022}. In this setup, a beam of atoms is split into two branches that exhibit correlation with the spin component of the atoms. These branches are subsequently recombined before the atoms exit the device. The induced decoherence in the relative phase between the two paths can then be measured.

Another approach to create a spatial superposition for a spin-1/2 particle involves utilizing free-falling nanodiamonds containing nitrogen-vacancy ($\text{NV}^-$) centers. In this method, a nanodiamond is placed in an electron spin superposition state, and an inhomogeneous magnetic field is employed to generate a spatial superposition. The behavior of free-falling nanodiamonds in the presence of an external magnetic field gradient has been extensively studied in various investigations \cite{Scala:2013,Pedernales2020,bose2018,Bennett2012,Asadian2014}. However, it is important to note that in both of these systems, the particles are subjected to a magnetic field, leading to deviations in their wave functions compared to those of free particles. Using an atom interferometer with fermionic atoms \cite{Roati2004,Modugno2004} is another way that does not have the mentioned weakness of other approaches.

\begin{figure}[t]
	\includegraphics[height=8cm]{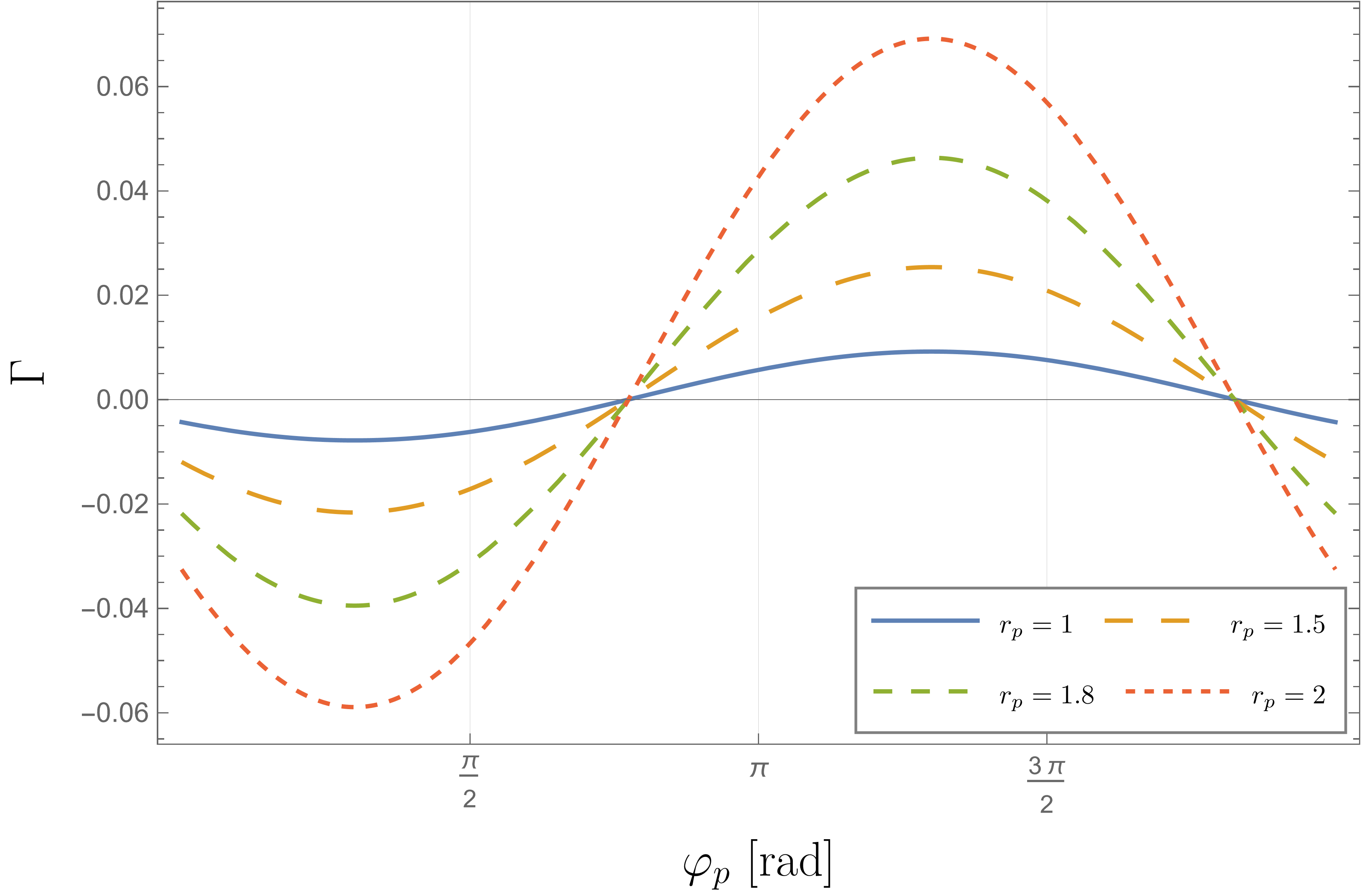}\centering
	\vspace{-.1cm}
	\caption{The decoherence rate of the spin-1/2 system as a function of squeezing angle in an experiment time of $\tau=1~\text{s}$ and by four squeezing strengths, $r_p=1$, $1.5$, $1.8$, and $2$. The mass of the fermion object is assumed to be $m=6.6\times 10^{-19}$~g.}
	\label{Fig:GammaVarphi}
\end{figure}

We consider an atom interferometer that utilizes $10^4$ fermionic atoms of ${}^{40}\text{K}$  ($m=6.6\times10^{-19}~\text{g}$) traveling at a velocity of $v=10^{-4}~\text{m/s}$ \cite{Roati2004}. As discussed in subsection \ref{section4.2}, we assume $f_{\text{ref}}=100~\text{Hz}$ and $A_{-1}=10^{-10}~M_\text{P}$. The resulting decoherence rate is depicted in Fig.~\eqref{Fig:S2Fermion} for $\varphi_p=\pi/2$. Within a measurement time of 1 s, a gravitational wave with a squeezing strength of $r_p\geq1.2$ can induce a 1\% decoherence in the spin-1/2 system. Higher squeezing strengths lead to faster decoherence, resulting in shorter experiment times. It is worth noting that the squeezing angle of the gravitational wave affects the decoherence rate, as shown in Fig.~\eqref{Fig:GammaVarphi}. Positive decoherence rate is observed for certain squeezing angles, which warrants deeper investigation in future studies.

Based on its constraint from the squeezing parameter, the frequency range of gravitational waves (GWs) that can be probed using this fermion-based scheme, as per Eq.~\eqref{squeezevsfreq}, is limited to $f\leq54~\text{MHz}$. However, we have further refined this range in Appendix~\ref{appendix:opfreq} and demonstrated that the influence of GWs on this fermion-massive object system is primarily concentrated within $1/\tau\leq f \leq 2/\tau$. This translates to a frequency span of $1~\text{Hz}\leq f \leq 2~\text{Hz}$ for a 1-second experiment.
%%%%%%%%%%%%%%%%%%%%%%

\section{Conclusions}
\label{section5}

Direct detection of gravitons is extremely difficult due to their weak interactions. This is why the idea of indirect detection has been suggested. One proposed method of indirect detection is through the process of decoherence. 
The theory of decoherence presents an experimental context where quantum mechanics and gravity may intertwine. Through our study utilizing QBE, we have demonstrated that a squeezed gravitational wave can induce decoherence in a system consisting of two spatial states. The existence of squeezing as a purely quantum mechanical trait in gravitational waves would provide compelling evidence of a fundamental connection between gravity and quantum mechanics.

In our first investigation, we applied open quantum system approach to obtain the decoherence rate in a system of two free-falling massive objects, with one of them initially placed in a coherent spatial superposition. Our findings using a simple interaction Hamiltonian are compatible with previous efforts and indicate that the coherence of the system is solely influenced by the squeezing strength, while the squeezing angle has no effect. To provide a detailed description, we considered a BEC composed of $10^8$ atoms of $^{87}$Rb. For an initial distance of $\xi=1~$mm between the two masses, our analysis revealed a 1\% decoherence within a time span of 10 seconds, resulting from the interaction with a gravitational wave possessing squeezing strengths of $r_p\geq2.1$.

Then, we extended our investigation to explore the decoherence effect caused by squeezed gravitational waves on a spin-1/2 particle system existing in a spatial superposition. Notably, in this fermion spatial superposition system, the decoherence effect is dependent on both the squeezing strength and the squeezing angle. Once again, we employed a fermionic atom interferometry containing a cloud of $10^4$ atoms of $^{40}$K, which is placed in a spatial superposition. Our results demonstrated a 1\% decoherence within an experimental time frame of 1 seconds, resulting from the interaction with gravitational waves characterized by squeezing strengths of $r_p\geq1.2$ and a squeezing angle of $\varphi_p=\pi/2$.

The efficiency of gravitational decoherence poses challenges in observing quantum coherence in measurements of primordial (inflationary) fluctuations. The quantum nature of the primordial GWs depends on various details of the system and the environment during inflation. It is possible that the inflation scale may be smaller than anticipated. Additionally, the modes under consideration might not spend as much time outside the Hubble scale during inflation, which can alter the degree of decoherence.
It is worth noting that the level of decoherence required to erase a particular quantum feature can vary. Some studies suggest that decohered states in a de Sitter universe can retain significant quantum discord if decoherence is slow enough. Although gravitational decoherence is highly effective, the complete erasure of quantum discord may not occur, potentially leaving room for detecting quantum signatures.
Exploring quantum effects at smaller scales, where fewer e-folds occur outside the Hubble radius during inflation, could also be a promising avenue of investigation. These smaller-scale structures might provide insights into quantum coherence that efficiently evade decoherence.

The current status regarding the quest for quantum features in the primordial gravitational wave background is not entirely settled. Therefore, the detection of observational evidence of quantum coherence among primordial GWs holds significant promise and could offer profound insights into our understanding of the physics of the early universe. Such measurement would be essential in clarifying the non-classicality criterion and the threshold for the emergence of classicality given by different criteria. This is exemplified by the possibility to put constraints on or to measure the level of squeezing of the GWs, which could bring precious information about the level of decoherence that might have taken place during inflation on certain scales.
If indirect evidence of the quantum nature of gravitational waves is uncovered, it will be essential to reassess the calculations regarding the classical-to-quantum transition during inflation and the dependence of decoherence on the parameters of the underlying models used for the calculations.

Our proposed setup holds interesting potential for investigating the strong equivalence principle (SEP). To conduct this exploration, it is necessary to devise two distinct setups—one involving spin 1/2 particles and the other spinless scalar particles. In addition to examining the interaction of spin-1/2 system, careful consideration must be given to the interaction between spinless scalar particles and squeezed GWs. Subsequently, the induced decoherence in both systems will be compared, allowing us to assess the validity of the SEP in quantum systems. The detailed comparison of the differences in induced decoherence will serve as a means to confirm the validity of SEP.  We leave the investigation of this possibility to future work.

\section*{Acknowledgments}

	We would like to thank M. Peloso for useful discussions and comments. M. Z. would like to thank INFN and the Department of Physics and Astronomy ``G. Galilei” at the University of Padova and also the CERN Theory Division for their warm hospitality while this work was done. N.B. and S.M. thank the partial support for this work by the MUR Departments of Excellence grant 2023-2027 ``Quantum Frontiers".
\appendix
\section{The Intensity of GWs In Terms of Spectral Density} \label{appendixA}
In this appendix, we will calculate the expectation value of tensor modes in terms of gravitational Stokes parameter $I^{(g)}$. Starting from \eqref{fourierh}, we can find
\begin{eqnarray}
	\left< h_{\mu \nu}  \,\, h^{\mu \nu}\right > &=& \left< \int\frac{d^{3}q}{(2\pi)^{3}2q^0} \int\frac{d^{3}q'}{(2\pi)^{3}2q'^{0}}\sum_{r,r'}\left[b^{(r)}_{\mathbf{q}}\,h^{(r)}_{\mu\nu}\,e^{iqx}+
	b^{(r)\,\dag}_{\mathbf{q}}\,h^{(r)\,\ast}_{\mu\nu}\,e^{-iqx}\right] 
	\right.\nonumber \\ & \times& \left.
	\left[b^{(r')}_{\mathbf{q'}}\,h^{\mu\nu(r')}\,e^{iq'x}+
	b^{(r')\,\dag}_{\mathbf{q'}}\,h^{ \mu\nu(r')\ast}\,e^{-iq'x}\right]\right > ~,
\end{eqnarray}
which can be further simplified as
\begin{eqnarray}
	\left<h_{\mu \nu}  \,\, h^{\mu \nu}\right>&=& \int\frac{d^{3}q}{(2\pi)^{3}2q^0} \int\frac{d^{3}q'}{(2\pi)^{3}2q'^0}\sum_{r,r'}\left\{h^{(r)}_{\mu\nu}\,h^{\mu\nu(r')}\,e^{i(q+q')(x)}\,\left < b^{(r)}_{\mathbf{q}}\,b^{(r')}_{\mathbf{q'}}\right >
	\right.\nonumber\\
	&&\left.
	+ h^{(r)}_{\mu\nu}\,h^{\mu\nu (r') \ast }\,e^{i(q-q')(x)}\,\left<b^{(r)}_{\mathbf{q}}\,b^{(r')\,\dag}_{\mathbf{q'}}\right >+h^{(r)\,\ast}_{\mu\nu}\,h^{\mu\nu(r')}\,e^{i(q'-q)(x)}\,\left < b^{(r) \dag}_{\mathbf{q}}\,b^{(r')}_{\mathbf{q'}}\right >
	\right.\nonumber\\
	&&\left. 
	+ h^{(r)\,\ast}_{\mu\nu}\,h^{\mu\nu(r') \ast }\,e^{-i(q'+q)(x)}\,\left <b^{(r) \dag}_{\mathbf{q}}\,b^{(r')\,\dag}_{\mathbf{q'}}\right >\right\} \nonumber\\
	&=&\int\frac{d^{3}q}{(2\pi)^{3}2q^0} \int\frac{d^{3}q'}{(2\pi)^{3}2q'^0}\sum_{r,r'} h^{(r)\,\ast}_{\mu\nu}\,h^{\mu\nu(r')}\,e^{i(q'-q)(x)}\,\left< b^{(r) \dag}_{\mathbf{q}}\,b^{(r')\,}_{\mathbf{q'}}\right > \nonumber\\
	&=&\int\frac{d^{3}q}{(2\pi)^{3}2q^0} \int\frac{d^{3}q'}{(2\pi)^{3}2q'^0}\nonumber\\
	&&\times\sum_{r,r'}  h^{(r)\,\ast}_{\mu\nu}\,h^{\mu\nu (r')}\,e^{i(q'-q)(x)}\, \left(2q'^0 (2\pi)^3 \delta^{(3)}(\mathbf{q'}-\mathbf{q}) \rho^\mathcal{S}_{r r'}(\mathbf{q'})\right)~.
\end{eqnarray}
Now, using Eq. \eqref{IStokes}, we get
\begin{eqnarray}
	\left< h_{\mu \nu}  \,\, h^{\mu \nu}\right >&=& \int\frac{d^{3}q}{(2\pi)^{3}2q^0} \sum_{r,r'}   \rho^\mathcal{S}_{r r'}(\mathbf{q}) \delta_{r,r'}= \int\frac{d^{3}q}{(2\pi)^{3}2q^0}\,I^{(g)}(\mathbf{q})~.
\end{eqnarray}
We can expand $I^{(g)}(\mathbf{q})$ in terms of spherical harmonics as \cite{Bartolo:2018igk}
\begin{eqnarray} \label{harmonic}
	I^{(g)}(\mathbf{q})=I^{(g)}(q^0)\sum_{l,m}c^I_{lm}Y^{m}_{l}(\theta',\phi')~.
\end{eqnarray}
Therefore, one can write
\begin{eqnarray}
	\left< h_{\mu \nu}  \,\, h^{\mu \nu}\right >&=&\frac{1}{4\pi} \int df f I^{(g)}(f)\sum_{l,m}\int d^{2}\hat{\mathbf{q}}\,c^I_{lm}Y^{m}_{l}(\theta',\phi')~,
\end{eqnarray}
where $q^0=2\pi f$. The isotropic monopole part is given by
\begin{eqnarray}
	\left < h_{\mu \nu}  \,\, h^{\mu \nu}\right >= \int df f I^{(g)}(f)~,
\end{eqnarray}
where we have normalized the monopole moment as $c^I_{00}=\sqrt{4\pi}$.
Comparing with \cite{Allen:1997ad,Romano:2016dpx}
\begin{eqnarray}
	\left <h_{\mu \nu}  \,\, h^{\mu \nu}\right >= 4\int df S_h(f)~,
	\end{eqnarray}
and assuming $S_h(f)$ as the GW strain power spectral density, we can derive the following relationship
\begin{eqnarray}
	I^{(g)}(f)= \frac{1}{4f}S_h(f)~.
\end{eqnarray}
The GW strain power spectral density $S_h(f)$ is related to the fractional energy density spectrum in GWs $\Omega_{\textrm{gw}}(f)$ by a simple relation
\begin{eqnarray}
	S_h(f)=\frac{3H_0^2}{2\pi^2}\frac{\Omega_{\textrm{gw}}( f )}{f^3}~,
\end{eqnarray}
where $H_0$ is the Hubble constant. Therefore we have, 
\begin{eqnarray}
	I^{(g)}(f)= \frac{3H_0^2}{8\pi^2}\frac{\Omega_{\textrm{gw}}( f )}{f^4}~.
\end{eqnarray}
One can also express $\Omega_{\textrm{gw}}( f )$ in the form of a power law, as given by \cite{Allen:1997ad,Romano:2016dpx}
\begin{eqnarray}
	\Omega_{\textrm{gw}}( f )= \Omega_{\beta}\left(\frac{f}{f_{\textrm{ref}}}\right)^\beta~,
\end{eqnarray}
where 
\begin{eqnarray}
	\Omega_{\beta}=\frac{2\pi^2}{3H_0^2}f^2_{\textrm{ref}}A^2_{\alpha}~,~~~~~~~ \beta=2(\alpha+1)~.
\end{eqnarray}
Therefore, we find
\begin{eqnarray}
	I^{(g)}(f)=f^2_{\textrm{ref}}A^2_{\alpha} \frac{1}{4f^4}\left(\frac{f}{f_{\textrm{ref}}}\right)^\beta~.
\end{eqnarray}
For inflationary backgrounds, it is often assumed that $\beta =0$ and $\alpha=-1$, thus
\begin{eqnarray}
	\label{IGW0}
	I^{(g)}(f)= \frac{f^2_{\textrm{ref}}A^2_{-1}}{4f^4}~.
\end{eqnarray}
\section{Massive Object System Couplings}
\label{appendix:MassiveGammas}
In this appendix, we aim to calculate $\Gamma_{ij}$ for the closed loop example discussed in Section \ref{section4.2}. To do so, we substitute trajectories \eqref{xi1} and \eqref{xi2} into \eqref{gammaQQ}-\eqref{gammaUU} and integrate over $\p_2$. We can express $|\p_2|$ as $2\pi f$, where $f$ represents the frequency of the gravitational wave (GW). By utilizing \eqref{IGW0}, we can relate the intensity of the GW to its frequency. Therefore, we obtain the coefficients $\Gamma_{ij}$ as follows
\begin{eqnarray} \label{gammaQQ2}
	\Gamma_{11}(\tmes)&=&
	\pi^3\kappa^2m^2 f^2_{\textrm{ref}}A^2_{-1}
	\int_0^{\Omega_\text{UV}} f df \int_{0}^{\tau}dt_{\textrm{mic}}\, \xi_{1}(\tmes)\xi_{2}(\tmes)\xi_{1}(\tmic)\xi_{2}(\tmic)
	\nonumber \\&& \times 
	\big(\cos[2\pi f(\tmic+\tmes)-\varphi_p]\sinh[2r_p]-\cos[2\pi f(\tmic-\tmes)]\cosh[2r_p]\big)
	\nonumber \\&=&
	\frac{1}{4}\pi\kappa^2 m^2 f_{\text{ref}}^2 A^2_{-1}\Bigg\{\sinh[2r_p]\bigg(\frac{\xi^2}{\tmes(\tmes+\tau)}\big(-\tau\cos[\varphi_p]\nonumber\\
	&&+\;(\tmes+\tau)\cos[\varphi_p-2\pi\tmes\Omega_{\text{UV}}]-\tmes\cos[\varphi_p-2\pi(\tmes+\tau)\Omega_{\text{UV}}]\big)\nonumber\\
	&&+\;2v^2\cos[\varphi_p]\Big(\tmes\ln\left[\frac{\tmes}{\tmes+\tau}\right]+\tau\ln\left[\frac{2\,\tmes+\tau}{2(\tmes+\tau)}\right]\Big)\bigg)\nonumber\\
	&&+\;\cosh[2r_p]\left(\frac{1}{\tmes(\tmes-\tau)}\right)\Bigg(\tau\big(\xi^2+2\tmes\,v^2(\tau-\tmes)\big)\nonumber\\&&-\;\tmes\xi^2\cos\left[2\pi(\tmes-\tau)\Omega_{\text{UV}}\right]+(\tmes-\tau)\bigg(2\tmes v^2 \Big(\tmes\ln\left[\frac{\tmes}{\tmes-\tau}\right]\nonumber\\
	&&+\;\tau\ln\left[\frac{\tmes-\tau}{\tmes-\tau/2}\right]\Big)+\xi^2\cos[2\pi\tmes\Omega_{\text{UV}}]\bigg)\Bigg)\Bigg\} \xi_1(\tmes) \xi_2(\tmes) 
	~,\nonumber\\
\end{eqnarray}
\begin{eqnarray} \label{gammaQU2}
	\Gamma_{12}(\tmes)&=&
	-\frac{1}{2}\pi^3\kappa^2m^2 f^2_{\textrm{ref}}A^2_{-1} 
	\int_0^{\Omega_\text{UV}} f df \int_{0}^{\tau}dt_{\textrm{mic}} \,[ \xi^2_{1}(\tmes)-\xi^2_{2}(\tmes)]\xi_{1}(\tmic)\xi_{2}(\tmic)
	\nonumber \\&& \times\;\big(\cos[2\pi f(\tmic+\tmes)-\varphi_p]\sinh[2r_p]-\cos[2\pi f(\tmic-\tmes)]\cosh[2r_p]\big)
	\nonumber \\& = &
	-\frac{1}{2}\pi\kappa^2 m^2 f_{\text{ref}}^2 A^2_{-1}\Bigg\{\sinh[2r_p]\bigg(\frac{\xi^2}{\tmes(\tmes+\tau)}\big(-\tau\cos[\varphi_p]\nonumber\\
	&&+\;(\tmes+\tau)\cos[\varphi_p-2\pi\tmes\Omega_{\text{UV}}]-\tmes\cos[\varphi_p-2\pi(\tmes+\tau)\Omega_{\text{UV}}]\big)\nonumber\\
	&&+\;2v^2\cos[\varphi_p]\Big(\tmes\ln\left[\frac{\tmes}{\tmes+\tau}\right]+\tau\ln\left[\frac{2\,\tmes+\tau}{2(\tmes+\tau)}\right]\Big)\bigg)\nonumber\\
	&&+\;\cosh[2r_p]\left(\frac{1}{\tmes(\tmes-\tau)}\right)\Bigg(\tau\big(\xi^2+2\tmes\,v^2(\tau-\tmes)\big)\nonumber\\&&-\;\tmes\xi^2\cos\left[2\pi(\tmes-\tau)\Omega_{\text{UV}}\right]+(\tmes-\tau)\bigg(2\tmes v^2 \Big(\tmes\ln\left[\frac{\tmes}{\tmes-\tau}\right]\nonumber\\
	&&+\;\tau\ln\left[\frac{\tmes-\tau}{\tmes-\tau/2}\right]\Big)+\xi^2\cos[2\pi\tmes\Omega_{\text{UV}}]\bigg)\Bigg)\Bigg\} \left[(\xi_{1}(\tmes))^2- (\xi_{2}(\tmes))^2\right]
	~,\nonumber\\
\end{eqnarray}
\begin{eqnarray} \label{gammaUQ2}
	\Gamma_{21}(\tmes)&=&
	-\frac{1}{2}\pi^3\kappa^2m^2 f^2_{\textrm{ref}}A^2_{-1}
	\int_0^{\infty} f df \int_{0}^{\tau}dt_{\textrm{mic}}
	\left[(\xi_{1}(\tmic))^2- (\xi_{2}(\tmic))^2\right]\xi_{1}(\tmes)\xi_{2}(\tmes)
	\nonumber \\&&\times\;
	\big(\cos[2\pi f(\tmic+\tmes)-\varphi_p]\sinh[2r_p]-\cos[2\pi f(\tmic-\tmes)]\cosh[2r_p]\big)
	\nonumber \\& = &
	\frac{1}{2}\pi\kappa^2 m^2 f_{\text{ref}}^2 A^2_{-1}\,\xi\,v\,\left\{\sinh[2r_p]\cos[\varphi_p]\ln\left[\frac{4\tmes(\tmes+\tau)}{(2\tmes+\tau)^2}\right]\right.\nonumber\\
	&&-\;\left.\cosh[2r_p]\ln\left[\frac{4\tmes(\tmes-\tau)}{(2\tmes-\tau)^2}\right]\right\}\xi_1(\tmes) \xi_2(\tmes) 
	~,\nonumber\\
\end{eqnarray}
and 
\begin{eqnarray} \label{gammaUU2}
	\Gamma_{22}(\tmes)&=&
	\frac{1}{4}\pi^3\kappa^2m^2 f^2_{\textrm{ref}}A^2_{-1} 
	 \int_{0}^{\tau}dt_{\textrm{mic}} \left[(\xi_{1}(\tmic))^2- (\xi_{2}(\tmic))^2\right]\left[(\xi_{1}(\tmes))^2- (\xi_{2}(\tmes))^2\right]
	\nonumber \\&\times&
	\int_{0}^{\Omega_{\text{UV}}} f df\big(\cos[2\pi f(\tmic+\tmes)-\varphi_p]\sinh[2r_p]-\cos[2\pi f(\tmic-\tmes)]\cosh[2r_p]\big)
	\nonumber \\&=&
	\frac{1}{4}\pi\kappa^2m^2 f^2_{\textrm{ref}}A^2_{-1}\,\xi\,v \Bigg\{
	\sinh[2r_p]\cos[\varphi_p]\ln\left[\frac{4\,\tmes(\tmes+\tau)}{(2\tmes+\tau)^2}\right]\nonumber \\
	&& -\;\cosh[2r_p]\ln\left[\frac{4\,\tmes(\tmes-\tau)}{(2\tmes-\tau)^2}\right]\Bigg\} \left[(\xi_{1}(\tmes))^2- (\xi_{2}(\tmes))^2\right]
	~,\nonumber\\
\end{eqnarray}
where the coefficient expressions for $\Gamma_{11}$ and $\Gamma_{12}$ involve a UV cutoff $\Omega_{\text{UV}}$ for the frequency of graviton modes. In the calculation of $\Gamma_{21}$ and $\Gamma_{22}$, we have used the following approximation for  cosine integrals 
\begin{eqnarray}
	\text{Ci}(f)\simeq \ln{(f)},
\end{eqnarray}
In experimental scenarios, a characteristic acceleration time usually aligns with the peak frequency within the power spectrum of the force exerted on the massive object. This inherent upper limit, known as the natural UV cutoff, is defined by the acceleration time, which is proportional to the minimal wavelength of radiation and corresponds to the Compton wavelength of the object \cite{Breuer:2002pc,toros2020}. However, it's noteworthy that no such UV cutoff exists in $\Gamma_{22}$ as utilized within the context of the text.

\section{Exploring the Predominant Domain of $\Gamma_{22}$}\label{appendix:dom}

 \begin{figure}[t]
 	\vspace{-1cm}
 	\includegraphics[height=9cm]{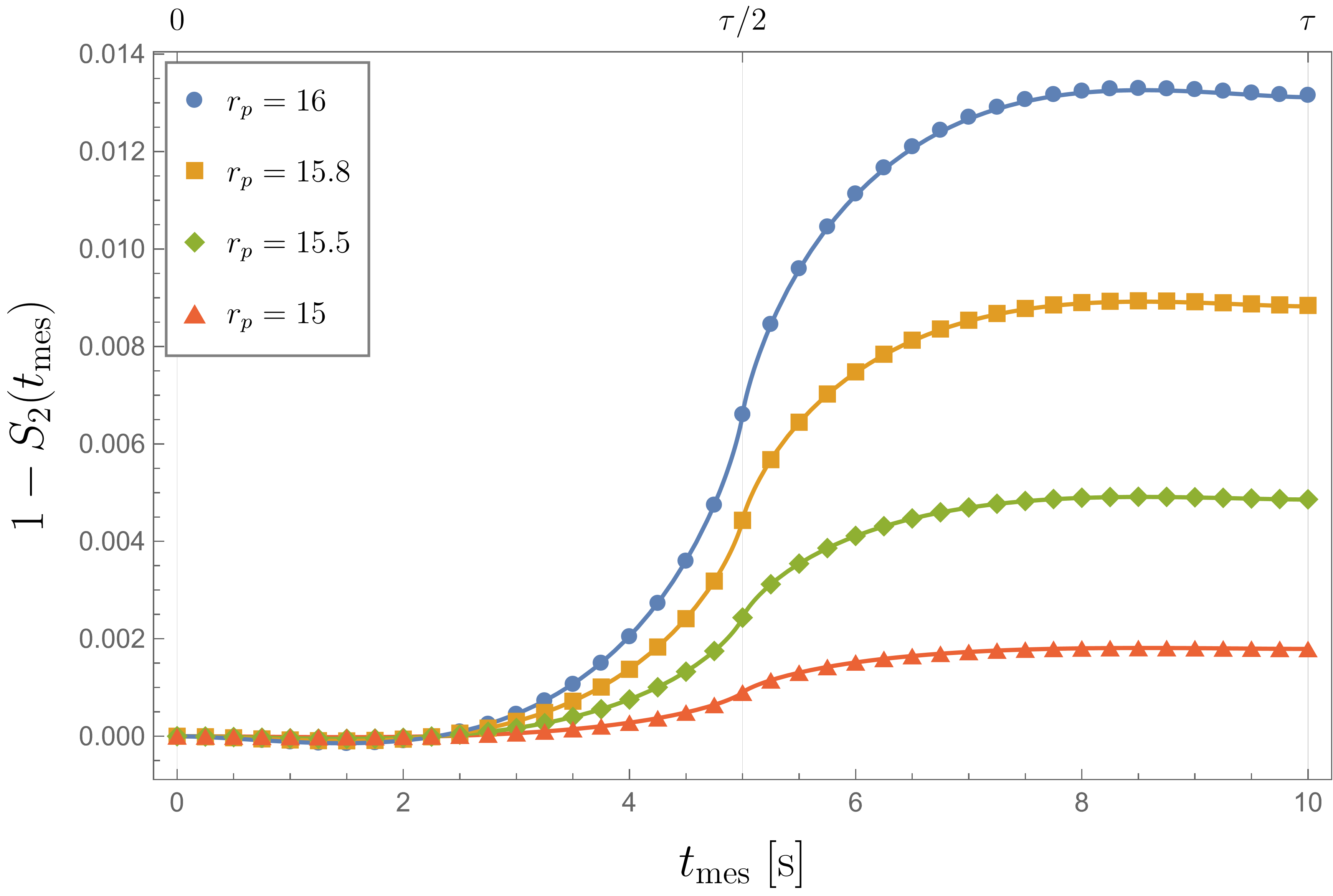}\centering
 	\caption{Numerical Comparison of $S_2(\tmes)$ under the condition of Eq.~\eqref{eq:condition} in two scenarios: with and without $\Gamma_{21}$ depicted by solid lines and markers, respectively. The red, green, orange, and blue plots represent $r_p=15,~15.5,~15.8,~\text{and}~16$, respectively. The presented results are obtained using the following parameters: $v=10^{-9}$~m/s, $\tau=10$~s, and $\xi=2.05~\text{nm}$. The remaining parameters match those used in Fig.~\eqref{Fig:MassiveObject}.}	\label{Fig:S2}
  \end{figure}
This section elucidates the circumstances under which $\Gamma_{22}$ takes precedence over $\Gamma_{21}$ and when Eq.~\eqref{dotU} can be streamlined to Eq.~\eqref{dotS2}. By employing the trajectory of the the massive object as given in Eqs.~\eqref{xi1} and \eqref{xi2}, and juxtaposing Eq.~\eqref{gammaUQ} with Eq.~\eqref{gammaUU}, the outcome is obtained as follows
\begin{eqnarray}\label{eq:gammacompare}
	\frac{\Gamma_{22}}{\Gamma_{21}}=\frac{2\xi_{1}(\tmes)\xi_{2}(\tmes)}{(\xi_{2}(\tmes))^2-(\xi_{1}(\tmes))^2}=
	\begin{cases}
		\frac{v\tmes}{2\xi}-\frac{\xi}{2v\tmes} & {\rm for} \quad 0<\tmes\leq \tau/2 \,,\\
		\frac{(v\tmes+\xi-v\tau)(-v\tmes+\xi+v\tau)}{2v\xi(\tmes-\tau)} & {\rm for} \quad \tau/2<\tmes<\tau \,.
		\end{cases} \nonumber\\
\end{eqnarray}
To ensure that $\Gamma_{22}/\Gamma_{21}$ is greater than 1, the subsequent condition for $\xi$ must be satisfied
\begin{eqnarray}\label{eq:condition}
	\xi<\frac{v\tau}{2(1+\sqrt{2})}\quad\text{for}\quad(1+\sqrt{2})\frac{\xi}{v}<\tmes<\tau-(1+\sqrt{2})\frac{\xi}{v}\,.
\end{eqnarray}
This condition demonstrates that $\xi$ should be bounded. To provide a numerical perspective, we opt for $r_p=15$-16, $v=10^{-9}~\text{m/s}$, and $\tau=10~\text{s}$. The remaining parameters are akin to those employed in subsection~\ref{massivebehavior}. Using Eq.~\eqref{eq:condition}, we find $\xi=\frac{0.99v\tau}{2(1+\sqrt{2})}=2.05~\text{nm}$. Subsequently, we numerically solve Eqs.~\eqref{dotQ}~\&~\eqref{dotU} by employing Mathematica's \verb|NDSolve| function and compare $S_2(\tmes)$ with the numeric result of \eqref{dotS2}. Additionally, we have employed the Compton frequency of the massive object as the ultraviolet cutoff in $\Gamma_{11}$ and $\Gamma_{12}$, as referenced in \cite{Breuer:2002pc}. The outcome of $S_2(t_{\text{mes}})$ is illustrated in Fig.~\eqref{Fig:S2}, showcasing a seamless solution of the two differential equations. The parameters utilized in the main text are deliberately selected to fulfill the requirement of Eq.~\eqref{eq:condition}. Therefore, in the condition of \eqref{eq:condition} we can shorten the coupled differential equations of $S_1(\tmes)$ and $S_2(\tmes)$ into one independent differential equation in the form of Eq.~\eqref{dotS2}.
  
\section{Density Matrix Calculations for the Spin-1/2 System}\label{appendix:ferden}

Here we want to calculate the density matrix elements for the fermionic system of section \ref{sec::fermion}. Begining with the  spinor fields given by Eqs.~\eqref{psi1} and \eqref{psi2}, and the graviton field expressed in Eq.~\eqref{fourierh} and put them into \eqref{Hint1}, we obtain
\begin{eqnarray} \label{hgnu2}
	H_{\textrm{abs}}(t)&=& -\frac{i}{2}\kappa \int d^3x\, \frac{d^3q}{(2\pi)^3}\frac{d^3q'}{(2\pi)^3}\, \frac{d^3 p}{(2\pi)^32p^0}\nonumber\\
	&\times&\sum_{s,r,r'}
	e^s_{ij}(\p)u^s_{\p}(t) \bar{\chi}_{r'}(t)q^{i}\gamma^{j}\chi_r(t)
	\,b_s(p)\hat{a}^{\dag}_{r'}\hat{a}_{r}\,e^{i(q^0-q'^0)t}e^{-i(\bf{q}-\q'-\p)\cdot\bf{x}}\,
	~,
\end{eqnarray}
where we have approximated $\partial^j \psi^+ \sim -iq^j \psi^+$, assuming $q^j$ represents the momentum of the particle.
Next, we express the density matrix $\rho^\mathcal{S}$ of the spin-1/2 system in terms of the Bloch vectors \eqref{rhoS}. The evolution of the density matrix elements is given by \cite{Zarei:2021dpb}
\begin{eqnarray}
	(2\pi)^3\delta^3(0)\frac{d}{dt_{\textrm{mes}}}\rho^\mathcal{S}_{ij}(t_{\textrm{mes}})= D_{ij}[\rho^\mathcal{S}(t_{\textrm{mes}})]~.
\end{eqnarray}
The dissipator is obtained by substituting the spin-1/2 interaction Hamiltonian (sum of Eqs.~\eqref{Hint1} \&~\eqref{Hemifermion}) into Eq.~\eqref{damping2} and can be expressed in the following form
\begin{eqnarray}
	&&D_{ij}[\rho^\mathcal{S}(t_{\textrm{mes}})]=-\int_{0}^{\tau}dt_{\textrm{mic}}\left<\left[H_{\textrm{int}}(t_{\textrm{mes}}),\left[H^{0\dag}_{\textrm{int}}(t_{\textrm{mic}}) ,\hat{\mathcal{N}}^S_{ij}(t_{\textrm{mes}}-t_{\textrm{mic}})\right]\right]\right>_\textrm{c} \nonumber \\&=&
	-\frac{\kappa^2}{2} \sum_{s_1,r_1,r'_1}\sum_{s_2,r_2,r'_2}\int_{0}^{\tau}dt_{\textrm{mic}}\int d^3x_1\,d^3x_2\,\frac{d^3q_1}{(2\pi)^3}\frac{d^3q'_1}{(2\pi)^3}\frac{d^3q_2}{(2\pi)^3}\frac{d^3q'_2}{(2\pi)^3}\frac{d^3 p_1}{(2\pi)^32p_1^0}\, \frac{d^3 p_2}{(2\pi)^32p_2^0}\,\,
	\nonumber \\&&\times\,\textrm{Re} \left [u^{\rm sq}_{\p_1}(\tmes)  u^{\rm sq\,\ast}_{\p_2}(\tmic) \right] e^{-i (\q_1-\q'_1-\p_1) \cdot {\x_1}}e^{-i (\q_2-\q'_2-\p_2) \cdot {\x_2}}e^{i (q^0_1-q'^0_1-p^0_1) \cdot t_{\text mes}}e^{i (q^0_2-q'^0_2-p^0_2) \cdot t_{\text mic}}\nonumber\\
	&&\times e_{m_1n_1}^{(s_1)}(\p_1)e_{m_2n_2}^{(s_2)\ast}(\p_2)\bar{\chi}_{r_1}(\tmes)\gamma^{m_1}q^{n_1}_{\text{mes}}\,   \chi_{r'_1}(\tmes)    \bar{\chi}_{r_2}(\tmic)\gamma^{m_2}q^{n_2}_{\text{mic}}\,   \chi_{r'_2}(\tmic) 
	\nonumber \\&& \times\,  \left< b^{(s_2\dag)}(\mathbf{ p_2})b^{(s_1)}(\mathbf{ p_1})\right>_\textrm{c}\left[
	\left<\hat{a}^{(r_1)\dag} \hat{a}^{(r'_1)}\hat{a}^{(r'_2)\dag}\hat{a}^{(r_2)} \hat{a}^{(i)\dag} \hat{a}^{(j)} \right> _\textrm{c}- \left<\hat{a}^{(r_1)\dag} \hat{a}^{(r'_1)}\hat{a}^{(i)\dag} \hat{a}^{(j)}\hat{a}^{(r'_2)\dag}\hat{a}^{(r_2)}  \right>_\textrm{c}
	\right. \nonumber  \\  &&  \left.  -
	\left<\hat{a}^{(r'_2)\dag}\hat{a}^{(r_2)} \hat{a}^{(i)\dag} \hat{a}^{(j)}\hat{a}^{(r_1)\dag} \hat{a}^{(r'_1)} \right>_\textrm{c} +\left< \hat{a}^{(i)\dag} \hat{a}^{(j)}\hat{a}^{(r'_2)\dag}\hat{a}^{(r_2)} \hat{a}^{(r_1)\dag} \hat{a}^{(r'_1)} \right>_\textrm{c}~
	\right ]~.
\end{eqnarray}
By utilizing expectation values of Eqs. \eqref{EVGW} and \eqref{EVS}, integrating over $x$, and substituting the obtained dissipator into \eqref{boltzmanneq3}, we arrive at
\begin{eqnarray}
	&&(2\pi)^3\delta^3(0)\frac{d}{dt_{\textrm{mes}}}\rho^\mathcal{S}_{ij}(t_{\textrm{mes}})= \nonumber \\&=&
	-\frac{\kappa^2}{2} \sum_{s_1,r_1,r'_1}\sum_{s_2,r_2,r'_2}\int_{0}^{\tau}dt_{\textrm{mic}}\int \,\frac{d^3q_1}{(2\pi)^3}\frac{d^3q'_1}{(2\pi)^3}\frac{d^3q_2}{(2\pi)^3}\frac{d^3q'_2}{(2\pi)^3}\frac{d^3 p_1}{(2\pi)^32p_1^0}\, \frac{d^3 p_2}{(2\pi)^32p_2^0}\,\,
	\nonumber \\&&\times\,\textrm{Re} \left [u^{\rm sq}_{\p_1}(\tmes)  u^{\rm sq\,\ast}_{\p_2}(\tmic) \right] (2\pi)^3\delta^3(\q_1-\q'_1-\p_1)(2\pi)^3\delta^3(\q_2-\q'_2-\p_2)\nonumber\\&& \times\: \bar{\chi}_{r_1}(\tmes)\gamma^{m_1}q^{n_1}_{\text{mes}}\,\chi_{r'_1}(\tmes) \bar{\chi}_{r_2}(\tmic)\gamma^{m_2}q^{n_2}_{\text{mic}}\,\chi_{r'_2}(\tmic) 
	e_{m_1n_1}^{(s_1)}(\p_1)e_{m_2n_2}^{(s_2)\ast}(\p_2)
	\nonumber \\&& \times(2\pi)^32p^0_1\delta^3(\p_1-\p_2)\rho^g_{s_1s_2}(\p_1)\,e^{i (q^0_1-q'^0_1-p^0_1) \cdot t_{\text mes}}e^{i (q^0_2-q'^0_2-p^0_2) \cdot t_{\text mic}}\nonumber\\ &&\times\Big[ \rho^\mathcal{S}_{jr_1}(2\pi)^3\delta^3(\k-\q_1)\delta^{ir_2}(2\pi)^3\delta^3(\k-\q_2)\delta^{r'_2r'_1}(2\pi)^3\delta^3(\q'_2-\q'_1)\nonumber\\
	&&\;-\rho^\mathcal{S}_{r_2r_1}(2\pi)^3\delta^3(\q_2-\q_1)\delta^{ir'_1}(2\pi)^3\delta^3(\k-\q'_1)\delta^{r'_2j}(2\pi)^3\delta^3(\k-\q'_2)\nonumber\\
	&&\;-\rho^\mathcal{S}_{r'_1r'_2}(2\pi)^3\delta^3(\q'_2-\q'_1)\delta^{r_1j}(2\pi)^3\delta^3(\k-\q_1)\delta^{ir_2}(2\pi)^3\delta^3(\k-\q_2)
	\nonumber\\
	&&\;+\rho^\mathcal{S}_{r'_1i}(2\pi)^3\delta^3(\q'_1-\k)\delta^{r_1r_2}(2\pi)^3\delta^3(\q_1-\q_2)\delta^{r'_2j}(2\pi)^3\delta^3(\k-\q'_2)
	\Big].
\end{eqnarray}
Integrating over $q_1, q_2, q'_1, q'_2,$ and $p_1$, we obtain
\begin{eqnarray}\label{eq:rhofer}
	\frac{d}{dt_{\textrm{mes}}}\rho^\mathcal{S}_{ij}(t_{\textrm{mes}})&=&-\frac{\kappa^2}{2} \sum_{s_1,r_1,r'_1}\sum_{s_2,r_2,r'_2}
	\,\int \frac{d^3 p_2}{(2\pi)^32p_2^0}\,\int_{0}^{\tau}dt_{\textrm{mic}}\,
	\textrm{Re} \left[u^{\rm sq}_{\p_2}(\tmes)  u^{\rm sq\,\ast}_{\p_2}(\tmic) \right] \nonumber \\&\times& e_{m_1n_1}^{(s_1)}(\p_2)e_{m_2n_2}^{(s_2)\ast}(\p_2)\,\bar{\chi}_{r_1}(\tmes)\gamma^{m_1}k^{n_1}_{\text{mes}}   \chi_{r'_1}(\tmes)    \bar{\chi}_{r_2}(\tmic)\gamma^{m_2}k^{n_2}_{\text{mic}}     \chi_{r'_2}(\tmic) 
	\nonumber  \\&\times& \rho^g_{s_1s_2}(\p_2) \left[ \rho^\mathcal{S}_{jr_1}\delta^{ir_2}\delta^{r'_2r'_1}
	-\rho^\mathcal{S}_{r_2r_1}\delta^{ir'_1}\delta^{r'_2j}
	-\rho^\mathcal{S}_{r'_1r'_2}\delta^{r_1j}\delta^{ir_2}
	+ \rho^\mathcal{S}_{r'_1i}\delta^{r_1r_2}\delta^{r'_2j}
	\right]~.\nonumber\\
\end{eqnarray}
The time evolution of the off-diagonal components of the density matrix, described by $S_1$ and $S_2$ can be evaluated using differential equations.~\eqref{dotQ} and \eqref{dotU}, similar to the case of decoherence in a massive object discussed in Section \ref{sec::massivedecoherence}.

\section{Characterizing the GW Frequencies Detectable by the New Scheme}\label{appendix:opfreq}
\begin{figure}[t]
	\vspace{-1cm}
	\includegraphics[height=8cm]{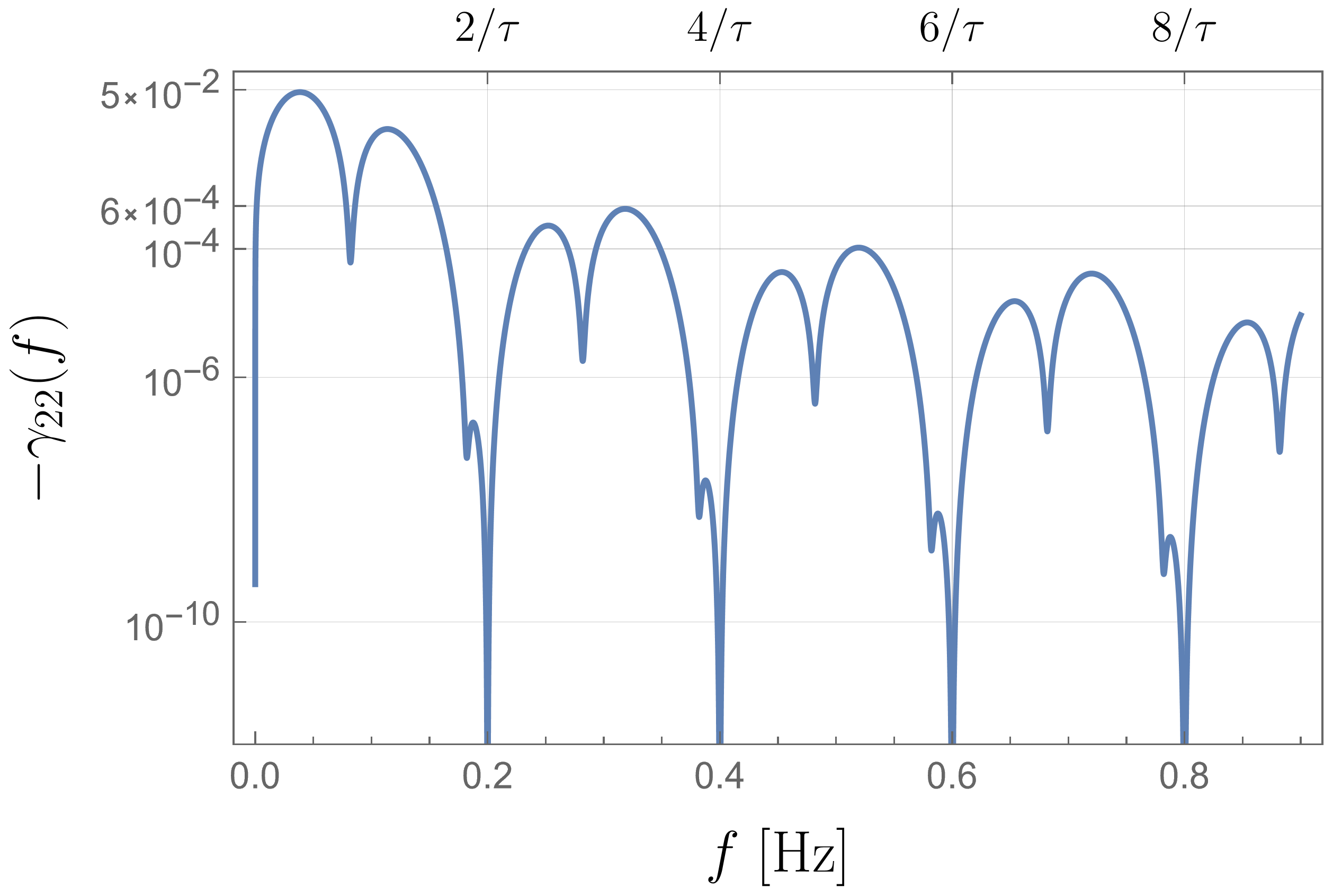}\centering
	\caption{The integrand of decoherence rate in massive object system in terms of the frequency. This figure was generated using identical parameters as those in Fig.~\eqref{Fig:MassiveObject}, with the following values: $r_p=24$ and $\xi=10~\text{nm}$.}
	\label{Fig:GammaIntegrand1}
\end{figure}
In this appendix, our aim is to identify the frequency band of the SGWB in which the schemes operate with the highest efficiency.
\subsection{Massive Object}
Commencing with Eq.~\eqref{gammaUU} and performing integrations over $\tmes$ and $\tmic$, the resultant integrand, $\gamma_{22}(f)$, within the $f$ integration, $\Gamma=\int\gamma_{22}(f)df$, is obtained as follows
\begin{eqnarray}\label{eq:integrand1}
	\gamma_{22}(f)=-4 \kappa^2f_{\text{ref}}^2A^2_{-1}v^2\,m^2\xi^2\frac{\sin^4[\pi f\tau/2]}{\pi f^3}\big(\cosh[2r_p]-\cos[2\pi f\tau-\varphi_p]\sinh[2r_p]\big)~.
\end{eqnarray}
This function has zeros at $f=2n/\tau$ where $n=1, 2, 3,\dots$, as illustrated in Fig.~\eqref{Fig:GammaIntegrand1}. Integrating $\gamma_{22}(f)$ over two regions in frequency
\begin{eqnarray}
	\int_{0}^{6/\tau}df\,\gamma_{22}(f)&=&\pi\kappa^2m^2 f^2_{\textrm{ref}}A^2_{-1} \xi^2 v^2\tau^2e^{2r_p}\big(-0.691-0.0921\cos[\varphi_p]+7\times10^{-5}\sin[\varphi_p]\big)~,\nonumber\\
	\int_{0}^{\infty}df\,\gamma_{22}(f)&=&\pi\kappa^2m^2 f^2_{\textrm{ref}}A^2_{-1} \xi^2 v^2\tau^2e^{2r_p}\big(-0.693-0.0920\cos[\varphi_p]\big)~,
\end{eqnarray}
shows that the dominant contribution to the integral arises from the range where $0\leq f\leq6/\tau$. As a result, the effect on the massive object system is notably pronounced within the frequency spectrum up to $f=6/\tau$. For instance, an experimental duration of 10 seconds is most significantly influenced by frequencies in the range of $0\leq f \leq 0.6~\text{Hz}$.
\subsection{Massive Spin-1/2 Object}
\begin{figure}[t]
	\vspace{-1cm}
	\includegraphics[height=8cm]{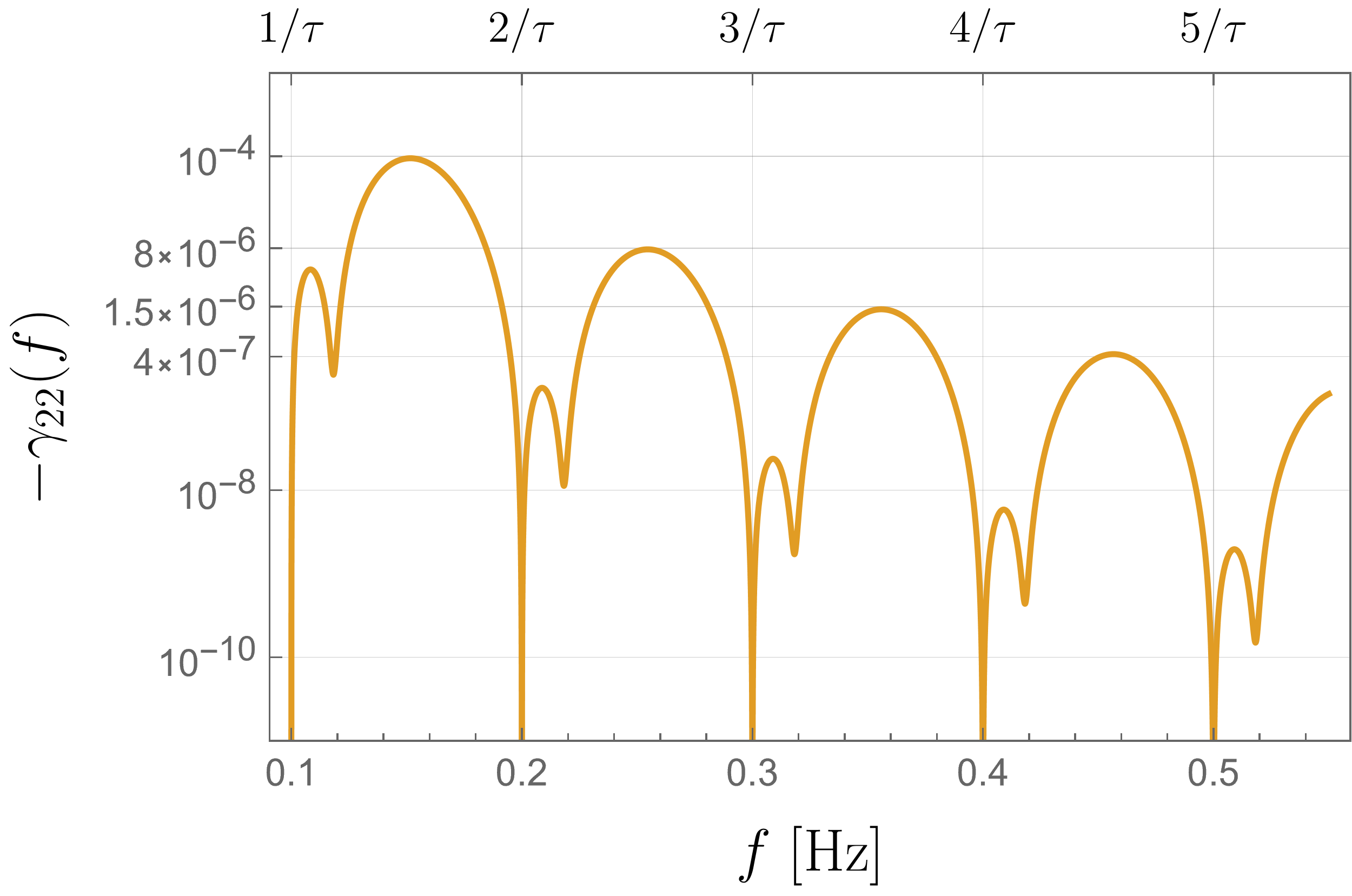}\centering
	\vspace{-.1cm}
	\caption{
		The integrand of decoherence rate in spin-1/2 system in terms of the frequency. The parameters utilized to create this figure remain in accordance with those employed in Fig.~\eqref{Fig:S2Fermion}, which encompass $r_p=20$ and $\varphi_p=\pi/2$.}
	\label{Fig:GammaIntegrand2}
\end{figure}

The integrand $\gamma_{22}(f)$ for the spin-1/2 system is acquired through a method akin to that described in Eq.~\eqref{eq:integrand1}.
%The integrand $\gamma_{22}(f)$ for the spin-1/2 system is obtained using a similar approach as given by Eq.~\eqref{eq:fermionGamma}
\begin{eqnarray}
	\gamma_{22}(f)=- \kappa^2f_{\text{ref}}^2A^2_{-1}v^4\,m^2\,\frac{\sin^2[\pi f\tau]}{4\pi^3 f^3}\big(\cosh[2r_p]-\cos[2\pi f\tau-\varphi_p]\sinh[2r_p]\big)~,
\end{eqnarray}
which is illustrated over the interval of $1/\tau \leq f \leq 5.5/\tau$ in Fig.~\eqref{Fig:GammaIntegrand2}, exhibiting zeros at $f=n/\tau$ where $n=1, 2, 3,\dots$. In the following, we demonstrate that the principal effective range of the integral is situated between the first and second zeros,
\begin{eqnarray}
	\int_{\frac{1}{\tau}}^{\frac{2}{\tau}}df\,\gamma_{22}(f)&\simeq&\frac{-1}{\pi}\kappa^2m^2 f^2_{\textrm{ref}}A^2_{-1}  v^4\tau^4e^{2r_p}\big(0.0021+7.7\times10^{-4}\cos[\varphi_p]-8\times10^{-4}\sin[\varphi_p]\big)~,\nonumber\\
	\int_{\frac{1}{\tau}}^{\infty}df\,\gamma_{22}(f)&\simeq&\frac{-1}{\pi}\kappa^2m^2 f^2_{\textrm{ref}}A^2_{-1}  v^4\tau^4e^{2r_p}\big(0.0022+8.5\times10^{-4}\cos[\varphi_p]-8.4\times10^{-4}\sin[\varphi_p]\big)~.\nonumber\\
\end{eqnarray}
The system is most profoundly influenced by the primary frequency band of $\frac{1}{\tau}\leq f \leq \frac{2}{\tau}$. For instance, GWs within the range of $1~\text{Hz}\leq f \leq 2~\text{Hz}$ exert the strongest impact on a fermion experiment with a duration of $\tau=1$~s. 

%%%%%%%%%%%%%%%%%%%%%%%
%	Written by Mohammad Sharifian
\nocite{apsrev41Control} % for adding title of articles to the references
\bibliographystyle{apsrev4-1}
\bibliography{refcontrol,references} % add your references to the file references.bib

\begin{thebibliography}{73}%
\makeatletter
\providecommand \@ifxundefined [1]{%
 \@ifx{#1\undefined}
}%
\providecommand \@ifnum [1]{%
 \ifnum #1\expandafter \@firstoftwo
 \else \expandafter \@secondoftwo
 \fi
}%
\providecommand \@ifx [1]{%
 \ifx #1\expandafter \@firstoftwo
 \else \expandafter \@secondoftwo
 \fi
}%
\providecommand \natexlab [1]{#1}%
\providecommand \enquote  [1]{``#1''}%
\providecommand \bibnamefont  [1]{#1}%
\providecommand \bibfnamefont [1]{#1}%
\providecommand \citenamefont [1]{#1}%
\providecommand \href@noop [0]{\@secondoftwo}%
\providecommand \href [0]{\begingroup \@sanitize@url \@href}%
\providecommand \@href[1]{\@@startlink{#1}\@@href}%
\providecommand \@@href[1]{\endgroup#1\@@endlink}%
\providecommand \@sanitize@url [0]{\catcode `\\12\catcode `\$12\catcode
  `\&12\catcode `\#12\catcode `\^12\catcode `\_12\catcode `\%12\relax}%
\providecommand \@@startlink[1]{}%
\providecommand \@@endlink[0]{}%
\providecommand \url  [0]{\begingroup\@sanitize@url \@url }%
\providecommand \@url [1]{\endgroup\@href {#1}{\urlprefix }}%
\providecommand \urlprefix  [0]{URL }%
\providecommand \Eprint [0]{\href }%
\providecommand \doibase [0]{http://dx.doi.org/}%
\providecommand \selectlanguage [0]{\@gobble}%
\providecommand \bibinfo  [0]{\@secondoftwo}%
\providecommand \bibfield  [0]{\@secondoftwo}%
\providecommand \translation [1]{[#1]}%
\providecommand \BibitemOpen [0]{}%
\providecommand \bibitemStop [0]{}%
\providecommand \bibitemNoStop [0]{.\EOS\space}%
\providecommand \EOS [0]{\spacefactor3000\relax}%
\providecommand \BibitemShut  [1]{\csname bibitem#1\endcsname}%
\let\auto@bib@innerbib\@empty
%</preamble>
\bibitem [{\citenamefont {{Christensen}}(2019)}]{Christensen:2018iqi}%
  \BibitemOpen
  \bibfield  {author} {\bibinfo {author} {\bibfnamefont {N.}~\bibnamefont
  {{Christensen}}},\ }\bibfield  {title} {\enquote {\bibinfo {title}
  {{Stochastic gravitational wave backgrounds}},}\ }\href {\doibase
  10.1088/1361-6633/aae6b5} {\bibfield  {journal} {\bibinfo  {journal} {Reports
  on Progress in Physics}\ }\textbf {\bibinfo {volume} {82}},\ \bibinfo {eid}
  {016903} (\bibinfo {year} {2019})},\ \Eprint
  {http://arxiv.org/abs/1811.08797} {arXiv:1811.08797 [gr-qc]} \BibitemShut
  {NoStop}%
\bibitem [{\citenamefont {{Chiara Guzzetti}}\ \emph {et~al.}(2016)\citenamefont
  {{Chiara Guzzetti}}, \citenamefont {{Bartolo}}, \citenamefont {{Liguori}},\
  and\ \citenamefont {{Matarrese}}}]{Guzzetii2016}%
  \BibitemOpen
  \bibfield  {author} {\bibinfo {author} {\bibfnamefont {M.}~\bibnamefont
  {{Chiara Guzzetti}}}, \bibinfo {author} {\bibfnamefont {N.}~\bibnamefont
  {{Bartolo}}}, \bibinfo {author} {\bibfnamefont {M.}~\bibnamefont
  {{Liguori}}}, \ and\ \bibinfo {author} {\bibfnamefont {S.}~\bibnamefont
  {{Matarrese}}},\ }\bibfield  {title} {\enquote {\bibinfo {title}
  {{Gravitational waves from inflation}},}\ }\href {\doibase
  10.48550/arXiv.1605.01615} {\bibfield  {journal} {\bibinfo  {journal} {arXiv
  e-prints}\ ,\ \bibinfo {eid} {arXiv:1605.01615}} (\bibinfo {year} {2016})},\
  \Eprint {http://arxiv.org/abs/1605.01615} {arXiv:1605.01615 [astro-ph.CO]}
  \BibitemShut {NoStop}%
\bibitem [{\citenamefont {{Caprini}}\ and\ \citenamefont
  {{Figueroa}}(2018)}]{Caprini2018}%
  \BibitemOpen
  \bibfield  {author} {\bibinfo {author} {\bibfnamefont {C.}~\bibnamefont
  {{Caprini}}}\ and\ \bibinfo {author} {\bibfnamefont {D.~G.}\ \bibnamefont
  {{Figueroa}}},\ }\bibfield  {title} {\enquote {\bibinfo {title}
  {{Cosmological backgrounds of gravitational waves}},}\ }\href {\doibase
  10.1088/1361-6382/aac608} {\bibfield  {journal} {\bibinfo  {journal}
  {Classical and Quantum Gravity}\ }\textbf {\bibinfo {volume} {35}},\ \bibinfo
  {eid} {163001} (\bibinfo {year} {2018})},\ \Eprint
  {http://arxiv.org/abs/1801.04268} {arXiv:1801.04268 [astro-ph.CO]}
  \BibitemShut {NoStop}%
\bibitem [{\citenamefont {{Allen}}\ \emph {et~al.}(1999)\citenamefont
  {{Allen}}, \citenamefont {{Flanagan}},\ and\ \citenamefont
  {{Papa}}}]{Allen1999}%
  \BibitemOpen
  \bibfield  {author} {\bibinfo {author} {\bibfnamefont {B.}~\bibnamefont
  {{Allen}}}, \bibinfo {author} {\bibfnamefont {{\'E}.~{\'E}.}\ \bibnamefont
  {{Flanagan}}}, \ and\ \bibinfo {author} {\bibfnamefont {M.~A.}\ \bibnamefont
  {{Papa}}},\ }\bibfield  {title} {\enquote {\bibinfo {title} {{Is the
  squeezing of relic gravitational waves produced by inflation detectable?}}}\
  }\href {\doibase 10.1103/PhysRevD.61.024024} {\bibfield  {journal} {\bibinfo
  {journal} {\prd}\ }\textbf {\bibinfo {volume} {61}},\ \bibinfo {eid} {024024}
  (\bibinfo {year} {1999})},\ \Eprint {http://arxiv.org/abs/gr-qc/9906054}
  {arXiv:gr-qc/9906054 [gr-qc]} \BibitemShut {NoStop}%
\bibitem [{\citenamefont {{Parikh}}\ \emph {et~al.}(2020)\citenamefont
  {{Parikh}}, \citenamefont {{Wilczek}},\ and\ \citenamefont
  {{Zahariade}}}]{Parikh:2020nrd}%
  \BibitemOpen
  \bibfield  {author} {\bibinfo {author} {\bibfnamefont {M.}~\bibnamefont
  {{Parikh}}}, \bibinfo {author} {\bibfnamefont {F.}~\bibnamefont {{Wilczek}}},
  \ and\ \bibinfo {author} {\bibfnamefont {G.}~\bibnamefont {{Zahariade}}},\
  }\bibfield  {title} {\enquote {\bibinfo {title} {{The noise of gravitons}},}\
  }\href {\doibase 10.1142/S0218271820420018} {\bibfield  {journal} {\bibinfo
  {journal} {International Journal of Modern Physics D}\ }\textbf {\bibinfo
  {volume} {29}},\ \bibinfo {eid} {2042001-359} (\bibinfo {year} {2020})},\
  \Eprint {http://arxiv.org/abs/2005.07211} {arXiv:2005.07211 [hep-th]}
  \BibitemShut {NoStop}%
\bibitem [{\citenamefont {{Parikh}}\ \emph
  {et~al.}(2021{\natexlab{a}})\citenamefont {{Parikh}}, \citenamefont
  {{Wilczek}},\ and\ \citenamefont {{Zahariade}}}]{Parikh:2020kfh}%
  \BibitemOpen
  \bibfield  {author} {\bibinfo {author} {\bibfnamefont {M.}~\bibnamefont
  {{Parikh}}}, \bibinfo {author} {\bibfnamefont {F.}~\bibnamefont {{Wilczek}}},
  \ and\ \bibinfo {author} {\bibfnamefont {G.}~\bibnamefont {{Zahariade}}},\
  }\bibfield  {title} {\enquote {\bibinfo {title} {{Quantum Mechanics of
  Gravitational Waves}},}\ }\href {\doibase 10.1103/PhysRevLett.127.081602}
  {\bibfield  {journal} {\bibinfo  {journal} {\prl}\ }\textbf {\bibinfo
  {volume} {127}},\ \bibinfo {eid} {081602} (\bibinfo {year}
  {2021}{\natexlab{a}})},\ \Eprint {http://arxiv.org/abs/2010.08205}
  {arXiv:2010.08205 [hep-th]} \BibitemShut {NoStop}%
\bibitem [{\citenamefont {{Parikh}}\ \emph
  {et~al.}(2021{\natexlab{b}})\citenamefont {{Parikh}}, \citenamefont
  {{Wilczek}},\ and\ \citenamefont {{Zahariade}}}]{Parikh:2020fhy}%
  \BibitemOpen
  \bibfield  {author} {\bibinfo {author} {\bibfnamefont {M.}~\bibnamefont
  {{Parikh}}}, \bibinfo {author} {\bibfnamefont {F.}~\bibnamefont {{Wilczek}}},
  \ and\ \bibinfo {author} {\bibfnamefont {G.}~\bibnamefont {{Zahariade}}},\
  }\bibfield  {title} {\enquote {\bibinfo {title} {{Signatures of the
  quantization of gravity at gravitational wave detectors}},}\ }\href {\doibase
  10.1103/PhysRevD.104.046021} {\bibfield  {journal} {\bibinfo  {journal}
  {\prd}\ }\textbf {\bibinfo {volume} {104}},\ \bibinfo {eid} {046021}
  (\bibinfo {year} {2021}{\natexlab{b}})},\ \Eprint
  {http://arxiv.org/abs/2010.08208} {arXiv:2010.08208 [hep-th]} \BibitemShut
  {NoStop}%
\bibitem [{\citenamefont {{Kanno}}\ \emph
  {et~al.}(2021{\natexlab{a}})\citenamefont {{Kanno}}, \citenamefont {{Soda}},\
  and\ \citenamefont {{Tokuda}}}]{Kanno:2020usf}%
  \BibitemOpen
  \bibfield  {author} {\bibinfo {author} {\bibfnamefont {S.}~\bibnamefont
  {{Kanno}}}, \bibinfo {author} {\bibfnamefont {J.}~\bibnamefont {{Soda}}}, \
  and\ \bibinfo {author} {\bibfnamefont {J.}~\bibnamefont {{Tokuda}}},\
  }\bibfield  {title} {\enquote {\bibinfo {title} {{Noise and decoherence
  induced by gravitons}},}\ }\href {\doibase 10.1103/PhysRevD.103.044017}
  {\bibfield  {journal} {\bibinfo  {journal} {\prd}\ }\textbf {\bibinfo
  {volume} {103}},\ \bibinfo {eid} {044017} (\bibinfo {year}
  {2021}{\natexlab{a}})},\ \Eprint {http://arxiv.org/abs/2007.09838}
  {arXiv:2007.09838 [hep-th]} \BibitemShut {NoStop}%
\bibitem [{\citenamefont {{Bassi}}\ \emph {et~al.}(2017)\citenamefont
  {{Bassi}}, \citenamefont {{Gro{\ss}ardt}},\ and\ \citenamefont
  {{Ulbricht}}}]{bassi2017gravitational}%
  \BibitemOpen
  \bibfield  {author} {\bibinfo {author} {\bibfnamefont {A.}~\bibnamefont
  {{Bassi}}}, \bibinfo {author} {\bibfnamefont {A.}~\bibnamefont
  {{Gro{\ss}ardt}}}, \ and\ \bibinfo {author} {\bibfnamefont {H.}~\bibnamefont
  {{Ulbricht}}},\ }\bibfield  {title} {\enquote {\bibinfo {title}
  {{Gravitational decoherence}},}\ }\href {\doibase 10.1088/1361-6382/aa864f}
  {\bibfield  {journal} {\bibinfo  {journal} {Classical and Quantum Gravity}\
  }\textbf {\bibinfo {volume} {34}},\ \bibinfo {eid} {193002} (\bibinfo {year}
  {2017})},\ \Eprint {http://arxiv.org/abs/1706.05677} {arXiv:1706.05677
  [quant-ph]} \BibitemShut {NoStop}%
\bibitem [{\citenamefont {{Anastopoulos}}\ and\ \citenamefont
  {{Hu}}(2013)}]{anastopoulos2013master}%
  \BibitemOpen
  \bibfield  {author} {\bibinfo {author} {\bibfnamefont {C.}~\bibnamefont
  {{Anastopoulos}}}\ and\ \bibinfo {author} {\bibfnamefont {B.~L.}\
  \bibnamefont {{Hu}}},\ }\bibfield  {title} {\enquote {\bibinfo {title} {{A
  master equation for gravitational decoherence: probing the textures of
  spacetime}},}\ }\href {\doibase 10.1088/0264-9381/30/16/165007} {\bibfield
  {journal} {\bibinfo  {journal} {Classical and Quantum Gravity}\ }\textbf
  {\bibinfo {volume} {30}},\ \bibinfo {eid} {165007} (\bibinfo {year}
  {2013})},\ \Eprint {http://arxiv.org/abs/1305.5231} {arXiv:1305.5231 [gr-qc]}
  \BibitemShut {NoStop}%
\bibitem [{\citenamefont {{Jess Riedel}}(2013)}]{Riedel:2013yca}%
  \BibitemOpen
  \bibfield  {author} {\bibinfo {author} {\bibfnamefont {C.}~\bibnamefont
  {{Jess Riedel}}},\ }\bibfield  {title} {\enquote {\bibinfo {title} {{Evidence
  for gravitons from decoherence by bremsstrahlung}},}\ }\href@noop {} {\
  (\bibinfo {year} {2013})},\ \Eprint {http://arxiv.org/abs/1310.6347}
  {arXiv:1310.6347 [quant-ph]} \BibitemShut {NoStop}%
\bibitem [{\citenamefont {{Suzuki}}\ and\ \citenamefont
  {{Queisser}}(2015)}]{Suzuki:2015nva}%
  \BibitemOpen
  \bibfield  {author} {\bibinfo {author} {\bibfnamefont {F.}~\bibnamefont
  {{Suzuki}}}\ and\ \bibinfo {author} {\bibfnamefont {F.}~\bibnamefont
  {{Queisser}}},\ }\bibfield  {title} {\enquote {\bibinfo {title}
  {{Environmental gravitational decoherence and a tensor noise model}},}\ \
  }(\bibinfo {year} {2015})\ p.\ \bibinfo {pages} {012039},\ \Eprint
  {http://arxiv.org/abs/1502.01386} {arXiv:1502.01386 [gr-qc]} \BibitemShut
  {NoStop}%
\bibitem [{\citenamefont {{Guerreiro}}(2020)}]{Guerreiro:2019vbq}%
  \BibitemOpen
  \bibfield  {author} {\bibinfo {author} {\bibfnamefont {T.}~\bibnamefont
  {{Guerreiro}}},\ }\bibfield  {title} {\enquote {\bibinfo {title} {{Quantum
  effects in gravity waves}},}\ }\href {\doibase 10.1088/1361-6382/ab9d5d}
  {\bibfield  {journal} {\bibinfo  {journal} {Classical and Quantum Gravity}\
  }\textbf {\bibinfo {volume} {37}},\ \bibinfo {eid} {155001} (\bibinfo {year}
  {2020})},\ \Eprint {http://arxiv.org/abs/1911.11593} {arXiv:1911.11593
  [quant-ph]} \BibitemShut {NoStop}%
\bibitem [{\citenamefont {{Blencowe}}(2013)}]{blencowe2013effective}%
  \BibitemOpen
  \bibfield  {author} {\bibinfo {author} {\bibfnamefont {M.~P.}\ \bibnamefont
  {{Blencowe}}},\ }\bibfield  {title} {\enquote {\bibinfo {title} {{Effective
  Field Theory Approach to Gravitationally Induced Decoherence}},}\ }\href
  {\doibase 10.1103/PhysRevLett.111.021302} {\bibfield  {journal} {\bibinfo
  {journal} {\prl}\ }\textbf {\bibinfo {volume} {111}},\ \bibinfo {eid}
  {021302} (\bibinfo {year} {2013})},\ \Eprint {http://arxiv.org/abs/1211.4751}
  {arXiv:1211.4751 [quant-ph]} \BibitemShut {NoStop}%
\bibitem [{\citenamefont {{Lamine}}\ \emph {et~al.}(2006)\citenamefont
  {{Lamine}}, \citenamefont {{Herv{\'e}}}, \citenamefont {{Lambrecht}},\ and\
  \citenamefont {{Reynaud}}}]{lamine2006ultimate}%
  \BibitemOpen
  \bibfield  {author} {\bibinfo {author} {\bibfnamefont {B.}~\bibnamefont
  {{Lamine}}}, \bibinfo {author} {\bibfnamefont {R.}~\bibnamefont
  {{Herv{\'e}}}}, \bibinfo {author} {\bibfnamefont {A.}~\bibnamefont
  {{Lambrecht}}}, \ and\ \bibinfo {author} {\bibfnamefont {S.}~\bibnamefont
  {{Reynaud}}},\ }\bibfield  {title} {\enquote {\bibinfo {title} {{Ultimate
  Decoherence Border for Matter-Wave Interferometry}},}\ }\href {\doibase
  10.1103/PhysRevLett.96.050405} {\bibfield  {journal} {\bibinfo  {journal}
  {\prl}\ }\textbf {\bibinfo {volume} {96}},\ \bibinfo {eid} {050405} (\bibinfo
  {year} {2006})},\ \Eprint {http://arxiv.org/abs/quant-ph/0505074}
  {arXiv:quant-ph/0505074 [quant-ph]} \BibitemShut {NoStop}%
\bibitem [{\citenamefont {{Kanno}}\ \emph
  {et~al.}(2021{\natexlab{b}})\citenamefont {{Kanno}}, \citenamefont {{Soda}},\
  and\ \citenamefont {{Tokuda}}}]{Kanno:2021gpt}%
  \BibitemOpen
  \bibfield  {author} {\bibinfo {author} {\bibfnamefont {S.}~\bibnamefont
  {{Kanno}}}, \bibinfo {author} {\bibfnamefont {J.}~\bibnamefont {{Soda}}}, \
  and\ \bibinfo {author} {\bibfnamefont {J.}~\bibnamefont {{Tokuda}}},\
  }\bibfield  {title} {\enquote {\bibinfo {title} {{Indirect detection of
  gravitons through quantum entanglement}},}\ }\href {\doibase
  10.1103/PhysRevD.104.083516} {\bibfield  {journal} {\bibinfo  {journal}
  {\prd}\ }\textbf {\bibinfo {volume} {104}},\ \bibinfo {eid} {083516}
  (\bibinfo {year} {2021}{\natexlab{b}})},\ \Eprint
  {http://arxiv.org/abs/2103.17053} {arXiv:2103.17053 [gr-qc]} \BibitemShut
  {NoStop}%
\bibitem [{\citenamefont {{Feynman}}\ and\ \citenamefont
  {{Vernon}}(1963)}]{Feynman:1963fq}%
  \BibitemOpen
  \bibfield  {author} {\bibinfo {author} {\bibfnamefont {R.~P.}\ \bibnamefont
  {{Feynman}}}\ and\ \bibinfo {author} {\bibfnamefont {J.}~\bibnamefont
  {{Vernon}}, \bibfnamefont {F.~L.}},\ }\bibfield  {title} {\enquote {\bibinfo
  {title} {{The theory of a general quantum system interacting with a linear
  dissipative system}},}\ }\href {\doibase 10.1016/0003-4916(63)90068-X}
  {\bibfield  {journal} {\bibinfo  {journal} {Annals of Physics}\ }\textbf
  {\bibinfo {volume} {24}},\ \bibinfo {pages} {118--173} (\bibinfo {year}
  {1963})}\BibitemShut {NoStop}%
\bibitem [{\citenamefont {Scully}\ and\ \citenamefont
  {Zubairy}(1997)}]{Scully}%
  \BibitemOpen
  \bibfield  {author} {\bibinfo {author} {\bibfnamefont {M.~O.}\ \bibnamefont
  {Scully}}\ and\ \bibinfo {author} {\bibfnamefont {M.~S.}\ \bibnamefont
  {Zubairy}},\ }\href@noop {} {\emph {\bibinfo {title} {Quantum optics}}}\
  (\bibinfo  {publisher} {Cambridge University Press},\ \bibinfo {year}
  {1997})\BibitemShut {NoStop}%
\bibitem [{\citenamefont {{Grishchuk}}\ and\ \citenamefont
  {{Sidorov}}(1990{\natexlab{a}})}]{Grishchuk:1990bj}%
  \BibitemOpen
  \bibfield  {author} {\bibinfo {author} {\bibfnamefont {L.~P.}\ \bibnamefont
  {{Grishchuk}}}\ and\ \bibinfo {author} {\bibfnamefont {Y.~V.}\ \bibnamefont
  {{Sidorov}}},\ }\bibfield  {title} {\enquote {\bibinfo {title} {{Squeezed
  quantum states of relic gravitons and primordial density fluctuations}},}\
  }\href {\doibase 10.1103/PhysRevD.42.3413} {\bibfield  {journal} {\bibinfo
  {journal} {\prd}\ }\textbf {\bibinfo {volume} {42}},\ \bibinfo {pages}
  {3413--3421} (\bibinfo {year} {1990}{\natexlab{a}})}\BibitemShut {NoStop}%
\bibitem [{\citenamefont {{Lesgourgues}}\ \emph {et~al.}(1997)\citenamefont
  {{Lesgourgues}}, \citenamefont {{Polarski}},\ and\ \citenamefont
  {{Starobinsky}}}]{Lesgourgues1997}%
  \BibitemOpen
  \bibfield  {author} {\bibinfo {author} {\bibfnamefont {J.}~\bibnamefont
  {{Lesgourgues}}}, \bibinfo {author} {\bibfnamefont {D.}~\bibnamefont
  {{Polarski}}}, \ and\ \bibinfo {author} {\bibfnamefont {A.~A.}\ \bibnamefont
  {{Starobinsky}}},\ }\bibfield  {title} {\enquote {\bibinfo {title}
  {{Quantum-to-classical transition of cosmological perturbations for
  non-vacuum initial states}},}\ }\href {\doibase
  10.1016/S0550-3213(97)00224-1} {\bibfield  {journal} {\bibinfo  {journal}
  {Nuclear Physics B}\ }\textbf {\bibinfo {volume} {497}},\ \bibinfo {pages}
  {479--508} (\bibinfo {year} {1997})},\ \Eprint
  {http://arxiv.org/abs/gr-qc/9611019} {arXiv:gr-qc/9611019 [gr-qc]}
  \BibitemShut {NoStop}%
\bibitem [{\citenamefont {{Sudarsky}}(2011)}]{sudarsky2011}%
  \BibitemOpen
  \bibfield  {author} {\bibinfo {author} {\bibfnamefont {D.}~\bibnamefont
  {{Sudarsky}}},\ }\bibfield  {title} {\enquote {\bibinfo {title}
  {{Shortcomings in the Understanding of why Cosmological Perturbations Look
  Classical}},}\ }\href {\doibase 10.1142/S0218271811018937} {\bibfield
  {journal} {\bibinfo  {journal} {International Journal of Modern Physics D}\
  }\textbf {\bibinfo {volume} {20}},\ \bibinfo {pages} {509--552} (\bibinfo
  {year} {2011})},\ \Eprint {http://arxiv.org/abs/0906.0315} {arXiv:0906.0315
  [gr-qc]} \BibitemShut {NoStop}%
\bibitem [{\citenamefont {{Kiefer}}\ and\ \citenamefont
  {{Polarski}}(2008)}]{Kiefer2008}%
  \BibitemOpen
  \bibfield  {author} {\bibinfo {author} {\bibfnamefont {C.}~\bibnamefont
  {{Kiefer}}}\ and\ \bibinfo {author} {\bibfnamefont {D.}~\bibnamefont
  {{Polarski}}},\ }\bibfield  {title} {\enquote {\bibinfo {title} {{Why do
  cosmological perturbations look classical to us?}}}\ }\href {\doibase
  10.48550/arXiv.0810.0087} {\bibfield  {journal} {\bibinfo  {journal} {arXiv
  e-prints}\ ,\ \bibinfo {eid} {arXiv:0810.0087}} (\bibinfo {year} {2008})},\
  \Eprint {http://arxiv.org/abs/0810.0087} {arXiv:0810.0087 [astro-ph]}
  \BibitemShut {NoStop}%
\bibitem [{\citenamefont {{Kiefer}}\ \emph
  {et~al.}(2007{\natexlab{a}})\citenamefont {{Kiefer}}, \citenamefont
  {{Lohmar}}, \citenamefont {{Polarski}},\ and\ \citenamefont
  {{Starobinsky}}}]{Kiefer2007}%
  \BibitemOpen
  \bibfield  {author} {\bibinfo {author} {\bibfnamefont {C.}~\bibnamefont
  {{Kiefer}}}, \bibinfo {author} {\bibfnamefont {I.}~\bibnamefont {{Lohmar}}},
  \bibinfo {author} {\bibfnamefont {D.}~\bibnamefont {{Polarski}}}, \ and\
  \bibinfo {author} {\bibfnamefont {A.~A.}\ \bibnamefont {{Starobinsky}}},\
  }\bibfield  {title} {\enquote {\bibinfo {title} {{Origin of classical
  structure in the Universe}},}\ }\href {\doibase
  10.1088/1742-6596/67/1/012023} {\bibfield  {journal} {\bibinfo  {journal}
  {Journal of Physics, Conference Series}\ }\textbf {\bibinfo {volume} {67}},\
  \bibinfo {eid} {012023} (\bibinfo {year} {2007}{\natexlab{a}})}\BibitemShut
  {NoStop}%
\bibitem [{\citenamefont {{Kiefer}}\ \emph
  {et~al.}(2007{\natexlab{b}})\citenamefont {{Kiefer}}, \citenamefont
  {{Lohmar}}, \citenamefont {{Polarski}},\ and\ \citenamefont
  {{Starobinsky}}}]{Kiefer2007P}%
  \BibitemOpen
  \bibfield  {author} {\bibinfo {author} {\bibfnamefont {C.}~\bibnamefont
  {{Kiefer}}}, \bibinfo {author} {\bibfnamefont {I.}~\bibnamefont {{Lohmar}}},
  \bibinfo {author} {\bibfnamefont {D.}~\bibnamefont {{Polarski}}}, \ and\
  \bibinfo {author} {\bibfnamefont {A.~A.}\ \bibnamefont {{Starobinsky}}},\
  }\bibfield  {title} {\enquote {\bibinfo {title} {{Pointer states for
  primordial fluctuations in inflationary cosmology}},}\ }\href {\doibase
  10.1088/0264-9381/24/7/002} {\bibfield  {journal} {\bibinfo  {journal}
  {Classical and Quantum Gravity}\ }\textbf {\bibinfo {volume} {24}},\ \bibinfo
  {pages} {1699--1718} (\bibinfo {year} {2007}{\natexlab{b}})},\ \Eprint
  {http://arxiv.org/abs/astro-ph/0610700} {arXiv:astro-ph/0610700 [astro-ph]}
  \BibitemShut {NoStop}%
\bibitem [{\citenamefont {{Campo}}\ and\ \citenamefont
  {{Parentani}}(2006)}]{Campo2006}%
  \BibitemOpen
  \bibfield  {author} {\bibinfo {author} {\bibfnamefont {D.}~\bibnamefont
  {{Campo}}}\ and\ \bibinfo {author} {\bibfnamefont {R.}~\bibnamefont
  {{Parentani}}},\ }\bibfield  {title} {\enquote {\bibinfo {title}
  {{Inflationary spectra and violations of Bell inequalities}},}\ }\href
  {\doibase 10.1103/PhysRevD.74.025001} {\bibfield  {journal} {\bibinfo
  {journal} {\prd}\ }\textbf {\bibinfo {volume} {74}},\ \bibinfo {eid} {025001}
  (\bibinfo {year} {2006})},\ \Eprint {http://arxiv.org/abs/astro-ph/0505376}
  {arXiv:astro-ph/0505376 [astro-ph]} \BibitemShut {NoStop}%
\bibitem [{\citenamefont {{Martin}}\ and\ \citenamefont
  {{Vennin}}(2016)}]{Martin2016}%
  \BibitemOpen
  \bibfield  {author} {\bibinfo {author} {\bibfnamefont {J.}~\bibnamefont
  {{Martin}}}\ and\ \bibinfo {author} {\bibfnamefont {V.}~\bibnamefont
  {{Vennin}}},\ }\bibfield  {title} {\enquote {\bibinfo {title} {{Leggett-Garg
  inequalities for squeezed states}},}\ }\href {\doibase
  10.1103/PhysRevA.94.052135} {\bibfield  {journal} {\bibinfo  {journal}
  {\pra}\ }\textbf {\bibinfo {volume} {94}},\ \bibinfo {eid} {052135} (\bibinfo
  {year} {2016})},\ \Eprint {http://arxiv.org/abs/1611.01785} {arXiv:1611.01785
  [quant-ph]} \BibitemShut {NoStop}%
\bibitem [{\citenamefont {{Martin}}\ and\ \citenamefont
  {{Vennin}}(2017)}]{Martin2017}%
  \BibitemOpen
  \bibfield  {author} {\bibinfo {author} {\bibfnamefont {J.}~\bibnamefont
  {{Martin}}}\ and\ \bibinfo {author} {\bibfnamefont {V.}~\bibnamefont
  {{Vennin}}},\ }\bibfield  {title} {\enquote {\bibinfo {title} {{Obstructions
  to Bell CMB experiments}},}\ }\href {\doibase 10.1103/PhysRevD.96.063501}
  {\bibfield  {journal} {\bibinfo  {journal} {\prd}\ }\textbf {\bibinfo
  {volume} {96}},\ \bibinfo {eid} {063501} (\bibinfo {year} {2017})},\ \Eprint
  {http://arxiv.org/abs/1706.05001} {arXiv:1706.05001 [astro-ph.CO]}
  \BibitemShut {NoStop}%
\bibitem [{\citenamefont {{Martin}}\ and\ \citenamefont
  {{Vennin}}(2018{\natexlab{a}})}]{Martin2018}%
  \BibitemOpen
  \bibfield  {author} {\bibinfo {author} {\bibfnamefont {J.}~\bibnamefont
  {{Martin}}}\ and\ \bibinfo {author} {\bibfnamefont {V.}~\bibnamefont
  {{Vennin}}},\ }\bibfield  {title} {\enquote {\bibinfo {title} {{Observational
  constraints on quantum decoherence during inflation}},}\ }\href {\doibase
  10.1088/1475-7516/2018/05/063} {\bibfield  {journal} {\bibinfo  {journal}
  {Journal of Cosmology and Astroparticle Physics}\ }\textbf {\bibinfo {volume}
  {2018}},\ \bibinfo {eid} {063} (\bibinfo {year} {2018}{\natexlab{a}})},\
  \Eprint {http://arxiv.org/abs/1801.09949} {arXiv:1801.09949 [astro-ph.CO]}
  \BibitemShut {NoStop}%
\bibitem [{\citenamefont {{Martin}}\ and\ \citenamefont
  {{Vennin}}(2018{\natexlab{b}})}]{Martin2018JCAP}%
  \BibitemOpen
  \bibfield  {author} {\bibinfo {author} {\bibfnamefont {J.}~\bibnamefont
  {{Martin}}}\ and\ \bibinfo {author} {\bibfnamefont {V.}~\bibnamefont
  {{Vennin}}},\ }\bibfield  {title} {\enquote {\bibinfo {title} {{Non
  Gaussianities from quantum decoherence during inflation}},}\ }\href {\doibase
  10.1088/1475-7516/2018/06/037} {\bibfield  {journal} {\bibinfo  {journal}
  {Journal of Cosmology and Astroparticle Physics}\ }\textbf {\bibinfo {volume}
  {2018}},\ \bibinfo {eid} {037} (\bibinfo {year} {2018}{\natexlab{b}})},\
  \Eprint {http://arxiv.org/abs/1805.05609} {arXiv:1805.05609 [astro-ph.CO]}
  \BibitemShut {NoStop}%
\bibitem [{\citenamefont {{Martin}}\ \emph {et~al.}(2023)\citenamefont
  {{Martin}}, \citenamefont {{Micheli}},\ and\ \citenamefont
  {{Vennin}}}]{Martin2023}%
  \BibitemOpen
  \bibfield  {author} {\bibinfo {author} {\bibfnamefont {J.}~\bibnamefont
  {{Martin}}}, \bibinfo {author} {\bibfnamefont {A.}~\bibnamefont {{Micheli}}},
  \ and\ \bibinfo {author} {\bibfnamefont {V.}~\bibnamefont {{Vennin}}},\
  }\bibfield  {title} {\enquote {\bibinfo {title} {{Comparing quantumness
  criteria}},}\ }\href {\doibase 10.1209/0295-5075/acc3be} {\bibfield
  {journal} {\bibinfo  {journal} {EPL (Europhysics Letters)}\ }\textbf
  {\bibinfo {volume} {142}},\ \bibinfo {eid} {18001} (\bibinfo {year}
  {2023})},\ \Eprint {http://arxiv.org/abs/2211.10114} {arXiv:2211.10114
  [quant-ph]} \BibitemShut {NoStop}%
\bibitem [{\citenamefont {{Berera}}\ \emph {et~al.}(2021)\citenamefont
  {{Berera}}, \citenamefont {{Brahma}}, \citenamefont {{Brandenberger}},
  \citenamefont {{Calder{\'o}n-Figueroa}},\ and\ \citenamefont
  {{Heavens}}}]{Berera2021}%
  \BibitemOpen
  \bibfield  {author} {\bibinfo {author} {\bibfnamefont {A.}~\bibnamefont
  {{Berera}}}, \bibinfo {author} {\bibfnamefont {S.}~\bibnamefont {{Brahma}}},
  \bibinfo {author} {\bibfnamefont {R.}~\bibnamefont {{Brandenberger}}},
  \bibinfo {author} {\bibfnamefont {J.}~\bibnamefont
  {{Calder{\'o}n-Figueroa}}}, \ and\ \bibinfo {author} {\bibfnamefont
  {A.}~\bibnamefont {{Heavens}}},\ }\bibfield  {title} {\enquote {\bibinfo
  {title} {{Quantum coherence of photons to cosmological distances}},}\ }\href
  {\doibase 10.1103/PhysRevD.104.063519} {\bibfield  {journal} {\bibinfo
  {journal} {\prd}\ }\textbf {\bibinfo {volume} {104}},\ \bibinfo {eid}
  {063519} (\bibinfo {year} {2021})},\ \Eprint
  {http://arxiv.org/abs/2107.06914} {arXiv:2107.06914 [hep-ph]} \BibitemShut
  {NoStop}%
\bibitem [{\citenamefont {{Berera}}\ and\ \citenamefont
  {{Calder{\'o}n-Figueroa}}(2022)}]{Berera2022}%
  \BibitemOpen
  \bibfield  {author} {\bibinfo {author} {\bibfnamefont {A.}~\bibnamefont
  {{Berera}}}\ and\ \bibinfo {author} {\bibfnamefont {J.}~\bibnamefont
  {{Calder{\'o}n-Figueroa}}},\ }\bibfield  {title} {\enquote {\bibinfo {title}
  {{Viability of quantum communication across interstellar distances}},}\
  }\href {\doibase 10.1103/PhysRevD.105.123033} {\bibfield  {journal} {\bibinfo
   {journal} {\prd}\ }\textbf {\bibinfo {volume} {105}},\ \bibinfo {eid}
  {123033} (\bibinfo {year} {2022})},\ \Eprint
  {http://arxiv.org/abs/2205.11816} {arXiv:2205.11816 [quant-ph]} \BibitemShut
  {NoStop}%
\bibitem [{\citenamefont {{Colas}}\ \emph {et~al.}(2023)\citenamefont
  {{Colas}}, \citenamefont {{Grain}},\ and\ \citenamefont
  {{Vennin}}}]{Colas2023}%
  \BibitemOpen
  \bibfield  {author} {\bibinfo {author} {\bibfnamefont {T.}~\bibnamefont
  {{Colas}}}, \bibinfo {author} {\bibfnamefont {J.}~\bibnamefont {{Grain}}}, \
  and\ \bibinfo {author} {\bibfnamefont {V.}~\bibnamefont {{Vennin}}},\
  }\bibfield  {title} {\enquote {\bibinfo {title} {{Quantum recoherence in the
  early universe}},}\ }\href {\doibase 10.1209/0295-5075/acdd94} {\bibfield
  {journal} {\bibinfo  {journal} {EPL (Europhysics Letters)}\ }\textbf
  {\bibinfo {volume} {142}},\ \bibinfo {eid} {69002} (\bibinfo {year}
  {2023})},\ \Eprint {http://arxiv.org/abs/2212.09486} {arXiv:2212.09486
  [gr-qc]} \BibitemShut {NoStop}%
\bibitem [{\citenamefont {{Burgess}}\ \emph {et~al.}(2022)\citenamefont
  {{Burgess}}, \citenamefont {{Holman}}, \citenamefont {{Kaplanek}},
  \citenamefont {{Martin}},\ and\ \citenamefont {{Vennin}}}]{Burgess2022}%
  \BibitemOpen
  \bibfield  {author} {\bibinfo {author} {\bibfnamefont {C.~P.}\ \bibnamefont
  {{Burgess}}}, \bibinfo {author} {\bibfnamefont {R.}~\bibnamefont {{Holman}}},
  \bibinfo {author} {\bibfnamefont {G.}~\bibnamefont {{Kaplanek}}}, \bibinfo
  {author} {\bibfnamefont {J.}~\bibnamefont {{Martin}}}, \ and\ \bibinfo
  {author} {\bibfnamefont {V.}~\bibnamefont {{Vennin}}},\ }\bibfield  {title}
  {\enquote {\bibinfo {title} {{Minimal decoherence from inflation}},}\ }\href
  {\doibase 10.48550/arXiv.2211.11046} {\bibfield  {journal} {\bibinfo
  {journal} {arXiv e-prints}\ ,\ \bibinfo {eid} {arXiv:2211.11046}} (\bibinfo
  {year} {2022})},\ \Eprint {http://arxiv.org/abs/2211.11046} {arXiv:2211.11046
  [hep-th]} \BibitemShut {NoStop}%
\bibitem [{\citenamefont {{Daddi Hammou}}\ and\ \citenamefont
  {{Bartolo}}(2023)}]{Daddi2023}%
  \BibitemOpen
  \bibfield  {author} {\bibinfo {author} {\bibfnamefont {A.}~\bibnamefont
  {{Daddi Hammou}}}\ and\ \bibinfo {author} {\bibfnamefont {N.}~\bibnamefont
  {{Bartolo}}},\ }\bibfield  {title} {\enquote {\bibinfo {title} {{Cosmic
  decoherence: primordial power spectra and non-Gaussianities}},}\ }\href
  {\doibase 10.1088/1475-7516/2023/04/055} {\bibfield  {journal} {\bibinfo
  {journal} {Journal of Cosmology and Astroparticle Physics}\ }\textbf
  {\bibinfo {volume} {2023}},\ \bibinfo {eid} {055} (\bibinfo {year} {2023})},\
  \Eprint {http://arxiv.org/abs/2211.07598} {arXiv:2211.07598 [astro-ph.CO]}
  \BibitemShut {NoStop}%
\bibitem [{\citenamefont {{Micheli}}\ and\ \citenamefont
  {{Peter}}(2022)}]{Micheli2022}%
  \BibitemOpen
  \bibfield  {author} {\bibinfo {author} {\bibfnamefont {A.}~\bibnamefont
  {{Micheli}}}\ and\ \bibinfo {author} {\bibfnamefont {P.}~\bibnamefont
  {{Peter}}},\ }\bibfield  {title} {\enquote {\bibinfo {title} {{Quantum
  cosmological gravitational waves?}}}\ }\href {\doibase
  10.48550/arXiv.2211.00182} {\bibfield  {journal} {\bibinfo  {journal} {arXiv
  e-prints}\ ,\ \bibinfo {eid} {arXiv:2211.00182}} (\bibinfo {year} {2022})},\
  \Eprint {http://arxiv.org/abs/2211.00182} {arXiv:2211.00182 [gr-qc]}
  \BibitemShut {NoStop}%
\bibitem [{\citenamefont {{Anastopoulos}}\ and\ \citenamefont
  {{Hu}}(2020)}]{Anastopolos2020}%
  \BibitemOpen
  \bibfield  {author} {\bibinfo {author} {\bibfnamefont {C.}~\bibnamefont
  {{Anastopoulos}}}\ and\ \bibinfo {author} {\bibfnamefont {B.~L.}\
  \bibnamefont {{Hu}}},\ }\bibfield  {title} {\enquote {\bibinfo {title}
  {{Quantum superposition of two gravitational cat states}},}\ }\href {\doibase
  10.1088/1361-6382/abbe6f} {\bibfield  {journal} {\bibinfo  {journal}
  {Classical and Quantum Gravity}\ }\textbf {\bibinfo {volume} {37}},\ \bibinfo
  {eid} {235012} (\bibinfo {year} {2020})},\ \Eprint
  {http://arxiv.org/abs/2007.06446} {arXiv:2007.06446 [quant-ph]} \BibitemShut
  {NoStop}%
\bibitem [{\citenamefont {{Lagouvardos}}\ and\ \citenamefont
  {{Anastopoulos}}(2021)}]{Anastopolos2021}%
  \BibitemOpen
  \bibfield  {author} {\bibinfo {author} {\bibfnamefont {M.}~\bibnamefont
  {{Lagouvardos}}}\ and\ \bibinfo {author} {\bibfnamefont {C.}~\bibnamefont
  {{Anastopoulos}}},\ }\bibfield  {title} {\enquote {\bibinfo {title}
  {{Gravitational decoherence of photons}},}\ }\href {\doibase
  10.1088/1361-6382/abf2f3} {\bibfield  {journal} {\bibinfo  {journal}
  {Classical and Quantum Gravity}\ }\textbf {\bibinfo {volume} {38}},\ \bibinfo
  {eid} {115012} (\bibinfo {year} {2021})},\ \Eprint
  {http://arxiv.org/abs/2011.08270} {arXiv:2011.08270 [gr-qc]} \BibitemShut
  {NoStop}%
\bibitem [{\citenamefont {{Giacomini}}\ and\ \citenamefont
  {{Kempf}}(2022)}]{Giacomini2022}%
  \BibitemOpen
  \bibfield  {author} {\bibinfo {author} {\bibfnamefont {F.}~\bibnamefont
  {{Giacomini}}}\ and\ \bibinfo {author} {\bibfnamefont {A.}~\bibnamefont
  {{Kempf}}},\ }\bibfield  {title} {\enquote {\bibinfo {title}
  {{Second-quantized Unruh-DeWitt detectors and their quantum reference frame
  transformations}},}\ }\href {\doibase 10.1103/PhysRevD.105.125001} {\bibfield
   {journal} {\bibinfo  {journal} {\prd}\ }\textbf {\bibinfo {volume} {105}},\
  \bibinfo {eid} {125001} (\bibinfo {year} {2022})},\ \Eprint
  {http://arxiv.org/abs/2201.03120} {arXiv:2201.03120 [quant-ph]} \BibitemShut
  {NoStop}%
\bibitem [{\citenamefont {{Belenchia}}\ \emph {et~al.}(2018)\citenamefont
  {{Belenchia}}, \citenamefont {{Wald}}, \citenamefont {{Giacomini}},
  \citenamefont {{Castro-Ruiz}}, \citenamefont {{Brukner}},\ and\ \citenamefont
  {{Aspelmeyer}}}]{Belenchia2018}%
  \BibitemOpen
  \bibfield  {author} {\bibinfo {author} {\bibfnamefont {A.}~\bibnamefont
  {{Belenchia}}}, \bibinfo {author} {\bibfnamefont {R.~M.}\ \bibnamefont
  {{Wald}}}, \bibinfo {author} {\bibfnamefont {F.}~\bibnamefont {{Giacomini}}},
  \bibinfo {author} {\bibfnamefont {E.}~\bibnamefont {{Castro-Ruiz}}}, \bibinfo
  {author} {\bibfnamefont {{\v{C}}.}~\bibnamefont {{Brukner}}}, \ and\ \bibinfo
  {author} {\bibfnamefont {M.}~\bibnamefont {{Aspelmeyer}}},\ }\bibfield
  {title} {\enquote {\bibinfo {title} {{Quantum superposition of massive
  objects and the quantization of gravity}},}\ }\href {\doibase
  10.1103/PhysRevD.98.126009} {\bibfield  {journal} {\bibinfo  {journal}
  {\prd}\ }\textbf {\bibinfo {volume} {98}},\ \bibinfo {eid} {126009} (\bibinfo
  {year} {2018})},\ \Eprint {http://arxiv.org/abs/1807.07015} {arXiv:1807.07015
  [quant-ph]} \BibitemShut {NoStop}%
\bibitem [{\citenamefont {{Streiter}}\ \emph {et~al.}(2021)\citenamefont
  {{Streiter}}, \citenamefont {{Giacomini}},\ and\ \citenamefont
  {{Brukner}}}]{Streiter2021}%
  \BibitemOpen
  \bibfield  {author} {\bibinfo {author} {\bibfnamefont {L.~F.}\ \bibnamefont
  {{Streiter}}}, \bibinfo {author} {\bibfnamefont {F.}~\bibnamefont
  {{Giacomini}}}, \ and\ \bibinfo {author} {\bibfnamefont
  {{\v{C}}.}~\bibnamefont {{Brukner}}},\ }\bibfield  {title} {\enquote
  {\bibinfo {title} {{Relativistic Bell Test within Quantum Reference
  Frames}},}\ }\href {\doibase 10.1103/PhysRevLett.126.230403} {\bibfield
  {journal} {\bibinfo  {journal} {\prl}\ }\textbf {\bibinfo {volume} {126}},\
  \bibinfo {eid} {230403} (\bibinfo {year} {2021})},\ \Eprint
  {http://arxiv.org/abs/2008.03317} {arXiv:2008.03317 [quant-ph]} \BibitemShut
  {NoStop}%
\bibitem [{\citenamefont {{Matsumura}}\ and\ \citenamefont
  {{Yamamoto}}(2020)}]{Matsumura2020}%
  \BibitemOpen
  \bibfield  {author} {\bibinfo {author} {\bibfnamefont {A.}~\bibnamefont
  {{Matsumura}}}\ and\ \bibinfo {author} {\bibfnamefont {K.}~\bibnamefont
  {{Yamamoto}}},\ }\bibfield  {title} {\enquote {\bibinfo {title}
  {{Gravity-induced entanglement in optomechanical systems}},}\ }\href
  {\doibase 10.1103/PhysRevD.102.106021} {\bibfield  {journal} {\bibinfo
  {journal} {\prd}\ }\textbf {\bibinfo {volume} {102}},\ \bibinfo {eid}
  {106021} (\bibinfo {year} {2020})},\ \Eprint
  {http://arxiv.org/abs/2010.05161} {arXiv:2010.05161 [gr-qc]} \BibitemShut
  {NoStop}%
\bibitem [{\citenamefont {{Matsumura}}(2021)}]{Matsumura2021}%
  \BibitemOpen
  \bibfield  {author} {\bibinfo {author} {\bibfnamefont {A.}~\bibnamefont
  {{Matsumura}}},\ }\bibfield  {title} {\enquote {\bibinfo {title}
  {{Field-induced entanglement in spatially superposed objects}},}\ }\href
  {\doibase 10.1103/PhysRevD.104.046001} {\bibfield  {journal} {\bibinfo
  {journal} {\prd}\ }\textbf {\bibinfo {volume} {104}},\ \bibinfo {eid}
  {046001} (\bibinfo {year} {2021})},\ \Eprint
  {http://arxiv.org/abs/2102.10792} {arXiv:2102.10792 [quant-ph]} \BibitemShut
  {NoStop}%
\bibitem [{\citenamefont {{Bose}}\ \emph {et~al.}(2017)\citenamefont {{Bose}},
  \citenamefont {{Mazumdar}}, \citenamefont {{Morley}}, \citenamefont
  {{Ulbricht}}, \citenamefont {{Toro{\v{s}}}}, \citenamefont {{Paternostro}},
  \citenamefont {{Geraci}}, \citenamefont {{Barker}}, \citenamefont {{Kim}},\
  and\ \citenamefont {{Milburn}}}]{Bose:2017nin}%
  \BibitemOpen
  \bibfield  {author} {\bibinfo {author} {\bibfnamefont {S.}~\bibnamefont
  {{Bose}}}, \bibinfo {author} {\bibfnamefont {A.}~\bibnamefont {{Mazumdar}}},
  \bibinfo {author} {\bibfnamefont {G.~W.}\ \bibnamefont {{Morley}}}, \bibinfo
  {author} {\bibfnamefont {H.}~\bibnamefont {{Ulbricht}}}, \bibinfo {author}
  {\bibfnamefont {M.}~\bibnamefont {{Toro{\v{s}}}}}, \bibinfo {author}
  {\bibfnamefont {M.}~\bibnamefont {{Paternostro}}}, \bibinfo {author}
  {\bibfnamefont {A.~A.}\ \bibnamefont {{Geraci}}}, \bibinfo {author}
  {\bibfnamefont {P.~F.}\ \bibnamefont {{Barker}}}, \bibinfo {author}
  {\bibfnamefont {M.~S.}\ \bibnamefont {{Kim}}}, \ and\ \bibinfo {author}
  {\bibfnamefont {G.}~\bibnamefont {{Milburn}}},\ }\bibfield  {title} {\enquote
  {\bibinfo {title} {{Spin Entanglement Witness for Quantum Gravity}},}\ }\href
  {\doibase 10.1103/PhysRevLett.119.240401} {\bibfield  {journal} {\bibinfo
  {journal} {\prl}\ }\textbf {\bibinfo {volume} {119}},\ \bibinfo {eid}
  {240401} (\bibinfo {year} {2017})},\ \Eprint
  {http://arxiv.org/abs/1707.06050} {arXiv:1707.06050 [quant-ph]} \BibitemShut
  {NoStop}%
\bibitem [{\citenamefont {{Marletto}}\ and\ \citenamefont
  {{Vedral}}(2017)}]{Marletto:2017kzi}%
  \BibitemOpen
  \bibfield  {author} {\bibinfo {author} {\bibfnamefont {C.}~\bibnamefont
  {{Marletto}}}\ and\ \bibinfo {author} {\bibfnamefont {V.}~\bibnamefont
  {{Vedral}}},\ }\bibfield  {title} {\enquote {\bibinfo {title}
  {{Gravitationally Induced Entanglement between Two Massive Particles is
  Sufficient Evidence of Quantum Effects in Gravity}},}\ }\href {\doibase
  10.1103/PhysRevLett.119.240402} {\bibfield  {journal} {\bibinfo  {journal}
  {\prl}\ }\textbf {\bibinfo {volume} {119}},\ \bibinfo {eid} {240402}
  (\bibinfo {year} {2017})},\ \Eprint {http://arxiv.org/abs/1707.06036}
  {arXiv:1707.06036 [quant-ph]} \BibitemShut {NoStop}%
\bibitem [{\citenamefont {{Krisnanda}}\ \emph {et~al.}(2020)\citenamefont
  {{Krisnanda}}, \citenamefont {{Tham}}, \citenamefont {{Paternostro}},\ and\
  \citenamefont {{Paterek}}}]{Krisnanda:2019glc}%
  \BibitemOpen
  \bibfield  {author} {\bibinfo {author} {\bibfnamefont {T.}~\bibnamefont
  {{Krisnanda}}}, \bibinfo {author} {\bibfnamefont {G.~Y.}\ \bibnamefont
  {{Tham}}}, \bibinfo {author} {\bibfnamefont {M.}~\bibnamefont
  {{Paternostro}}}, \ and\ \bibinfo {author} {\bibfnamefont {T.}~\bibnamefont
  {{Paterek}}},\ }\bibfield  {title} {\enquote {\bibinfo {title} {{Observable
  quantum entanglement due to gravity}},}\ }\href {\doibase
  10.1038/s41534-020-0243-y} {\bibfield  {journal} {\bibinfo  {journal} {npj
  Quantum Information}\ }\textbf {\bibinfo {volume} {6}},\ \bibinfo {eid} {12}
  (\bibinfo {year} {2020})},\ \Eprint {http://arxiv.org/abs/1906.08808}
  {arXiv:1906.08808 [quant-ph]} \BibitemShut {NoStop}%
\bibitem [{\citenamefont {{Rijavec}}\ \emph {et~al.}(2021)\citenamefont
  {{Rijavec}}, \citenamefont {{Carlesso}}, \citenamefont {{Bassi}},
  \citenamefont {{Vedral}},\ and\ \citenamefont
  {{Marletto}}}]{Rijavec:2020qxd}%
  \BibitemOpen
  \bibfield  {author} {\bibinfo {author} {\bibfnamefont {S.}~\bibnamefont
  {{Rijavec}}}, \bibinfo {author} {\bibfnamefont {M.}~\bibnamefont
  {{Carlesso}}}, \bibinfo {author} {\bibfnamefont {A.}~\bibnamefont {{Bassi}}},
  \bibinfo {author} {\bibfnamefont {V.}~\bibnamefont {{Vedral}}}, \ and\
  \bibinfo {author} {\bibfnamefont {C.}~\bibnamefont {{Marletto}}},\ }\bibfield
   {title} {\enquote {\bibinfo {title} {{Decoherence effects in
  non-classicality tests of gravity}},}\ }\href {\doibase
  10.1088/1367-2630/abf3eb} {\bibfield  {journal} {\bibinfo  {journal} {New
  Journal of Physics}\ }\textbf {\bibinfo {volume} {23}},\ \bibinfo {eid}
  {043040} (\bibinfo {year} {2021})},\ \Eprint
  {http://arxiv.org/abs/2012.06230} {arXiv:2012.06230 [quant-ph]} \BibitemShut
  {NoStop}%
\bibitem [{\citenamefont {{Guerreiro}}\ \emph {et~al.}(2022)\citenamefont
  {{Guerreiro}}, \citenamefont {{Coradeschi}}, \citenamefont {{Frassino}},
  \citenamefont {{West}},\ and\ \citenamefont
  {{Schioppa}}}]{Guerreiro:2021qgk}%
  \BibitemOpen
  \bibfield  {author} {\bibinfo {author} {\bibfnamefont {T.}~\bibnamefont
  {{Guerreiro}}}, \bibinfo {author} {\bibfnamefont {F.}~\bibnamefont
  {{Coradeschi}}}, \bibinfo {author} {\bibfnamefont {A.~M.}\ \bibnamefont
  {{Frassino}}}, \bibinfo {author} {\bibfnamefont {J.~R.}\ \bibnamefont
  {{West}}}, \ and\ \bibinfo {author} {\bibfnamefont {E.~J.}\ \bibnamefont
  {{Schioppa}}},\ }\bibfield  {title} {\enquote {\bibinfo {title} {{Quantum
  signatures in nonlinear gravitational waves}},}\ }\href {\doibase
  10.22331/q-2022-12-19-879} {\bibfield  {journal} {\bibinfo  {journal}
  {Quantum}\ }\textbf {\bibinfo {volume} {6}},\ \bibinfo {pages} {879}
  (\bibinfo {year} {2022})},\ \Eprint {http://arxiv.org/abs/2111.01779}
  {arXiv:2111.01779 [gr-qc]} \BibitemShut {NoStop}%
\bibitem [{\citenamefont {{Penrose}}(1996)}]{Penrose1996on}%
  \BibitemOpen
  \bibfield  {author} {\bibinfo {author} {\bibfnamefont {R.}~\bibnamefont
  {{Penrose}}},\ }\bibfield  {title} {\enquote {\bibinfo {title} {{On Gravity's
  role in Quantum State Reduction}},}\ }\href {\doibase 10.1007/BF02105068}
  {\bibfield  {journal} {\bibinfo  {journal} {General Relativity and
  Gravitation}\ }\textbf {\bibinfo {volume} {28}},\ \bibinfo {pages} {581--600}
  (\bibinfo {year} {1996})}\BibitemShut {NoStop}%
\bibitem [{\citenamefont {{Bassi}}\ and\ \citenamefont
  {{Ulbricht}}(2014)}]{Bassi2014collapse}%
  \BibitemOpen
  \bibfield  {author} {\bibinfo {author} {\bibfnamefont {A.}~\bibnamefont
  {{Bassi}}}\ and\ \bibinfo {author} {\bibfnamefont {H.}~\bibnamefont
  {{Ulbricht}}},\ }\bibfield  {title} {\enquote {\bibinfo {title} {{Collapse
  models: from theoretical foundations to experimental verifications}},}\ }in\
  \href {\doibase 10.1088/1742-6596/504/1/012023} {\emph {\bibinfo {booktitle}
  {Journal of Physics Conference Series}}},\ \bibinfo {series} {Journal of
  Physics Conference Series}, Vol.\ \bibinfo {volume} {504}\ (\bibinfo {year}
  {2014})\ p.\ \bibinfo {pages} {012023},\ \Eprint
  {http://arxiv.org/abs/1401.6314} {arXiv:1401.6314 [quant-ph]} \BibitemShut
  {NoStop}%
\bibitem [{\citenamefont {{Bronstein}}(2012)}]{Bronstein2012republication}%
  \BibitemOpen
  \bibfield  {author} {\bibinfo {author} {\bibfnamefont {M.}~\bibnamefont
  {{Bronstein}}},\ }\bibfield  {title} {\enquote {\bibinfo {title}
  {{Republication of: Quantum theory of weak gravitational fields}},}\ }\href
  {\doibase 10.1007/s10714-011-1285-4} {\bibfield  {journal} {\bibinfo
  {journal} {General Relativity and Gravitation}\ }\textbf {\bibinfo {volume}
  {44}},\ \bibinfo {pages} {267--283} (\bibinfo {year} {2012})}\BibitemShut
  {NoStop}%
\bibitem [{\citenamefont {{Gorelik}}(2005)}]{Gorelik2005from}%
  \BibitemOpen
  \bibfield  {author} {\bibinfo {author} {\bibfnamefont {G.~E.}\ \bibnamefont
  {{Gorelik}}},\ }\bibfield  {title} {\enquote {\bibinfo {title} {{Matvei
  Bronstein and quantum gravity: 70th anniversary of the unsolved problem}},}\
  }\href {\doibase 10.1070/PU2005v048n10ABEH005820} {\bibfield  {journal}
  {\bibinfo  {journal} {Physics Uspekhi}\ }\textbf {\bibinfo {volume} {48}},\
  \bibinfo {pages} {1039--1053} (\bibinfo {year} {2005})}\BibitemShut {NoStop}%
\bibitem [{\citenamefont {{Gunnink}}\ \emph {et~al.}(2023)\citenamefont
  {{Gunnink}}, \citenamefont {{Mazumdar}}, \citenamefont {{Schut}},\ and\
  \citenamefont {{Toro{\v{s}}}}}]{Gunnink2023gravitational}%
  \BibitemOpen
  \bibfield  {author} {\bibinfo {author} {\bibfnamefont {F.}~\bibnamefont
  {{Gunnink}}}, \bibinfo {author} {\bibfnamefont {A.}~\bibnamefont
  {{Mazumdar}}}, \bibinfo {author} {\bibfnamefont {M.}~\bibnamefont {{Schut}}},
  \ and\ \bibinfo {author} {\bibfnamefont {M.}~\bibnamefont {{Toro{\v{s}}}}},\
  }\bibfield  {title} {\enquote {\bibinfo {title} {{Gravitational decoherence
  by the apparatus in the quantum-gravity-induced entanglement of masses}},}\
  }\href {\doibase 10.1088/1361-6382/ad0496} {\bibfield  {journal} {\bibinfo
  {journal} {Classical and Quantum Gravity}\ }\textbf {\bibinfo {volume}
  {40}},\ \bibinfo {eid} {235006} (\bibinfo {year} {2023})},\ \Eprint
  {http://arxiv.org/abs/2210.16919} {arXiv:2210.16919 [quant-ph]} \BibitemShut
  {NoStop}%
\bibitem [{\citenamefont {{Torrieri}}(2022)}]{Torrieri2022the}%
  \BibitemOpen
  \bibfield  {author} {\bibinfo {author} {\bibfnamefont {G.}~\bibnamefont
  {{Torrieri}}},\ }\bibfield  {title} {\enquote {\bibinfo {title} {{The
  equivalence principle and inertial-gravitational decoherence}},}\ }\href
  {\doibase 10.48550/arXiv.2210.08586} {\bibfield  {journal} {\bibinfo
  {journal} {arXiv e-prints}\ ,\ \bibinfo {eid} {arXiv:2210.08586}} (\bibinfo
  {year} {2022})},\ \Eprint {http://arxiv.org/abs/2210.08586} {arXiv:2210.08586
  [gr-qc]} \BibitemShut {NoStop}%
\bibitem [{\citenamefont {{Breuer}}\ and\ \citenamefont
  {{Petruccione}}(2001)}]{Breuer2001Destruction}%
  \BibitemOpen
  \bibfield  {author} {\bibinfo {author} {\bibfnamefont {H.-P.}\ \bibnamefont
  {{Breuer}}}\ and\ \bibinfo {author} {\bibfnamefont {F.}~\bibnamefont
  {{Petruccione}}},\ }\bibfield  {title} {\enquote {\bibinfo {title}
  {{Destruction of quantum coherence through emission of bremsstrahlung}},}\
  }\href {\doibase 10.1103/PhysRevA.63.032102} {\bibfield  {journal} {\bibinfo
  {journal} {\pra}\ }\textbf {\bibinfo {volume} {63}},\ \bibinfo {eid} {032102}
  (\bibinfo {year} {2001})}\BibitemShut {NoStop}%
\bibitem [{\citenamefont {Breuer}\ \emph {et~al.}(2002)\citenamefont {Breuer},
  \citenamefont {Petruccione} \emph {et~al.}}]{Breuer:2002pc}%
  \BibitemOpen
  \bibfield  {author} {\bibinfo {author} {\bibfnamefont {H.~P.}\ \bibnamefont
  {Breuer}}, \bibinfo {author} {\bibfnamefont {F.}~\bibnamefont {Petruccione}},
   \emph {et~al.},\ }\href@noop {} {\emph {\bibinfo {title} {The theory of open
  quantum systems}}}\ (\bibinfo  {publisher} {Oxford University Press},\
  \bibinfo {year} {2002})\BibitemShut {NoStop}%
\bibitem [{\citenamefont {{Zarei}}\ \emph {et~al.}(2021)\citenamefont
  {{Zarei}}, \citenamefont {{Bartolo}}, \citenamefont {{Bertacca}},
  \citenamefont {{Matarrese}},\ and\ \citenamefont
  {{Ricciardone}}}]{Zarei:2021dpb}%
  \BibitemOpen
  \bibfield  {author} {\bibinfo {author} {\bibfnamefont {M.}~\bibnamefont
  {{Zarei}}}, \bibinfo {author} {\bibfnamefont {N.}~\bibnamefont {{Bartolo}}},
  \bibinfo {author} {\bibfnamefont {D.}~\bibnamefont {{Bertacca}}}, \bibinfo
  {author} {\bibfnamefont {S.}~\bibnamefont {{Matarrese}}}, \ and\ \bibinfo
  {author} {\bibfnamefont {A.}~\bibnamefont {{Ricciardone}}},\ }\bibfield
  {title} {\enquote {\bibinfo {title} {{Non-Markovian open quantum system
  approach to the early Universe: Damping of gravitational waves by matter}},}\
  }\href {\doibase 10.1103/PhysRevD.104.083508} {\bibfield  {journal} {\bibinfo
   {journal} {\prd}\ }\textbf {\bibinfo {volume} {104}},\ \bibinfo {eid}
  {083508} (\bibinfo {year} {2021})},\ \Eprint
  {http://arxiv.org/abs/2104.04836} {arXiv:2104.04836 [astro-ph.CO]}
  \BibitemShut {NoStop}%
\bibitem [{\citenamefont {{Grishchuk}}\ and\ \citenamefont
  {{Sidorov}}(1990{\natexlab{b}})}]{grishchuk1990squeezed}%
  \BibitemOpen
  \bibfield  {author} {\bibinfo {author} {\bibfnamefont {L.~P.}\ \bibnamefont
  {{Grishchuk}}}\ and\ \bibinfo {author} {\bibfnamefont {Y.~V.}\ \bibnamefont
  {{Sidorov}}},\ }\bibfield  {title} {\enquote {\bibinfo {title} {{Squeezed
  quantum states of relic gravitons and primordial density fluctuations}},}\
  }\href {\doibase 10.1103/PhysRevD.42.3413} {\bibfield  {journal} {\bibinfo
  {journal} {\prd}\ }\textbf {\bibinfo {volume} {42}},\ \bibinfo {pages}
  {3413--3421} (\bibinfo {year} {1990}{\natexlab{b}})}\BibitemShut {NoStop}%
\bibitem [{\citenamefont {{Albrecht}}\ \emph {et~al.}(1994)\citenamefont
  {{Albrecht}}, \citenamefont {{Ferreira}}, \citenamefont {{Joyce}},\ and\
  \citenamefont {{Prokopec}}}]{albrecht1994inflation}%
  \BibitemOpen
  \bibfield  {author} {\bibinfo {author} {\bibfnamefont {A.}~\bibnamefont
  {{Albrecht}}}, \bibinfo {author} {\bibfnamefont {P.}~\bibnamefont
  {{Ferreira}}}, \bibinfo {author} {\bibfnamefont {M.}~\bibnamefont {{Joyce}}},
  \ and\ \bibinfo {author} {\bibfnamefont {T.}~\bibnamefont {{Prokopec}}},\
  }\bibfield  {title} {\enquote {\bibinfo {title} {{Inflation and squeezed
  quantum states}},}\ }\href {\doibase 10.1103/PhysRevD.50.4807} {\bibfield
  {journal} {\bibinfo  {journal} {\prd}\ }\textbf {\bibinfo {volume} {50}},\
  \bibinfo {pages} {4807--4820} (\bibinfo {year} {1994})},\ \Eprint
  {http://arxiv.org/abs/astro-ph/9303001} {arXiv:astro-ph/9303001 [astro-ph]}
  \BibitemShut {NoStop}%
\bibitem [{\citenamefont {{Kanno}}\ and\ \citenamefont
  {{Soda}}(2022)}]{kanno2021squeezed}%
  \BibitemOpen
  \bibfield  {author} {\bibinfo {author} {\bibfnamefont {S.}~\bibnamefont
  {{Kanno}}}\ and\ \bibinfo {author} {\bibfnamefont {J.}~\bibnamefont
  {{Soda}}},\ }\bibfield  {title} {\enquote {\bibinfo {title} {{Squeezed
  quantum states of graviton and axion in the universe}},}\ }\href {\doibase
  10.1142/S0218271822500985} {\bibfield  {journal} {\bibinfo  {journal}
  {International Journal of Modern Physics D}\ }\textbf {\bibinfo {volume}
  {31}},\ \bibinfo {eid} {2250098} (\bibinfo {year} {2022})},\ \Eprint
  {http://arxiv.org/abs/2112.14496} {arXiv:2112.14496 [gr-qc]} \BibitemShut
  {NoStop}%
\bibitem [{\citenamefont {{Du}}\ \emph {et~al.}(2022)\citenamefont {{Du}},
  \citenamefont {{Murgui}}, \citenamefont {{Pardo}}, \citenamefont {{Wang}},\
  and\ \citenamefont {{Zurek}}}]{D2022}%
  \BibitemOpen
  \bibfield  {author} {\bibinfo {author} {\bibfnamefont {Y.}~\bibnamefont
  {{Du}}}, \bibinfo {author} {\bibfnamefont {C.}~\bibnamefont {{Murgui}}},
  \bibinfo {author} {\bibfnamefont {K.}~\bibnamefont {{Pardo}}}, \bibinfo
  {author} {\bibfnamefont {Y.}~\bibnamefont {{Wang}}}, \ and\ \bibinfo {author}
  {\bibfnamefont {K.~M.}\ \bibnamefont {{Zurek}}},\ }\bibfield  {title}
  {\enquote {\bibinfo {title} {{Atom interferometer tests of dark matter}},}\
  }\href {\doibase 10.1103/PhysRevD.106.095041} {\bibfield  {journal} {\bibinfo
   {journal} {\prd}\ }\textbf {\bibinfo {volume} {106}},\ \bibinfo {eid}
  {095041} (\bibinfo {year} {2022})},\ \Eprint
  {http://arxiv.org/abs/2205.13546} {arXiv:2205.13546 [hep-ph]} \BibitemShut
  {NoStop}%
\bibitem [{\citenamefont {{Thrane}}\ and\ \citenamefont
  {{Romano}}(2013)}]{thrane2013sensitivity}%
  \BibitemOpen
  \bibfield  {author} {\bibinfo {author} {\bibfnamefont {E.}~\bibnamefont
  {{Thrane}}}\ and\ \bibinfo {author} {\bibfnamefont {J.~D.}\ \bibnamefont
  {{Romano}}},\ }\bibfield  {title} {\enquote {\bibinfo {title} {{Sensitivity
  curves for searches for gravitational-wave backgrounds}},}\ }\href {\doibase
  10.1103/PhysRevD.88.124032} {\bibfield  {journal} {\bibinfo  {journal}
  {\prd}\ }\textbf {\bibinfo {volume} {88}},\ \bibinfo {eid} {124032} (\bibinfo
  {year} {2013})},\ \Eprint {http://arxiv.org/abs/1310.5300} {arXiv:1310.5300
  [astro-ph.IM]} \BibitemShut {NoStop}%
\bibitem [{\citenamefont {{Scala}}\ \emph {et~al.}(2013)\citenamefont
  {{Scala}}, \citenamefont {{Kim}}, \citenamefont {{Morley}}, \citenamefont
  {{Barker}},\ and\ \citenamefont {{Bose}}}]{Scala:2013}%
  \BibitemOpen
  \bibfield  {author} {\bibinfo {author} {\bibfnamefont {M.}~\bibnamefont
  {{Scala}}}, \bibinfo {author} {\bibfnamefont {M.~S.}\ \bibnamefont {{Kim}}},
  \bibinfo {author} {\bibfnamefont {G.~W.}\ \bibnamefont {{Morley}}}, \bibinfo
  {author} {\bibfnamefont {P.~F.}\ \bibnamefont {{Barker}}}, \ and\ \bibinfo
  {author} {\bibfnamefont {S.}~\bibnamefont {{Bose}}},\ }\bibfield  {title}
  {\enquote {\bibinfo {title} {{Matter-Wave Interferometry of a Levitated
  Thermal Nano-Oscillator Induced and Probed by a Spin}},}\ }\href {\doibase
  10.1103/PhysRevLett.111.180403} {\bibfield  {journal} {\bibinfo  {journal}
  {\prl}\ }\textbf {\bibinfo {volume} {111}},\ \bibinfo {eid} {180403}
  (\bibinfo {year} {2013})},\ \Eprint {http://arxiv.org/abs/1306.6579}
  {arXiv:1306.6579 [quant-ph]} \BibitemShut {NoStop}%
\bibitem [{\citenamefont {{Pedernales}}\ \emph {et~al.}(2020)\citenamefont
  {{Pedernales}}, \citenamefont {{Morley}},\ and\ \citenamefont
  {{Plenio}}}]{Pedernales2020}%
  \BibitemOpen
  \bibfield  {author} {\bibinfo {author} {\bibfnamefont {J.~S.}\ \bibnamefont
  {{Pedernales}}}, \bibinfo {author} {\bibfnamefont {G.~W.}\ \bibnamefont
  {{Morley}}}, \ and\ \bibinfo {author} {\bibfnamefont {M.~B.}\ \bibnamefont
  {{Plenio}}},\ }\bibfield  {title} {\enquote {\bibinfo {title} {{Motional
  Dynamical Decoupling for Interferometry with Macroscopic Particles}},}\
  }\href {\doibase 10.1103/PhysRevLett.125.023602} {\bibfield  {journal}
  {\bibinfo  {journal} {\prl}\ }\textbf {\bibinfo {volume} {125}},\ \bibinfo
  {eid} {023602} (\bibinfo {year} {2020})},\ \Eprint
  {http://arxiv.org/abs/1906.00835} {arXiv:1906.00835} \BibitemShut {NoStop}%
\bibitem [{\citenamefont {{Bose}}\ and\ \citenamefont
  {{Morley}}(2018)}]{bose2018}%
  \BibitemOpen
  \bibfield  {author} {\bibinfo {author} {\bibfnamefont {S.}~\bibnamefont
  {{Bose}}}\ and\ \bibinfo {author} {\bibfnamefont {G.~W.}\ \bibnamefont
  {{Morley}}},\ }\bibfield  {title} {\enquote {\bibinfo {title} {{Matter and
  spin superposition in vacuum experiment (MASSIVE)}},}\ }\href@noop {}
  {\bibfield  {journal} {\bibinfo  {journal} {arXiv e-prints}\ ,\ \bibinfo
  {eid} {arXiv:1810.07045}} (\bibinfo {year} {2018})},\ \Eprint
  {http://arxiv.org/abs/1810.07045} {arXiv:1810.07045 [quant-ph]} \BibitemShut
  {NoStop}%
\bibitem [{\citenamefont {{Bennett}}\ \emph {et~al.}(2012)\citenamefont
  {{Bennett}}, \citenamefont {{Kolkowitz}}, \citenamefont {{Unterreithmeier}},
  \citenamefont {{Rabl}}, \citenamefont {{Bleszynski Jayich}}, \citenamefont
  {{Harris}},\ and\ \citenamefont {{Lukin}}}]{Bennett2012}%
  \BibitemOpen
  \bibfield  {author} {\bibinfo {author} {\bibfnamefont {S.~D.}\ \bibnamefont
  {{Bennett}}}, \bibinfo {author} {\bibfnamefont {S.}~\bibnamefont
  {{Kolkowitz}}}, \bibinfo {author} {\bibfnamefont {Q.~P.}\ \bibnamefont
  {{Unterreithmeier}}}, \bibinfo {author} {\bibfnamefont {P.}~\bibnamefont
  {{Rabl}}}, \bibinfo {author} {\bibfnamefont {A.~C.}\ \bibnamefont
  {{Bleszynski Jayich}}}, \bibinfo {author} {\bibfnamefont {J.~G.~E.}\
  \bibnamefont {{Harris}}}, \ and\ \bibinfo {author} {\bibfnamefont {M.~D.}\
  \bibnamefont {{Lukin}}},\ }\bibfield  {title} {\enquote {\bibinfo {title}
  {{Measuring mechanical motion with a single spin}},}\ }\href {\doibase
  10.1088/1367-2630/14/12/125004} {\bibfield  {journal} {\bibinfo  {journal}
  {New Journal of Physics}\ }\textbf {\bibinfo {volume} {14}},\ \bibinfo {eid}
  {125004} (\bibinfo {year} {2012})},\ \Eprint {http://arxiv.org/abs/1205.6740}
  {arXiv:1205.6740 [cond-mat.mes-hall]} \BibitemShut {NoStop}%
\bibitem [{\citenamefont {{Asadian}}\ \emph {et~al.}(2014)\citenamefont
  {{Asadian}}, \citenamefont {{Brukner}},\ and\ \citenamefont
  {{Rabl}}}]{Asadian2014}%
  \BibitemOpen
  \bibfield  {author} {\bibinfo {author} {\bibfnamefont {A.}~\bibnamefont
  {{Asadian}}}, \bibinfo {author} {\bibfnamefont {C.}~\bibnamefont
  {{Brukner}}}, \ and\ \bibinfo {author} {\bibfnamefont {P.}~\bibnamefont
  {{Rabl}}},\ }\bibfield  {title} {\enquote {\bibinfo {title} {{Probing
  Macroscopic Realism via Ramsey Correlation Measurements}},}\ }\href {\doibase
  10.1103/PhysRevLett.112.190402} {\bibfield  {journal} {\bibinfo  {journal}
  {\prl}\ }\textbf {\bibinfo {volume} {112}},\ \bibinfo {eid} {190402}
  (\bibinfo {year} {2014})},\ \Eprint {http://arxiv.org/abs/1309.2229}
  {arXiv:1309.2229 [quant-ph]} \BibitemShut {NoStop}%
\bibitem [{\citenamefont {{Roati}}\ \emph {et~al.}(2004)\citenamefont
  {{Roati}}, \citenamefont {{de Mirandes}}, \citenamefont {{Ferlaino}},
  \citenamefont {{Ott}}, \citenamefont {{Modugno}},\ and\ \citenamefont
  {{Inguscio}}}]{Roati2004}%
  \BibitemOpen
  \bibfield  {author} {\bibinfo {author} {\bibfnamefont {G.}~\bibnamefont
  {{Roati}}}, \bibinfo {author} {\bibfnamefont {E.}~\bibnamefont {{de
  Mirandes}}}, \bibinfo {author} {\bibfnamefont {F.}~\bibnamefont
  {{Ferlaino}}}, \bibinfo {author} {\bibfnamefont {H.}~\bibnamefont {{Ott}}},
  \bibinfo {author} {\bibfnamefont {G.}~\bibnamefont {{Modugno}}}, \ and\
  \bibinfo {author} {\bibfnamefont {M.}~\bibnamefont {{Inguscio}}},\ }\bibfield
   {title} {\enquote {\bibinfo {title} {{Atom Interferometry with Trapped Fermi
  Gases}},}\ }\href {\doibase 10.1103/PhysRevLett.92.230402} {\bibfield
  {journal} {\bibinfo  {journal} {\prl}\ }\textbf {\bibinfo {volume} {92}},\
  \bibinfo {eid} {230402} (\bibinfo {year} {2004})},\ \Eprint
  {http://arxiv.org/abs/cond-mat/0402328} {arXiv:cond-mat/0402328
  [cond-mat.soft]} \BibitemShut {NoStop}%
\bibitem [{\citenamefont {{Modugno}}\ \emph {et~al.}(2004)\citenamefont
  {{Modugno}}, \citenamefont {{de Mirandes}}, \citenamefont {{Ferlaino}},
  \citenamefont {{Ott}}, \citenamefont {{Roati}},\ and\ \citenamefont
  {{Inguscio}}}]{Modugno2004}%
  \BibitemOpen
  \bibfield  {author} {\bibinfo {author} {\bibfnamefont {G.}~\bibnamefont
  {{Modugno}}}, \bibinfo {author} {\bibfnamefont {E.}~\bibnamefont {{de
  Mirandes}}}, \bibinfo {author} {\bibfnamefont {F.}~\bibnamefont
  {{Ferlaino}}}, \bibinfo {author} {\bibfnamefont {H.}~\bibnamefont {{Ott}}},
  \bibinfo {author} {\bibfnamefont {G.}~\bibnamefont {{Roati}}}, \ and\
  \bibinfo {author} {\bibfnamefont {M.}~\bibnamefont {{Inguscio}}},\ }\bibfield
   {title} {\enquote {\bibinfo {title} {{Atom interferometry in a vertical
  optical lattice}},}\ }\href {\doibase 10.1002/prop.200410187} {\bibfield
  {journal} {\bibinfo  {journal} {Fortschritte der Physik}\ }\textbf {\bibinfo
  {volume} {52}},\ \bibinfo {pages} {1173--1179} (\bibinfo {year} {2004})},\
  \Eprint {http://arxiv.org/abs/physics/0411097} {arXiv:physics/0411097
  [physics.atom-ph]} \BibitemShut {NoStop}%
\bibitem [{\citenamefont {{Bartolo}}\ \emph {et~al.}(2018)\citenamefont
  {{Bartolo}}, \citenamefont {{Hoseinpour}}, \citenamefont {{Orlando}},
  \citenamefont {{Matarrese}},\ and\ \citenamefont
  {{Zarei}}}]{Bartolo:2018igk}%
  \BibitemOpen
  \bibfield  {author} {\bibinfo {author} {\bibfnamefont {N.}~\bibnamefont
  {{Bartolo}}}, \bibinfo {author} {\bibfnamefont {A.}~\bibnamefont
  {{Hoseinpour}}}, \bibinfo {author} {\bibfnamefont {G.}~\bibnamefont
  {{Orlando}}}, \bibinfo {author} {\bibfnamefont {S.}~\bibnamefont
  {{Matarrese}}}, \ and\ \bibinfo {author} {\bibfnamefont {M.}~\bibnamefont
  {{Zarei}}},\ }\bibfield  {title} {\enquote {\bibinfo {title}
  {{Photon-graviton scattering: A new way to detect anisotropic gravitational
  waves?}}}\ }\href {\doibase 10.1103/PhysRevD.98.023518} {\bibfield  {journal}
  {\bibinfo  {journal} {\prd}\ }\textbf {\bibinfo {volume} {98}},\ \bibinfo
  {eid} {023518} (\bibinfo {year} {2018})},\ \Eprint
  {http://arxiv.org/abs/1804.06298} {arXiv:1804.06298 [gr-qc]} \BibitemShut
  {NoStop}%
\bibitem [{\citenamefont {{Allen}}\ and\ \citenamefont
  {{Romano}}(1999)}]{Allen:1997ad}%
  \BibitemOpen
  \bibfield  {author} {\bibinfo {author} {\bibfnamefont {B.}~\bibnamefont
  {{Allen}}}\ and\ \bibinfo {author} {\bibfnamefont {J.~D.}\ \bibnamefont
  {{Romano}}},\ }\bibfield  {title} {\enquote {\bibinfo {title} {{Detecting a
  stochastic background of gravitational radiation: Signal processing
  strategies and sensitivities}},}\ }\href {\doibase
  10.1103/PhysRevD.59.102001} {\bibfield  {journal} {\bibinfo  {journal}
  {\prd}\ }\textbf {\bibinfo {volume} {59}},\ \bibinfo {eid} {102001} (\bibinfo
  {year} {1999})},\ \Eprint {http://arxiv.org/abs/gr-qc/9710117}
  {arXiv:gr-qc/9710117 [gr-qc]} \BibitemShut {NoStop}%
\bibitem [{\citenamefont {{Romano}}\ and\ \citenamefont
  {{Cornish}}(2017)}]{Romano:2016dpx}%
  \BibitemOpen
  \bibfield  {author} {\bibinfo {author} {\bibfnamefont {J.~D.}\ \bibnamefont
  {{Romano}}}\ and\ \bibinfo {author} {\bibfnamefont {N.~J.}\ \bibnamefont
  {{Cornish}}},\ }\bibfield  {title} {\enquote {\bibinfo {title} {{Detection
  methods for stochastic gravitational-wave backgrounds: a unified
  treatment}},}\ }\href {\doibase 10.1007/s41114-017-0004-1} {\bibfield
  {journal} {\bibinfo  {journal} {Living Reviews in Relativity}\ }\textbf
  {\bibinfo {volume} {20}},\ \bibinfo {eid} {2} (\bibinfo {year} {2017})},\
  \Eprint {http://arxiv.org/abs/1608.06889} {arXiv:1608.06889 [gr-qc]}
  \BibitemShut {NoStop}%
\bibitem [{\citenamefont {{Toro{\v{s}}}}\ \emph {et~al.}(2021)\citenamefont
  {{Toro{\v{s}}}}, \citenamefont {{van de Kamp}}, \citenamefont {{Marshman}},
  \citenamefont {{Kim}}, \citenamefont {{Mazumdar}},\ and\ \citenamefont
  {{Bose}}}]{toros2020}%
  \BibitemOpen
  \bibfield  {author} {\bibinfo {author} {\bibfnamefont {M.}~\bibnamefont
  {{Toro{\v{s}}}}}, \bibinfo {author} {\bibfnamefont {T.~W.}\ \bibnamefont
  {{van de Kamp}}}, \bibinfo {author} {\bibfnamefont {R.~J.}\ \bibnamefont
  {{Marshman}}}, \bibinfo {author} {\bibfnamefont {M.~S.}\ \bibnamefont
  {{Kim}}}, \bibinfo {author} {\bibfnamefont {A.}~\bibnamefont {{Mazumdar}}}, \
  and\ \bibinfo {author} {\bibfnamefont {S.}~\bibnamefont {{Bose}}},\
  }\bibfield  {title} {\enquote {\bibinfo {title} {{Relative acceleration noise
  mitigation for nanocrystal matter-wave interferometry: Applications to
  entangling masses via quantum gravity}},}\ }\href {\doibase
  10.1103/PhysRevResearch.3.023178} {\bibfield  {journal} {\bibinfo  {journal}
  {Physical Review Research}\ }\textbf {\bibinfo {volume} {3}},\ \bibinfo {eid}
  {023178} (\bibinfo {year} {2021})},\ \Eprint
  {http://arxiv.org/abs/2007.15029} {arXiv:2007.15029 [gr-qc]} \BibitemShut
  {NoStop}%
\end{thebibliography}%

\end{document}